\documentclass[aps,twocolumn,secnumarabic,superscriptaddress,balancelastpage,amsmath,amssymb,nofootinbib,titlepage]{revtex4-2}
\usepackage[english]{babel}
\usepackage[utf8]{inputenc}
\usepackage{amsmath}
\usepackage{graphicx}
\usepackage[colorinlistoftodos]{todonotes}
\usepackage{bibentry}
\usepackage{url}
\usepackage{rotating}
\usepackage{float}
\usepackage{color}
\usepackage[sort&compress]{natbib}
\usepackage{hyperref}\hypersetup{colorlinks=true}
\usepackage[caption=false]{subfig}
\usepackage{empheq}

\newcommand{\Hh}{\hat{H}}
\newcommand{\hh}{\hat{h}}

\newcommand{\psih}{\hat{\psi}}
\newcommand{\br}{{\bf r}}

\newcommand{\bsigma}{\bm{\sigma}}

\newcommand{\sgn}{{\rm sgn}}

\newcommand{\be}{\begin{equation}}
\newcommand{\ee}{\end{equation}}
\newcommand{\bea}{\begin{eqnarray}}
\newcommand{\eea}{\end{eqnarray}}
\newcommand{\bse}{\begin{subequations}}
\newcommand{\ese}{\end{subequations}}

\DeclareMathOperator{\sech}{sech}
\begin{document}
\title{Unruh Effect and Takagi's Statistics Inversion in Strained Graphene}

\author{Anshuman Bhardwaj}
\email{abhard4@lsu.edu}
  \author{Daniel E. Sheehy}
\email{sheehy@lsu.edu}
  \affiliation{Department of Physics and Astronomy, Louisiana State University, Baton Rouge, LA 70803 USA}

\date{\today}

\begin{abstract}
We present a theoretical study of how a spatially-varying quasiparticle velocity
in honeycomb lattices, achievable using strained graphene or in engineered cold-atom optical lattices
that have a spatial dependence to the local tunneling amplitude, 
can yield the Rindler 
Hamiltonian embodying an observer
accelerating in Minkowski spacetime.  Within this setup, 
a sudden switch-on of the spatially-varying tunneling (or strain)
yields a spontaneous production of electron-hole pairs, an analogue
version of the Unruh effect characterized by the Unruh temperature.
We discuss how this thermal behavior, along with Takagi's statistics inversion, can manifest themselves in photo-emission and scanning tunneling microscopy experiments. 
We also calculate the average electronic conductivity and find that it grows linearly with frequency $\omega$.
Finally, we find that the total system energy at zero environment temperature looks like Planck's blackbody result for photons due to the aforementioned statistics inversion, whereas for an initial thermally excited state of fermions, the total internal energy undergoes stimulated particle reduction. 
\end{abstract}

\maketitle 

\section{Introduction}
Quantum field theory in curved spacetime \cite{Birrell:1982ix,Parker and Toms,Fulling:1989nb,Wald:1995yp,Mukhanov:2007zz,Fabbri:2005mw} is an exciting arena 
in which two cornerstones of modern physics, quantum field theory and general relativity, merge to produce surprising results. One classic prediction at this crossroads is that a quantum field in an initial vacuum state, under the influence of spacetime curvature (or gravity), leads to a spontaneous generation of particles associated with that field. This was first realized by Schrodinger \cite{Schrodinger:1939} in the context of relativistic quantum mechanics in an expanding universe and later by Parker \cite{Parker:1966,Parker:1968mv} who independently showed this in the context of general quantum fields in cosmological spacetimes. One such class of spacetimes is the one experienced by an accelerating observer: the Rindler spacetime {\cite{Rindler:1966zz}}. However, this spacetime is special because it creates particles with a thermal spectrum {\cite{Fulling:1972md,Davies:1974th,Unruh:1976db}}, i.e. an accelerating (or Rindler) observer sees the Minkowski (or flat) spacetime vacuum as a thermal bath of particles. This phenomenon is called the Fulling-Davies-Unruh effect (also known as the Unruh effect).  Here, the thermality emerges due to two reasons. The first is the appearance of a horizon that splits the entire spacetime into two mutually inaccessible regions (corresponding to observers accelerating in opposite directions) and thus vacuum expectation values in one region lead to tracing over the degrees of freedom of the other region, thus yielding a mixed state. The second reason is that the response function of an accelerating particle detector follows the principle of detailed balance, or in other words satisfies the Kubo-Martin-Schwinger (KMS) condition {\cite{Kubo:1957,Martin & Schwinger:1959}}, which is a sufficient condition for a spectrum to be called thermal. 

Similar horizons and therefore their associated thermal behavior also emerge in other spacetimes, such as black holes {\cite{Hawking:1974rv,Hawking:1974sw}} where this behavior is known as Hawking radiation, and the Gibbons-Hawking effect in de-Sitter cosmologies {\cite{Gibbons:1977mu}}. A surprising result that appears here is that the 
power spectrum, which depends on the density of states and the statistics, is sensitive to the dimensions of spacetime. In odd spacetime dimensions, the power spectrum of fermions has a Bose-Einstein distribution, whereas bosons follow a Fermi-Dirac
distribution. This is the well-known `apparent inversion of statistics' due to Takagi {\cite{Takagi:1986kn}} which is linked to the violation of Huygens' principle in odd spacetime dimensions {\cite{Ooguri:1985nv,Unruh:1986tc,Terashima:1999xp,Sriramkumar:2002nt,Sriramkumar:2002dn,Pascazio_Huygens,Arrechea:2021szl}}. 

There have been various proposals to detect the Unruh effect in accelerating systems~\cite{Kalinski:2005,Crispino:2007eb,Martin-Martinez:2010gnz,Nation:2011dka}, for example using Bose-Einstein condensates~\cite{Retzker:2008,Gooding:2020scc}. However, observing this effect is challenging as an acceleration of about $10^{21}{\rm m}/{\rm s}^2$ is required to generate a temperature of $1$K {\cite{Mukhanov:2007zz}} which is likely beyond the reach of current technology. In such a situation, analogue gravity~\cite{Barcelo:2005fc} offers an alternative arena for observing relativistic phenomena, in which  condensed matter or cold atom systems are
engineered to mimic the behavior of relativistic systems. This area emerged in 1981 when Unruh showed {\cite{Unruh:1980cg}} how water ripples in a draining bathtub can mimic the Klein-Gordon equation for a scalar field near a black hole horizon. This led to the prediction of analogue Hawking radiation which was realized in a series of experiments \cite{Philbin:2007ji,Belgiorno:2010wn,Weinfurtner:2010nu,Steinhauer:2015saa}. On the other hand, particle creation in the context of the inflationary early universe was recently observed in toroidal Bose-Einstein condensates \cite{Eckel:2017uqx,Banik:2021xjn} and studied theoretically in Refs.~\cite{Llorente:2019rbs,Bhardwaj:2020ndh,Eckel:2020qee}. More recently, it has been proposed in Refs.~\cite{Ghorashi:2020,Davis:2022}, that analog gravitational lensing could be realized in Dirac materials.


Such analogue platforms can be used to mimic the Unruh effect, as was recently  observed in Bose-Einstein condensates \cite{Hu:2018psq} by modulating the scattering length that determines the interactions between ultracold bosonic atoms. Various proposals have also been made to detect the  analogue Unruh effect in ultracold Fermi gases in square lattices {\cite{Boada:2010sh,Rodriguez-Laguna:2016kri,Kosior:2018vgx}}, in graphene {\cite{Iorio:2011yz,Iorio:2013ifa,Cvetic:2012vg}}, in quantum hall systems \cite{Hegde:2018xub,Subramanyan:2020fmx}, and in Weyl semi-metals \cite{Volovik:2016kid}.

Here our main interest is in exploring analogue Rindler physics, and the analogue Unruh effect, in 
graphene and related cold-atom systems (i.e., fermionic atoms in honeycomb lattices. Indeed,
the status of graphene as an analogue relativistic system has been long recognized~\cite{Wallace:1947,Semenoff}, and the fact that graphene's low-energy excitations obey the Dirac equation was established even from
the earliest experimental work on these systems~\cite{Novoselov 2004,Novoselov 2005}.  As is well
known, the effective
\lq\lq speed of light\rq\rq\ 
characterizing the Dirac quasiparticles in graphene takes a value $v\simeq c/300$ (with
$c$ the actual speed of light).  To achieve the Rindler Hamiltonian in graphene requires engineering 
a spatial variation in $v$ along one direction.

In this paper, our aim is to discuss how the Unruh effect would be manifested in honeycomb systems
such as mechanically strained graphene or in an appropriately engineered cold atom optical lattice
system \cite{Tarruell:2012zz,Soltan-Panahi 2011,Soltan-Panahi 2012,Lee:2009,Li:2016}. In either case what is needed is a spatial variation in the local tunneling matrix
elements between sites.  
The basic idea is to start with unstrained graphene, in equilibrium at low temperature $T$ 
(that we will usually assume to be $T=0$).  As mentioned above, fermionic excitations in unstrained graphene obey
the conventional Dirac equation, i.e., the Dirac equation in Minkowski (flat) spacetime.  
The next step is to suddenly switch on the strain field, changing the system Hamiltonian to
the Rindler Hamiltonian, with excitations described by a Rindler Dirac equation.  The Unruh effect
emerges because a vacuum initial (Minkowski) state becomes, after the strain, an effective
thermal distribution of Rindler quasiparticles characterized by the strain-dependent Unruh temperature.  


Earlier theoretical work by Rodr\'iguez-Laguna and collaborators showed~\cite{Rodriguez-Laguna:2016kri},
in the context of square optical lattices, that such a sudden quench should indeed yield the Unruh 
effect, provided that the timescale of the switching process is much faster than the timescale at which the electron dynamics operates (governed by the inverse tunneling rate). Here we assume the switching on is
sufficiently rapid so that, invoking the sudden approximation of quantum mechanics, the correct
procedure is to obtain observables by calculating the expectation values of operators in the strained system with respect to  states of the unstrained lattice (i.e., the Minkowski
vacuum or, at finite real temperature, a Fermi gas of Dirac quasiparticles and holes).


\begin{figure}[ht!]
	\begin{center}
		\includegraphics[width=1.0\columnwidth]{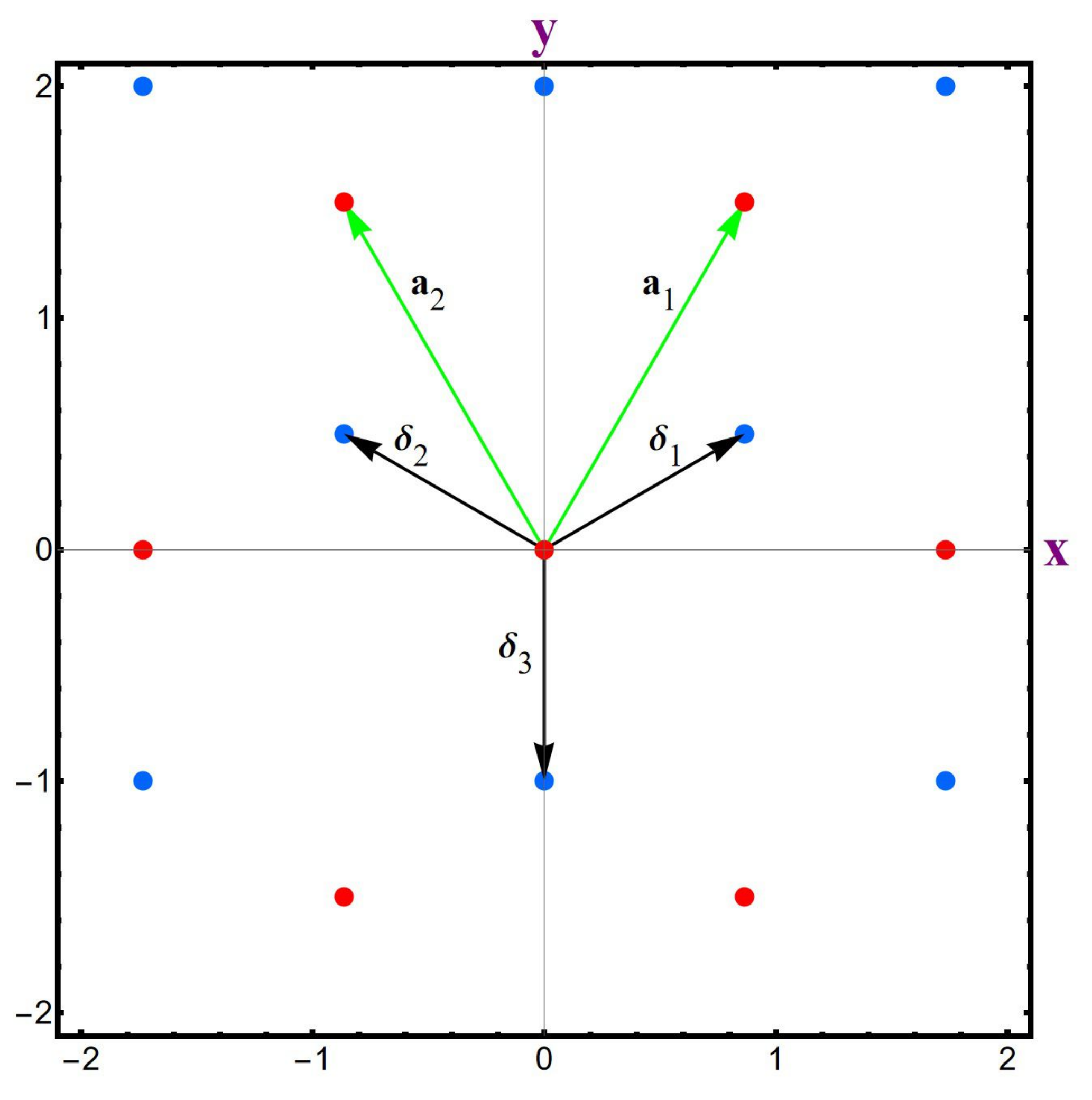}
	\end{center}
	\caption{(Color Online) The honeycomb lattice of graphene where the carbon atoms in red color for the Bravais lattice with primitive lattice vectors $\boldsymbol{a}_{1}=a(\sqrt{3}/2,3/2)$ and $\boldsymbol{a}_{2}=a(-\sqrt{3}/2,3/2)$, connected to the nearest neighbor carbon atoms shown in blue by $\boldsymbol{\delta}_{i}$, $i=1,2,3$, as defined in Eq.~(\ref{Nearest Neighbor Vectors}).}
	\label{Honeycomb}
\end{figure}

The rest of the paper is organized as follows. In Sec.~\ref{SEC:two}, we describe how the Rindler Hamiltonian can be realized for low-energy and long wavelength fermions in mechanically strained graphene.  
Since the basic effect relies only on engineering a spatially-varying tunneling matrix element, we expect it should be similarly possible to engineer the Rindler Hamiltonian in cold-atom systems.
In Sec.~\ref{SEC:three}, we revisit the Hamiltonian for fermions in flat spacetime (or flat graphene sheet) and identify the
normal modes of this system that correspond to particle and hole excitations.   In Sec.~\ref{SEC:four}, we derive the Dirac equation due to the Rindler Hamiltonian, obtaining a similar mode expansion for the strained case. In Sec.~\ref{SEC:five}, we use the mode expansions in flat and strained (Rindler) honeycomb lattices to 
derive how a sudden strain can induce spontaneous electron-hole creation with an emergent Fermi-Dirac distribution, which is the analogue Unruh effect. 
In Sec.~\ref{SEC:six} we analyze the Green's functions after such a sudden strain, showing how signatures
of the analogue Unruh effect may be measured in observables such as photoemission spectroscopy and
scanning tunneling microscopy and how 
the form of the emergent
thermality is connected to the violation of Huygens' principle. 
In Sec.~\ref{SEC:seven}, we study the frequency-dependent optical conductivity of this system, which we find to increase approximately linearly with 
increasing frequency, in contrast to flat graphene, where it is known to be nearly constant (i.e., frequency-independent)
~\cite{Nair2008,Mishchenko2008,Sheehy2009,Link2016}.
In Sec.~\ref{SEC:eight}, we discuss the effects of this sudden switching-on of the Rindler Hamiltonian on the total internal energy of fermions 
at finite environment temperature. 
In Sec.~\ref{SEC:nine} we provide brief concluding remarks. In Appendix \ref{SEC:Appendix Dirac Eqn}, we give details  on the Dirac equation in curved spacetime, and in Appendix \ref{SEC:Appendix Horizon Graphene}, we give details of how a Rindler horizon forms in strained graphene.

\section{Creating the Rindler Hamiltonian}
\label{SEC:two}
In this section, we will show how the Rindler Hamiltonian can be realized via graphene with a spatially-varying
strain that yields a Hamiltonian with a spatially-varying Fermi velocity.  
This is in contrast to 
the low-energy theory of conventional graphene that exhibits a spatially-uniform Fermi velocity.

To see how such a spatially-varying Fermi velocity can be engineered, we start with the tight binding
Hamiltonian for graphene which involves ($\pi$ orbital) electrons
hopping from carbon atoms in the $A$ sub-lattice to their nearest
neighboring $B$ carbon atoms (as shown in Fig.~\ref{Honeycomb}), and vice versa:
\begin{equation}
  \label{Tight Binding Hamiltonian}
  \hat{H} =
  -\sum_{\boldsymbol{R}_{j},n} t_{\boldsymbol{R}_{j},n} \Big[
    \hat{a}^{\dagger}_{\boldsymbol{R}_{j}}\hat{b}_{\boldsymbol{R}_{j}+\boldsymbol{\delta}_{n}} +
    \hat{b}^{\dagger}_{\boldsymbol{R}_{j}+\boldsymbol{\delta}_{n}}\hat{a}_{\boldsymbol{R}_{j}}\Big],
\end{equation}
where $\boldsymbol{R}_{j}$ labels the Bravais lattice points formed by
the $A$-atoms, and index $n$ denotes the three nearest neighboring $B$
atoms.  Here, the $\hat{a}$ and $\hat{b}$ operators annihilate fermions
on the $A$ and $B$ sublattices, respectively, with hopping amplitude
$t_{\boldsymbol{R}_{j},n}$ (that we have taken to be real).  The nearest neighbor vectors
$\boldsymbol{\delta}_{n}$ joining the $A$ and $B$ atoms are as
follows:
\begin{equation}\label{Nearest Neighbor Vectors}
  \boldsymbol{\delta}_{1}=a\bigg(\frac{\sqrt{3}}{2},\frac{1}{2}\bigg),
  \,\,\boldsymbol{\delta}_{2}=a\bigg(\frac{-\sqrt{3}}{2},\frac{1}{2}\bigg),
  \,\,\boldsymbol{\delta}_{3}=a\big(0,-1\big),
\end{equation}
 with $a$ the nearest-neighbor carbon distance.
When a graphene sheet undergoes a mechanical strain, with
$u_{ij}\equiv\frac{1}{2}(\partial_{i}u_{j}+\partial_{j}u_{i})$ being
the strain tensor, the distance between two carbon atoms changes and
thus the hopping amplitude gets adjusted accordingly. For perturbative
strains, we can then Taylor expand the hopping amplitude as follows
{\cite{deJuan:2012hxm}}:
\be
  \label{Taylor Expand Hopping}
  t_{\boldsymbol{R}_{j},n} = t_{0}\Big[1 - \beta\Delta u^{(1)}_{n} - \beta\Delta u^{(2)}_{n}\Big],
  \ee
  with
  \bea
  \Delta u^{(1)}_{n} &=& \frac{\delta_{n}^{i}\delta_{n}^{j}}{a^{2}}u_{ij},
\\
  \Delta u^{(2)}_{n} &=& \frac{\delta_{n}^{i}\delta_{n}^{j}\delta_{n}^{k}}{2a^{2}}\partial_{i}u_{jk}
\eea
where $\Delta u^{(1)}_{n}$ is the first order change due to strains alone, and $\Delta u^{(2)}_{n}$ denotes the first order change due to strains and their derivatives (which is a low energy approximation). Here, $a$ is the lattice spacing, and $\beta=|\frac{\partial\log t}{\partial\log a}|$ is the Gr\"{u}neisen parameter.  Note we also assume that the electrons cannot hop to the next nearest neighbors, i.e. $t'=0$.

With the aim of realizing the Rindler Hamiltonian, henceforth we choose the following components for the strain tensor:
\begin{eqnarray}\label{RindlerStrainPattern}	u_{xx} & = & u_{yy} = -\frac{|x|}{\beta\lambda},~~~~~~~~~~u_{xy}=0, \nonumber \\
	t_{1}(x) & = & 1 + \frac{|x|}{\lambda} + \frac{\sqrt{3}}{4}\frac{a}{\lambda}\text{sgn}(x), \nonumber \\
	t_{2}(x) & = & 1 + \frac{|x|}{\lambda} - \frac{\sqrt{3}}{4}\frac{a}{\lambda}\text{sgn}(x), \nonumber \\
	t_{3}(x) & = & 1 + \frac{|x|}{\lambda},
\end{eqnarray}
where $\lambda$
is the strain scale that measures the distance over which an appreciable
inhomogeneity develops in the honeycomb lattice. With this choice of strain tensor, the distance between atoms decreases with increasing distance from $x=0$. 
At low energies, the electron dynamics is governed by two nodes in the reciprocal space ${\bf K}=\Big(\frac{4\pi}{3a\sqrt{3}},0\Big)=-{\bf K}'$. We can thus write down the $a$ and $b$ operators localized near these nodes as {\cite{Castro Neto:2009}}:
\begin{eqnarray}\label{Opertors Near Nodes}
	\hat{a}_{\boldsymbol{R}_{j}} & = & e^{i\boldsymbol{K}\cdot\boldsymbol{R}_{j}}\hat{A}(\boldsymbol{R}_{j}) + e^{i\boldsymbol{K}'\cdot\boldsymbol{R}_{j}}\hat{A}'(\boldsymbol{R}_{j}), \\
	\hat{b}_{\boldsymbol{R}_{j}+\boldsymbol{\delta}_{n}} & = & e^{i\boldsymbol{K}\cdot(\boldsymbol{R}_{j}+\boldsymbol{\delta}_{n})}\hat{B}(\boldsymbol{R}_{j}+\boldsymbol{\delta}_{n}) \nonumber \\
	& + & e^{i\boldsymbol{K}'\cdot(\boldsymbol{R}_{j}+\boldsymbol{\delta}_{n})}\hat{B}'(\boldsymbol{R}_{j}+\boldsymbol{\delta}_{n}),
\end{eqnarray}
where the prime $'$ denotes operators associated to the ${\bf K}'$ node. For low energies, it suffices to Taylor expand the $\hat{b}_{\boldsymbol{R}+\boldsymbol{\delta}_{n}}$ operators to linear order in gradients of these operators {\cite{Castro Neto:2009}}:
\begin{eqnarray}\label{TaylorExpand B operators}
	\hat{B}(\boldsymbol{R}_{j}+\boldsymbol{\delta}_{n}) & \approx & \hat{B}(\boldsymbol{R}_{j}) + \boldsymbol{\delta}_n\cdot\boldsymbol{\nabla}\hat{B}(\boldsymbol{R}_{j}).
\end{eqnarray}

Plugging into the tight-binding Hamiltonian {(\ref{Tight Binding Hamiltonian})}, the expressions for operators near the nodes {(\ref{Opertors Near Nodes})}, and the Taylor expansions for the hopping amplitude {(\ref{Taylor Expand Hopping})} and for the operators on $B$ carbon atoms {(\ref{TaylorExpand B operators})}, gives us the following:
\begin{widetext}
\begin{eqnarray}
	\hat{H} & = & -t_{0}\sum_{\boldsymbol{R}_{j},n}\Big[1-\beta\Delta u^{(1)}_{n}-\beta\Delta u^{(2)}_{n}\Big]\cdot\Big[\hat{A}^{\dagger}(\boldsymbol{R}_{j})\Big\{\hat{B}(\boldsymbol{R}_{j})+\boldsymbol{\delta}_{n}\cdot\boldsymbol{\nabla}\hat{B}(\boldsymbol{R}_{j})\Big\}e^{i\boldsymbol{K}\cdot\boldsymbol{\delta}_{n}} + \text{h.c.} \Big] \nonumber \\
	& - & t_{0}\sum_{\boldsymbol{R}_{j},n}\Big[1-\beta\Delta u^{(1)}_{n}-\beta\Delta u^{(2)}_{n}\Big]\cdot\Big[\hat{A}^{'\dagger}(\boldsymbol{R}_{j})\Big\{\hat{B}'(\boldsymbol{R}_{j})+\boldsymbol{\delta}_{n}\cdot\boldsymbol{\nabla}\hat{B}'(\boldsymbol{R}_{j})\Big\}e^{i\boldsymbol{K}'\cdot\boldsymbol{\delta}_{n}} + \text{h.c.} \Big],
\end{eqnarray}
\end{widetext}
where second term in each line is the hermitian conjugate of the first, denoted by $\text{h.c.}$. Here we have ignored cross-terms between the two nodes like $\sim\sum_{\boldsymbol{R}_{j}}\hat{A}^{\dagger}(\boldsymbol{R}_{j})\hat{B}'(\boldsymbol{R}_{j})e^{i(\boldsymbol{K}-\boldsymbol{K}')\cdot\boldsymbol{R}_{j}}$, that destructively interfere and thus vanish. We now simplify this expression by using the Rindler strain pattern {(\ref{RindlerStrainPattern})} and 
keeping terms that are linear order in gradients, terms that are  linear order in strains and terms that are both linear in gradients as well as strains.  We also introduce two-component
field operators at the ${\bf K}$ and ${\bf K'}$ nodes:
\bea
\hat{\psi}_{\bf K}(\boldsymbol{R}_{j})=\begin{pmatrix}
	\hat{B}(\boldsymbol{R}_{j}) \\ \hat{A}(\boldsymbol{R}_{j})
\end{pmatrix},
\\
\hat{\psi}_{\bf K'} (\boldsymbol{R}_{j})=\begin{pmatrix}
	\hat{A}^{'}(\boldsymbol{R}_{j}) \\ \hat{B}^{'}(\boldsymbol{R}_{j})
\end{pmatrix}.
\eea
Upon approximating the sums over
Bravais lattice points $\boldsymbol{R}_{j}$ to spatial integrals over $\br$,
relabeling
the ${\bf K}$ and ${\bf K}'$ points to be the right ($R$) and left ($L$)
nodes, we finally arrive at the effective Hamiltonian:
\bea
\label{Eq:fullHAM}
\Hh &=& \sum_{i=R,L} \int d^2 r \psih_i^\dagger(\br) \hh_i \psih_i(\br),
\\ \label{Eq:fullHAM:Dirac}
\hh_R &\equiv & \sqrt{v(x)} 
   \bsigma\cdot ( \boldsymbol{\sigma}\cdot\hat{\boldsymbol{p}}\big) \sqrt{v(x)}
      = -\hh_L,
\eea
where $\boldsymbol{\sigma}=\big(\sigma_{x},\sigma_{y}\big)$ is the
vector of Pauli matrices,
$\hat{\boldsymbol{p}}=-i\hbar\boldsymbol{\nabla}$ is the momentum
operator, with
$\boldsymbol{\nabla}=\big(\partial_{x},\partial_{y}\big)$ being the
gradient.  Here, 
$v(x) =  v_{0}\big(1+\frac{|x|}{\lambda}\big)$
represents a spatially-varying Fermi velocity with $v_0  =
\frac{3t_{0}a}{2\hbar}$ being the Fermi velocity of the unstrained
Honeycomb lattice.  If we had instead
chosen a plus sign for the strain tensor components in {(\ref{RindlerStrainPattern})},
then we would get a spatially decreasing Fermi velocity
$v_{0}\big(1-\frac{|x|}{\lambda}\big)$. 
We emphasize that, although here we focus on strained graphene, for our purposes the essential 
goal is to achieve a spatially-varying hopping amplitude yielding a 2D Dirac Hamiltonian (\ref{Eq:fullHAM:Dirac}) with a spatially-varying velocity. Therefore, another method to realize an effective spatially-varying hopping will have similar behavior. For example, in Ref.~\cite{Jimenez-Galan:2020}, it was shown that the vector potential of a bi-circular laser field (in the long-pulse limit), can modify the hopping amplitude, providing another path to realizing Eq.~(\ref{Eq:fullHAM:Dirac}).


%

In the next step, we establish two different limiting cases of the Hamiltonian
Eq.~(\ref{Eq:fullHAM}): The unstrained case, $\lambda \to \infty$,
that yields the well known 2D Dirac Hamiltonian, and the
case of strong strains, $\lambda \to 0$, in which the system
Hamiltonian describes Dirac particles moving in a Rindler metric~\cite{Rindler:1966zz}.  In the strong-strain limit,
we can neglect the unit contribution in $v(x)$, leaving $v(x)=v_{0}|x|/\lambda$.  In fact, as we now argue,
this approximation also holds
in the long-wavelength limit.  Our argument relies on 
translation symmetry in the $y$-direction, which implies eigenfunctions of  
$\hh_R$ are plane waves in the $y$ direction, $\propto {\rm e}^{ik_y y}$ with wavevector $k_{y}$. Re-scaling the coordinates via $x\rightarrow x/|k_{y}|$ and $y\rightarrow y/|k_{y}|$ changes the spatially dependent Fermi velocity to $v(x)\rightarrow v_0\big(1+\frac{|x|}{|k_{y}|\lambda}\big)$ and the momentum operator becomes $\hat{\boldsymbol{p}}\rightarrow|k_{y}|\cdot\hat{\boldsymbol{p}}$. In the long-wavelength limit ($|k_{y}|\lambda\ll1$), the contribution of unity inside $v(x)$ becomes negligible and $|k_{y}|$ cancels out, giving us the 2D Rindler Hamiltonian which is just (\ref{Eq:fullHAM}) with the Fermi velocity $v(x)=v_{0}|x|/\lambda$. This shows the emergence of the effective Rindler Hamiltonian in the long-wavelength low-energy limit, exhibiting a horizon at $x=0$.  An alternate way to see the presence of this horizon, as shown in Appendix \ref{SEC:Appendix Horizon Graphene}, is to show that eigenfunctions of the strained Hamiltonian in Eq.~(\ref{Eq:fullHAM}) obey an effective Schrodinger-like equation with an infinite  potential barrier at $x=0$.


Having discussed how the Rindler Hamiltonian can be realized in strained honeycomb lattices, in the coming sections, we apply these ideas  to see how a sudden switch on of the system strain, suddenly changing the Hamiltonian from the 2D Dirac Hamiltonian to 
the 2D Rindler Hamiltonian can strongly modify low-energy 
and long-wavelength 
properties leading to the analogue Unruh effect. To begin with, in the next section, we start with a review of fermions in flat unstrained honeycomb lattices, i.e., the case of graphene.

\begin{figure*}[t]
\centering
\subfloat[Subfigure 1 list of figures text][Minkowski]{
		\includegraphics[width=0.45\textwidth]{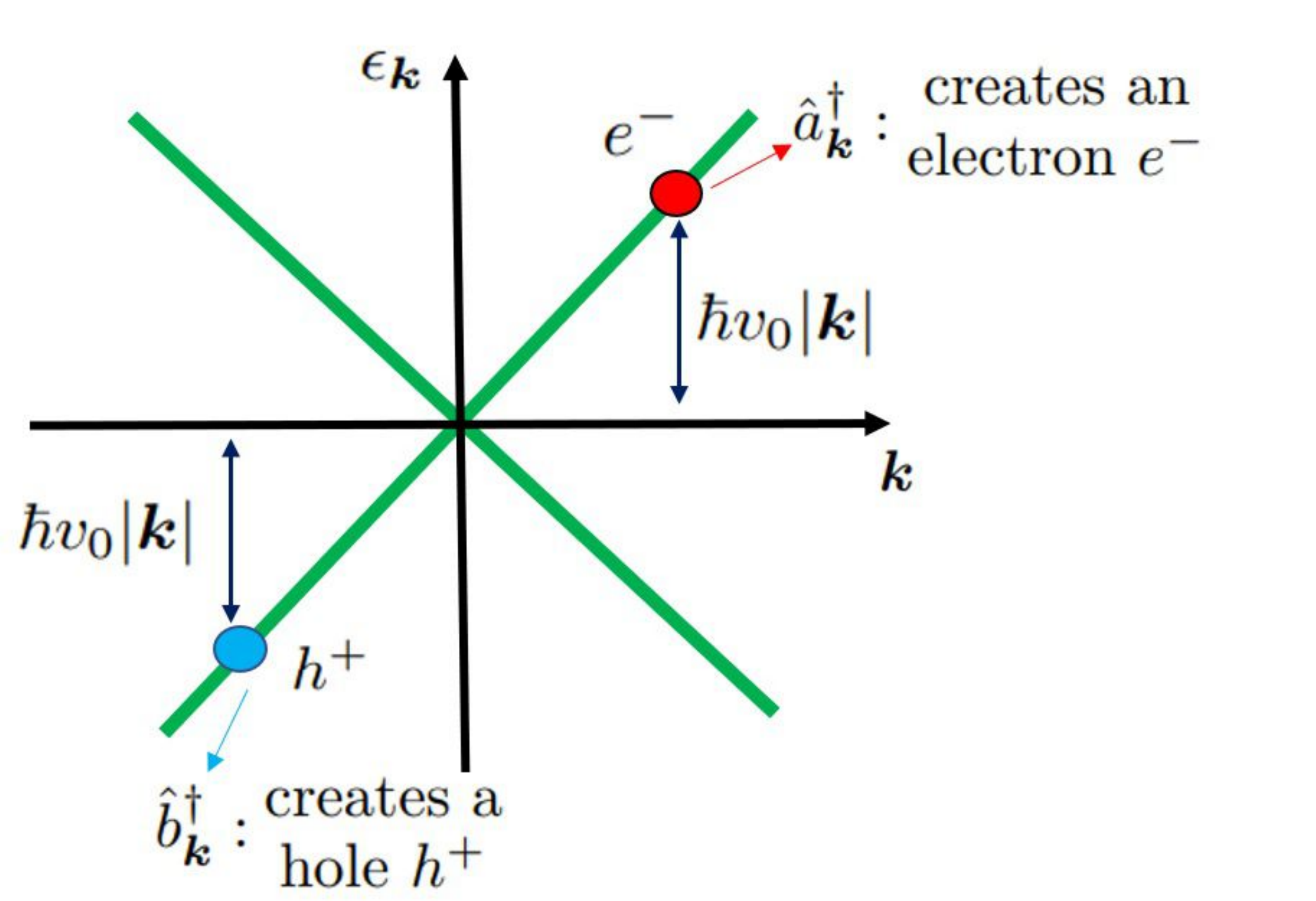}
		\label{Band Structure: Minkowski}}
\subfloat[Subfigure 2 list of figures text][Rindler]{
		\includegraphics[width=0.45\textwidth]{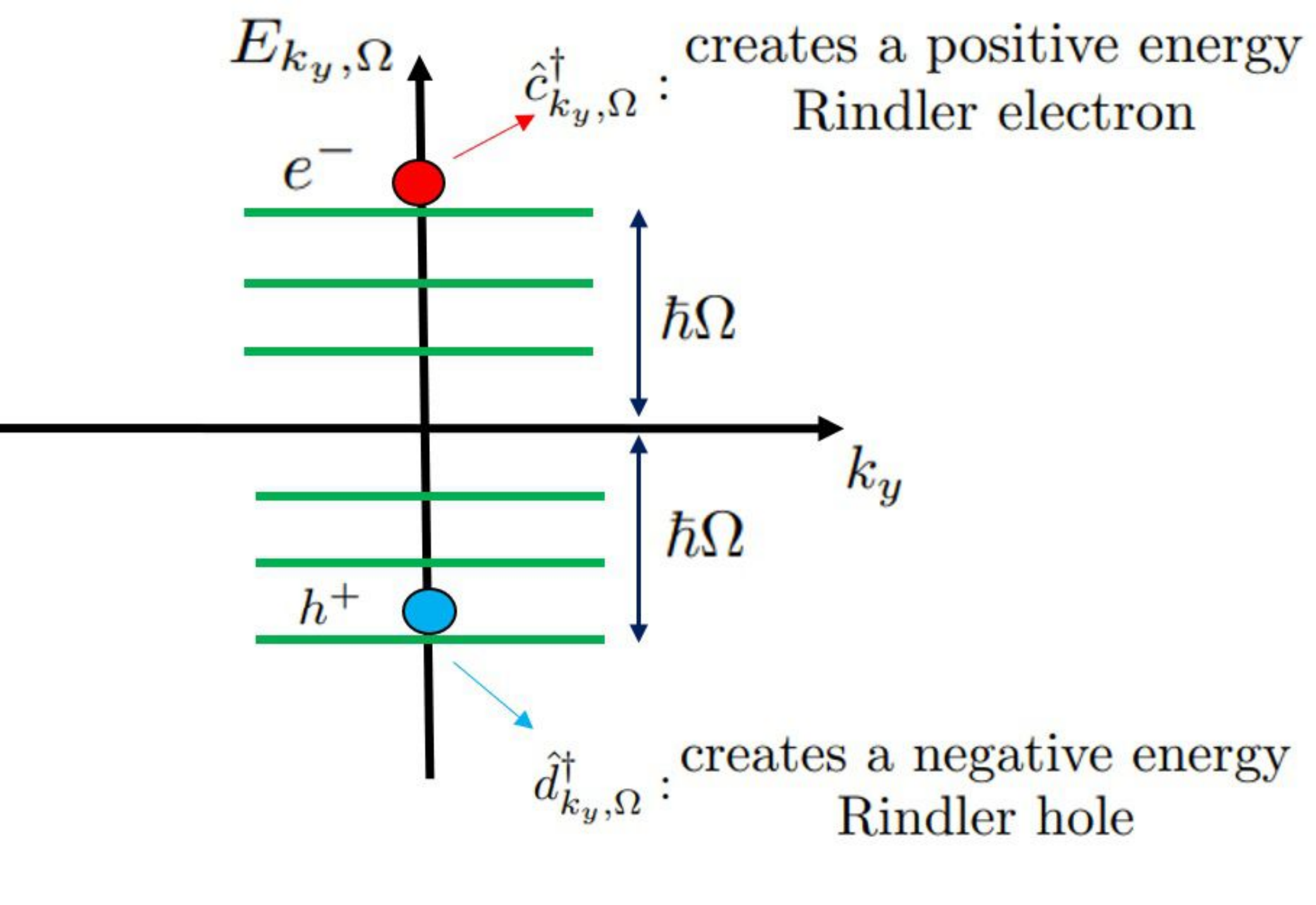}
		\label{Band Structure: Rindler}}
\caption{(Color Online) A schematic figure to depict the (a) Minkowski and (b) Rindler mode expansions. In flat graphene, the existence of translation symmetry yields a Dirac-like linear energy dispersion $\epsilon_{\boldsymbol{k}}=\hbar v_{0}|\boldsymbol{k}|$ (shown in green in panel a).  The electron and hole excitation energies are both positive ($\epsilon_{\boldsymbol{k}}>0$)
 with the operators $\hat{a}_{\boldsymbol{k}}|0_{\cal{M}}\rangle=0=\hat{b}_{\boldsymbol{k}}|0_{\cal{M}}\rangle$ annihilating the Minkowski vacuum. In strained graphene, the Rindler energy $E_{k_{y},\Omega}=\hbar\Omega>0$ (shown in green in panel b) and transverse momenta $\hbar k_{y}$ are decoupled, with their associated electron and hole operators annihilating the Rindler vacuum state $\hat{c}_{k_{y},\Omega}|0_{\cal{R}}\rangle=0=\hat{d}_{k_{y},\Omega}|0_{\cal{R}}\rangle$.}
\label{Band Structure}
\end{figure*}

\section{Mode expansion: Flat honeycomb lattice}
\label{SEC:three}
In this section, we review the Dirac equation for 
flat (unstrained) graphene 
and derive the resulting normal mode expansion
that describes electron and hole excitations. 
As we have already discussed, the low-energy Hamiltonian for fermions
hopping on a uniform
(unstrained) honeycomb lattice follows from taking the $\lambda \to \infty$
limit of Eq.~(\ref{Eq:fullHAM}), resulting in $\hat{H} = \hat{H}_{\rm R} + \hat{H}_{\rm L}$
with 
\begin{eqnarray}
&&\hat{H}_{\text{R}}
=  v_0 \int d^{2}r~\hat{\psi}^{\dagger}_{\text{R}}(\boldsymbol{r}) 
 \boldsymbol{\sigma}\cdot\hat{\boldsymbol{p}} \hat{\psi}_{\text{R}}(\boldsymbol{r}), \label{Dirac Hamiltonian}
\end{eqnarray}
where to get $\hat{H}_{\rm L}$ we simply replace $R\to L$ and take $v_0 \to -v_0$.
The field operators $\hat{\psi}_{i}$ ($i=L,R$) satisfy the anticommutation
relation
\be
\{\hat{\psi}_{i},\hat{\psi}_{j}^\dagger\} = \delta_{ij}  \delta(\boldsymbol{r}-\boldsymbol{r}'),
\label{Eq:acrelation}
\ee

In the following we focus on the right node, with results from the left
node easily following.  
The Heisenberg equation of motion for the field operators
$\hat{\psi}_{\text{R}}(\boldsymbol{r},t)$ is:  
\be
\label{Dirac Equation Flat Graphene}
i\hbar\partial_{t}\hat{\psi}_{\text{R}}(\boldsymbol{r},t)=[\hat{\psi}_{\text{R}}(\boldsymbol{r},t),\hat{H}] = v_{0}\boldsymbol{\sigma}\cdot\hat{\boldsymbol{p}}~\hat{\psi}_{\text{R}}(\boldsymbol{r},t),
\ee
the massless Dirac equation (Weyl equation) that describes how fermions (with zero rest mass) propagate in a flat spacetime with an emergent $(2+1)$-dimensional Minkowski line element labeled by the inertial coordinates $(T,X,Y)$: 
\begin{eqnarray}\label{Minkowski Metric}
	ds_{\text{Mink}}^{2} & = & -v_{0}^{2}dT^{2} + dX^{2} + dY^{2},
\end{eqnarray}
where the speed of light is now replaced by the Fermi velocity $c\rightarrow v_{0}$. In Appendix \ref{SEC:Appendix Dirac Eqn}, we describe how a metric expressed in inertial coordinates like {(\ref{Minkowski Metric})} (see Eq.~{(\ref{Inertial Coordinates})}) leads to a Dirac equation in inertial coordinates {(\ref{Dirac Equation Flat Graphene})} (see Eq.~{(\ref{Weyl Equations Inertial Coordinates})}). This metric describes the dynamical trajectories of inertial observers in a flat spacetime. Suppose two inertial frames $S$ and $S'$ moving with relative speed $v$, then the coordinates of an observer in frame $S'$ i.e. $(T',X',Y')$, are related to the ones in $S$ via Lorentz transformations:
\begin{eqnarray}\label{Lorentz Transformation}
v_{0}T' & = & v_{0}T\cosh\theta - x\sinh\theta, \nonumber \\
X' & = & x\cosh\theta - v_{0}T\sinh\theta, \nonumber \\
Y' & = & Y,
\end{eqnarray}
where $\cosh\theta=\gamma=\frac{1}{\sqrt{1-\beta^{2}}}$ is the Lorentz factor
with $\beta=\frac{v}{v_{0}}$, and $\sinh\theta=\gamma\beta$. The ratio
of these factors relate the velocity with rapidity
$\theta\in(-\infty,\infty)$: $\tanh\theta=\beta\in(-1,1)$. In either
frame, the trajectory of an inertial observer is of the form
$-v_{0}^{2}T^{2}+X^{2}+Y^{2}=\text{constant}$.

Thus, as one might expect, fermions hopping in an unstrained honeycomb
lattice obey an analogue Dirac equation with the Fermi velocity $v_0$
playing the role
of the speed of light.  Our next task is to expand the fermion field
operators into normal modes corresponding to positive energy
\lq \lq particle\rq\rq\ and negative energy \lq\lq hole\rq\rq\ excitations in graphene's Dirac band
structure.   
Since the system is homogeneous in space and time (or alternatively
the emergent metric components {(\ref{Minkowski
    Metric})} are constants), the Dirac equation solutions that
describe the evolution of fermions are plane waves of the form
$e^{\pm i(\boldsymbol{k}\cdot\boldsymbol{x}-\omega_{k}t)}$ and thus
the field operators on the right node can be expressed in terms of the
following mode expansion \cite{Das:2008zze}: 
\begin{eqnarray}\label{MinkowskiModeExpansionStandardRight1}
  \hat{\psi}_{\text{R}}(\boldsymbol{r})\! =\! \int
  \frac{d^{2}k}{2\pi}\Big(e^{i(\boldsymbol{k}\cdot\boldsymbol{r}-v_{0}kt)}u_{\boldsymbol{k}}\hat{a}_{\boldsymbol{k}}
  +
  e^{-i(\boldsymbol{k}\cdot\boldsymbol{r}-v_{0} k t)}v
  _{-\boldsymbol{k}}\hat{b}^{\dagger}_{\boldsymbol{k}}\Big), \nonumber \\
\end{eqnarray}
where the wave-vector $\boldsymbol{k}=(k_{x},k_{y})$ is related to the linear momenta in spatial directions via  $\boldsymbol{p}=\hbar\boldsymbol{k}$ and, thanks to  translation symmetry, is related to the energy $\epsilon_{k}=\hbar\omega_{k}$ ($\omega_{k}$ is the mode frequency), via the dispersion relations $\epsilon_{k}=\hbar v_{0}|\boldsymbol{k}|$ or $\omega_{k}=v_{0}k$ where $k  \equiv
|\boldsymbol{k}|   = 
\sqrt{k_{x}^{2}+k_{y}^{2}}$
is the wavevector magnitude.

This mode expansion for the right node ${\bf K}_{\text{R}}$
(right-handed Weyl fermions) should have positive
helicity, which is defined as the projection of the Pauli spin operator onto the
direction of the momentum vector
$h=\boldsymbol{\sigma}\cdot\hat{\boldsymbol{k}}$. Thus the flat spinors
used in the mode expansion (\ref{MinkowskiModeExpansionStandardRight1})
are defined as follows:
\begin{eqnarray}
u_{\boldsymbol{k}} = \frac{1}{\sqrt{2}}\begin{bmatrix}
1 \\
\frac{k_{x}+ik_{y}}{k}\end{bmatrix},~~~~~
v_{\boldsymbol{k}} = \frac{1}{\sqrt{2}}\begin{bmatrix}
-\big(\frac{k_{x}-ik_{y}}{k}\big) \\
1 \end{bmatrix}.
\end{eqnarray}
In the above definitions, $u_{\boldsymbol{k}}$ has positive helicity
$h = +1$, whereas $v_{\boldsymbol{k}}$ has negative helicity $h =
-1$. The particle
$\hat{a}$ and hole $\hat{b}$ operators satisfy anti-commutation
relations and annihilate the flat honeycomb (Minkowski) vacuum state
$|0_{\cal{M}}\rangle$:
\begin{eqnarray}\label{Minkowski Operators1}
	\{\hat{a}_{\boldsymbol{k}},\hat{a}^{\dagger}_{\boldsymbol{k}'}\} & = & \delta(\boldsymbol{k}-\boldsymbol{k}'),~~~~~ \{\hat{b}_{\boldsymbol{k}},\hat{b}^{\dagger}_{\boldsymbol{k}'}\}~=~\delta(\boldsymbol{k}-\boldsymbol{k}'), \nonumber \\
	\hat{a}_{\boldsymbol{k}}|0_{\cal{M}}\rangle & = & 0,~~~~~~~~~~~~~~~~~~~\hat{b}_{\boldsymbol{k}}|0_{\cal{M}}\rangle~=~0.
\end{eqnarray}
To obtain the mode expansion for the left node $\boldsymbol{K}_{\text{L}}$ (left-handed Weyl fermions), the particle and hole spinors $u_{\boldsymbol{k}}$ and $v_{-\boldsymbol{k}}$ in Eq.~(\ref{MinkowskiModeExpansionStandardRight1}) need to be switched with $v_{\boldsymbol{k}}$ and $u_{-\boldsymbol{k}}$, respectively, which means they both have negative helicities.

As is well known, the particle and hole fermionic excitations in 
graphene obey a linear dispersion relation, with $\omega_{k}\propto|\boldsymbol{k}|$.  In Fig.~\ref{Band Structure: Minkowski},
we depict this linear energy dispersion, with the system ground
state being a fully occupied valence band at negative energies
and a fully unoccupied conduction band at positive energies.  
This figure also depicts the positive energy particle (or electron)
and hole excitations that are captured by the mode expansion (\ref{MinkowskiModeExpansionStandardRight1}).

\section{Mode expansion: Rindler system}
\label{SEC:four}
In this section, we study the case of fermions hopping
in a honeycomb lattice in the presence of a strain field that 
leads to the Rindler low-energy Hamiltonian, obtained by approximating
$v(x) \simeq \frac{v_0}{\lambda}|x|$. 
As in the flat case,
the system Hamiltonian comprises terms from the left and right nodes,
$\hat{H} = \hat{H}_{\rm R} + \hat{H}_{\rm L}$,
with the right-node Hamiltonian:
\be
\hat{H}_{\text{R}}
=  \frac{v_0}{\lambda} \int d^{2}r~\hat{\psi}^{\dagger}_{\text{R}}(\boldsymbol{r})
\sqrt{|x|}
\boldsymbol{\sigma}\cdot\hat{\boldsymbol{p}}\sqrt{|x|}
\hat{\psi}_{\text{R}}(\boldsymbol{r}),
\label{Rindler Hamiltonian}
\ee
which we call the Rindler Hamiltonian by analogy with the well-known Rindler metric,
that describes how the flat Minkowski spacetime is seen by an accelerating observer \cite{Rindler:1966zz}.  Following
the discussion in the homogeneous case, we find the equation of motion
\begin{equation}\label{Dirac Equation Rindler Graphene}
	i\hbar\partial_{t}\hat{\psi}_{\text{R}}(\boldsymbol{r}) = \frac{v_0}{\lambda} 
	\sqrt{|x|} \boldsymbol{\sigma}\cdot\hat{\boldsymbol{p}} \sqrt{|x|} ~\hat{\psi}_{\text{R}}(\boldsymbol{r}),
\end{equation}
the Dirac equation for massless fermions in Rindler spacetime with Rindler coordinates
 $(t,x,y)$ {\cite{Wald:1984rg,Misner:1973prb,Rindler:2006km}} described by the line element
\begin{eqnarray}\label{Rindler Metric}
ds^{2} & = & -\Big(\frac{x}{\lambda}\Big)^{2}v_{0}^{2}dt^{2} + dx^{2} + dy^{2}.
\end{eqnarray}



In Appendix \ref{SEC:Appendix Dirac Eqn}, we describe how the  Rindler metric (see
Eq.~{(\ref{Rindler Coordinates})}) leads to a Dirac
equation for accelerating electrons (see
Eq.~{(\ref{Weyl Equations Rindler Coordinates})}). To
understand the role of this metric in the context of honeycomb
systems, we first need to understand its role in relativistic physics. Imagine a Rindler observer in the frame $S_{\text{R}}$, moving with some acceleration
$\boldsymbol{a}=a\hat{x}$ ($a>0$) with respect to an inertial frame $S$. The observer starts their journey far away at
$x=\infty$ at time $t=-\infty$ with velocity close to the speed of light $c$ moving
towards the origin $x=0$. Initially they decelerate, eventually stopping at a
certain distance from the origin $x_{\text{min}}=\frac{c^{2}}{a}$, and
then return to $x=+\infty$ at $t=+\infty$. Since at any one instant of time, the Rindler observer is moving at a certain velocity $v$, we expect a hyperbolic-like trajectory 
similar to the Minkowski case: $-v_{0}^{2}T^{2}+X^{2}+Y^{2}=\text{constant}$, and the transformation between inertial coordinates $(T,X)$ and Rindler $(t,x)$ coordinates to be similar to ${(\ref{Lorentz Transformation})}$. This is reminiscent of non-relativistic physics, where the trajectory of an accelerated observer is parabolic: $x=x_{0}+u_{0}t+\frac{1}{2}at^{2}$. However, relativistic accelerations need to be hyperbolic as motion also affects the rate at which the observer's clock ticks. Thus the relation between the inertial and Rindler coordinates are as follows {\cite{Wald:1984rg,Misner:1973prb,Rindler:2006km}}:
\begin{eqnarray}\label{Rindler Transformation}
	cT & = & x_{\text{min}}\sinh\frac{ct}{x_\text{min}}, \nonumber \\
	X & = & x_{\text{min}}\cosh\frac{ct}{x_\text{min}},
\end{eqnarray}
which gives us the trajectory of a Rindler observer viewed from an inertial frame $S$: $X^{2}-c^{2}T^{2} = x^{2}_{\text{min}}$. The above coordinates $(T,X)$ label the worldline of an accelerated observer from the perspective of an inertial frame. If the acceleration is changed to a different but constant value, then we get a family of Rindler observers, each with a different closest distance of approach $x_{\text{min}}$.
This family is parameterized using a new coordinate $x_{\text{min}}\rightarrow x$, giving us the Rindler metric in Eq.~{(\ref{Rindler Metric})}. If we set the spatial coordinates to zero, i.e. $dx=dy=0$ then $t$ behaves like the proper time as seen on the watch of a Rindler observer. Similar arguments hold for an observer accelerating in the opposite direction with $a<0$. Note that (\ref{Rindler Metric}) becomes degenerate at $x=0$, i.e. the time-time component of the metric tensor vanishes ($g_{tt}=0$) and hence has no inverse. This is known as the \emph{Rindler horizon}. Because of this horizon,  oppositely accelerating observers can never communicate with each other. 
Note that the connection between the coordinates $(T,X)$ and $(t,x)$ is just a switch of variables, therefore the metric {(\ref{Rindler Metric})}, is basically flat spacetime written in disguise, and thus the Riemann curvature of this spacetime is zero. Also note that the coordinates $(t,x,y)$ cover only two portions of the flat Minkowski spacetime: the right Rindler wedge $x>0$ for positive accelerations and the left Rindler wedge $x<0$ for negative accelerations.

In the context of strained graphene, the emergent metric in Eq.~(\ref{Rindler Metric}) tells us that similar Rindler physics is expected provided we replace the speed of light with the Fermi velocity $c\rightarrow v_{0}$, and the distance of closest approach with the strain scale $x_{\text{min}}\rightarrow\lambda$. Once we do this, then we can interpret the electron dynamics inside graphene as Rindler fermions where the analogue acceleration is given by $a=\frac{v_{0}^{2}}{\lambda}$, where a choice of strain $\lambda$ corresponds to choosing a unique Rindler observer with this acceleration. Such analogue accelerations are expected here because under the semiclassical model of electron dynamics, the strained graphene has an environment with broken translation symmetry that forces the Fermi velocity to be spatially dependent $v(x)=v_{0}(1+\frac{|x|}{\lambda})$. Moreover, the strain pattern in Eq.~(\ref{RindlerStrainPattern}) tells us that carbon atoms become closer with distance from the origin, thus enhancing electron hopping. This hopping from one carbon atom to another will be most difficult at the origin itself, especially for low-energy  and long-wavelength modes which cannot tunnel from one side to the other (see Appendix \ref{SEC:Appendix Horizon Graphene} for more details). Therefore, $x=0$ being a barrier for such modes acts as an analogue of the Rindler horizon, breaking the strained graphene into two disconnected pieces: the right side mimics the right Rindler wedge, and the left side mimics the left Rindler wedge.

Our next task is to identify the normal mode expansion for the field operator $\hat{\psi}_{\text{R}}(\boldsymbol{r})$ in
 the Rindler 
Dirac equation  (\ref{Dirac Equation Rindler Graphene}) \cite{Unruh:1974,Candelas:1978gg,Soffel:1980kx,Hughes:1983ch,Iyer:1985ufr,Jauregui:1991me,Crispino:2007eb,Takagi:1986kn}.   In doing this, we define
the frequency scale $\Omega>0$ and look for positive energy $(E= \hbar\Omega>0)$ 
solutions (corresponding to Rindler particles)  and
negative energy $(E= -\hbar\Omega<0)$ solutions (corresponding to Rindler holes).
Starting with the $E>0$ case, the solutions take the form $\psi^{+}_{\Omega}(x,k_{y})
e^{i(k_{y}y-\Omega t)}$,  where $p_{y}=\hbar k_{y}$ is the momentum in 
the $y$-direction.
If we define the components of the spinor part via
\be
 \psi^{+}_{\Omega}(x,k_{y}) = \begin{pmatrix}
	f(x) \\
	g(x)
\end{pmatrix}, 
\ee
then the functions $f(x)$ and $g(x)$ satisfy (henceforth we set $\hbar\to 1$):
\bse
\label{fandgee}
\bea
\label{Og-gives-f}
\bigg(|x|\frac{d}{dx}+k_{y}|x|+\frac{\text{sgn}(x)}{2}\bigg)g(x) & = & i\Omega f(x), \\ \label{Of-gives-g}
\bigg(|x|\frac{d}{dx}-k_{y}|x|+\frac{\text{sgn}(x)}{2}\bigg)f(x) & = & i\Omega g(x).
\eea
\ese
The dimensionless form of these equations came because we measured energy (or frequency, $\Omega$) relative to the scale
\be
\label{eq:omegacdef}
\omega_{c}=v_0/\lambda
\ee
characterizing the strain magnitude.  

Starting with the case of $x>0$ and $k_y>0$, and focusing on solutions that  are normalizable at $|x|\to \infty$, we find:
\bse
\label{Bessel Solutions}
\begin{eqnarray}
	f(x) & = & K_{\frac{1}{2}-i\Omega}\big(k_{y}x\big) - K_{\frac{1}{2}+i\Omega}\big(k_{y}x\big), 
	\\
	g(x) & = & K_{\frac{1}{2}-i\Omega}\big(k_{y}x\big) + K_{\frac{1}{2}+i\Omega}\big(k_{y}x\big),
\end{eqnarray}
\ese
where $K_{\nu}(x)$ is the modified Bessel function of the second kind, that diverges at the origin $x=0$ and for large negative arguments $x\rightarrow-\infty$. This divergence can be attributed to the form of the analogue Rindler metric {(\ref{Rindler Metric})}, whose time-time component vanishes at $x=0$, and contributes a non-smooth modulus function $|x|$ in the Weyl equations 
which leads to different solutions in the left and right spatial regions of the strained honeycomb lattice. As we have already discussed,  
this demarcation of the system at $x=0$ is known as the Rindler horizon. In analogy with relativity, the left spatial portion acts as the left Rindler wedge, and similarly for the right portion. There, an observer in right wedge will never be able to communicate with their counterpart in the left wedge. 
In the next section, we will see that this is an essential reason why a natural temperature emerges in this system.

The solutions for $f$ and $g$ above have Bessel functions with positive arguments. Therefore they are finite and vanish asymptotically for $k_{y}x\rightarrow\infty$. For the case $x>0$ and $k_{y}<0$, the equations {(\ref{Og-gives-f})} and {(\ref{Of-gives-g})} get interchanged, resulting in an exchange of the spinor components $f(x)\leftrightarrow g(x)$. The case of $x<0$ and $k_{y}>0$ effectively switches $\Omega\rightarrow-\Omega$ and $k_y \to -k_y$ relative to the $x>0$ and $k_y>0$ case,
while the case of $x<0$ and $k_{y}<0$ effectively switches $\Omega\rightarrow-\Omega$ relative to
the $x>0$ and $k_y>0$ case.  Taken together, these considerations imply the positive energy
spinor
%

\begin{equation*}
\psi^{+}_{\Omega}(x,k_{y}) = \begin{cases}
\begin{pmatrix}
K_{\frac{1}{2}-i\Omega} - \sgn(k_y) K_{\frac{1}{2}+i\Omega}\\
K_{\frac{1}{2}-i\Omega} + \sgn(k_y) K_{\frac{1}{2}+i\Omega} \\
\end{pmatrix} &\text{if $x>0$}\\
\begin{pmatrix}
K_{\frac{1}{2}+i\Omega} + \sgn(k_y) K_{\frac{1}{2}-i\Omega} \\
K_{\frac{1}{2}+i\Omega} - \sgn(k_y) K_{\frac{1}{2}-i\Omega}\\
\end{pmatrix} &\text{if $x<0$}
\end{cases}
\end{equation*}
where $ K_{\frac{1}{2}\pm i\Omega}$ is shorthand for $K_{\frac{1}{2}\pm i\Omega}(|k_y x|)$.  We emphasize here that the above two solutions come from solving the Rindler-Dirac equation separately for $x>0$ and $x<0$, pertaining to the two sides of the honeycomb lattice. Thus we define orthonormality separately in
the $x>0$ and $x<0$ regimes.  

Turning to the $E<0$ case, we take the solutions to have the form $\psi^{-}_{\Omega}(x,k_{y})
e^{-i(k_{y}y-\Omega t)}$,  which effectively changes the sign of $k_y$ and $\Omega$ relative to the
positive energy case.  This leads to the negative energy spinors:
%
%

\begin{equation*}
\psi^{-}_{\Omega}(x,k_{y}) = \begin{cases}
\begin{pmatrix}
K_{\frac{1}{2}+i\Omega} + \sgn(k_y) K_{\frac{1}{2}-i\Omega}\\
K_{\frac{1}{2}+i\Omega} - \sgn(k_y) K_{\frac{1}{2}-i\Omega} \\
\end{pmatrix} &\text{if $x>0$}\\
\begin{pmatrix}
K_{\frac{1}{2}-i\Omega} - \sgn(k_y) K_{\frac{1}{2}+i\Omega} \\
K_{\frac{1}{2}-i\Omega} + \sgn(k_y) K_{\frac{1}{2}+i\Omega}\\
\end{pmatrix} &\text{if $x<0$}
\end{cases}
\end{equation*}

The normal mode expansion then takes the form:
\cite{Unruh:1974,Candelas:1978gg,Soffel:1980kx,Hughes:1983ch,Iyer:1985ufr,Jauregui:1991me,Crispino:2007eb,Takagi:1986kn}
\bea
\label{ModeExpansionRightNode}
&&\hat{\psi}_{\text{R}}(\boldsymbol{r},t)  =  \int_{-\infty}^{\infty} \frac{dk_{y}}{\sqrt{2\pi}} \int_{0}^{\infty} d\Omega~N_{k_{y},\Omega} 
\\
&& \hspace{-0.5cm}\times \Big[\psi^{+}_{\Omega}(x,k_{y})e^{i(k_{y}y-\Omega t)}\hat{c}_{k_{y},\Omega}
+ \psi^{-}_{\Omega}(x,k_{y})e^{-i(k_{y}y-\Omega t)}\hat{d}^{\dagger}_{k_{y},\Omega} \Big],
\nonumber 
\eea
where the operators $\hat{c}_{k_{y},\Omega}$ annihilate positive energy Rindler
particles and the operator $\hat{d}^{\dagger}_{k_{y},\Omega}$ creates
a negative energy Rindler hole, as illustrated in Fig.~\ref{Band Structure: Rindler}.
These particle  and hole operators satisfy fermionic anticommutation relations:
\bea
\{\hat{c}_{k_{y},\Omega},\hat{c}^{\dagger}_{k'_{y},\Omega'}\} & = & \delta(k_{y}-k'_{y})\delta(\Omega-\Omega'),
\\
\{\hat{d}_{k_{y},\Omega},\hat{d}^{\dagger}_{k'_{y},\Omega'}\} & = & \delta(k_{y}-k'_{y})\delta(\Omega-\Omega').
\eea
We emphasize that, in our convention, the energy scale $\hbar \Omega>0$, so that
both particle and hole excitations have positive energy (although the latter
emerge from below the Fermi level).  Thus the Rindler vacuum $|0_{\cal{R}}\rangle$
is annihilated by both the electron and hole operators:
\bea
\hat{c}_{k_{y},\Omega}|0_{\cal{R}}\rangle & = & 0,
\\
\hat{d}_{k_{y},\Omega}|0_{\cal{R}}\rangle &=& 0.
\eea
%
%
For the left handed electrons, we need to solve the corresponding set of Weyl equations, which is the same as the equation for right-handed electrons, except for a minus sign associated with the time derivative. This amounts to saying that the fermions on ${\bf K}_{\text{L}}$ node will be described by the same mode expansion as {(\ref{ModeExpansionRightNode})}, except that the spinors will all change signs for the frequency i.e. $\psi^{\pm}_{\Omega}(x,k_{y})\rightarrow\psi^{\pm}_{-\Omega}(x,k_{y})$. Finally, to determine the normalization factor  $N_{k_{y},\Omega}=\sqrt{\frac{|k_{y}|}{2\pi^{2}}\cosh\pi\Omega}$ we make use of the inner product for Weyl spinors {\cite{Takagi:1986kn,Birrell:1982ix}}:
\begin{eqnarray}\label{Orthonormality}
&&\Big(\psi^{\sigma'}_{\Omega'}(x,k_{y}),\psi^{\sigma}_{\Omega}(x,k_{y})\Big) \equiv  \int_{0}^{\infty} dx~\psi^{\sigma'\dagger}_{\Omega'}(x,k_{y})
\psi^{\sigma}_{\Omega}(x,k_{y}) \nonumber \\
&&\qquad 
=  \delta^{\sigma\sigma'}\delta(\Omega-\Omega'),~~~~~~~
\end{eqnarray}
where $\sigma=\pm$ denotes the positive or negative energy spinors, and the following identity for Bessel functions {\cite{Jauregui:1991me,Gradshteyn}}: 
\bea\label{Normalization Bessel}
&&\int_0^\infty dx \, \Big[K_{\frac{1}{2}+i\Omega} (x)
K_{\frac{1}{2}-i\Omega'} (x)
+
K_{\frac{1}{2}-i\Omega} (x)
K_{\frac{1}{2}+i\Omega'} (x)\Big] \nonumber \\
& = & \pi^2 \sech (\pi\Omega)\delta(\Omega-\Omega').
\eea

Now that we have derived the mode expansion {(\ref{ModeExpansionRightNode})} in terms of Bessel functions that are singular at the horizon for the field operators in a strained graphene system (or in an ultracold honeycomb optical lattice that has a linear-in-position Fermi velocity), in the next section, we will describe how this leads to spontaneous creation of electron-hole pairs, which is equivalent to saying that a sudden change in the Fermi velocity $v_{0}\rightarrow v_{0}\frac{|x|}{\lambda}$ leads to a spontaneous jump of electrons from the valence to conduction band.

\section{Spontaneous Electron-Hole Pair Creation}\label{SEC:five}
In the last two sections, we discussed the Dirac Hamiltonian {(\ref{Dirac Hamiltonian})} and its solutions {(\ref{MinkowskiModeExpansionStandardRight1})} for a flat honeycomb system with homogeneous Fermi velocity $v(x)=v_{0}$, and the Rindler Hamiltonian {(\ref{Rindler Hamiltonian})} and its solutions {(\ref{ModeExpansionRightNode})} for an inhomogeneous honeycomb lattice with a spatially-varying Fermi velocity $v(x)=v_{0}\frac{|x|}{\lambda}$. The latter solutions are made out of spinors of Bessel functions that diverge at the horizon $x=0$,
with separate solutions at $x>0$ and $x<0$.
In this section, we will describe how this set-up leads to spontaneous creation of electron-hole pairs, with the spectrum of these excitations  described by an emergent Fermi-Dirac distribution that is a function of Rindler mode frequency $\Omega$ and the characteristic frequency $\omega_{c}$, 
defined in Eq.~(\ref{eq:omegacdef}), that is proportional to the Unruh temperature.

Since the Rindler $|0_{\cal{R}}\rangle$ and the Minkowski vacua $|0_{\cal{M}}\rangle$ are associated with strained and flat honeycomb lattices respectively, they are expected to be very different from each other, i.e. the notion of particles that one ascribes to with respect to the Minkowski vacuum cannot be same as the Rindler case, since in the
former case there exists translation symmetry, whereas in the latter, 
the mechanical strain strongly modifies the properties of system
eigenstates.  

We consider the situation where we start with the flat honeycomb Hamiltonian {(\ref{Dirac Hamiltonian})} described by the mode expansion {(\ref{MinkowskiModeExpansionStandardRight1})} for the field operators, and then suddenly switch on the linear-in-position Fermi velocity with a characteristic strain length $\lambda$, thereby invoking the Rindler Hamiltonian {(\ref{Rindler Hamiltonian})} and the corresponding mode expansion {(\ref{ModeExpansionRightNode})}. 
In the Heisenberg picture then, we expect that the mode expansion for the fermionic field operators $\hat{\psi}_{\text{R}}$ on the right node evolve from Eq.~{(\ref{MinkowskiModeExpansionStandardRight1})} to Eq.~{(\ref{ModeExpansionRightNode})}, whereas the state of the system will remain the Minkowski vacuum state $|0_{\cal{M}}\rangle$.
This is just the sudden approximation of quantum mechanics,  
where if a potential suddenly changes its shape, then the original ground state can be expressed as a linear combination of the eigenstates of the new Hamiltonian, and thus the observables can be found by taking expectation values of operators in the modified system with respect to the ground state of the original Hamiltonian. 

Thus, we shall treat the rapid strain of graphene within the sudden approximation.  Before embarking on this, 
we note some conditions for the validity of this approximation.  If the Rindler strains develop too quickly, then perturbations can grow exponentially with time, marking the onset of turbulence. This means electronic transitions to higher bands in graphene, and formation of vortices and solitons \cite{Hung:2012nc,Chen:2021xhd} in cold atom honeycomb setups.   Such effects are beyond the scope of the present low-energy description.  On the other hand, 
if the onset of the strain is too slow, then the system will remain adiabatically in the ground state
and will end up in the Rindler vacuum.  Although we assume the strain onset to be sufficiently
rapid such that the sudden approximation holds (while neglecting the abovementioned transitions to higher bands), it would be valuable in future work to study such effects.

In the present case, applying the sudden approximation requires knowing how the Rindler operators $\hat{c}$ and $\hat{d}$ of the strained system act on the Minkowski vacuum state $|0_{\cal{M}}\rangle$ (the initial system). For this, we need to find an expression of these Rindler operators in terms of the Minkowski annihilation operators $\hat{a}$ and $\hat{b}$.
%
%
%
%
To do this, we can simply equate the two mode expansions {(\ref{MinkowskiModeExpansionStandardRight1})} and (\ref{ModeExpansionRightNode}) as they describe the same quantum field operator $\hat{\psi}_{\text{R}}$. Then we take the inner product of the resulting equations with positive energy solutions $\big(\psi^{+}_{\Omega}(x,k_{y}),\hat{\psi}_{\text{R}}(x)\big)$ for electron, and negative energy solutions $\big(\psi^{-}_{\Omega}(x,k_{y}),\hat{\psi}_{\text{R}}(x)\big)$ for hole Rindler operators, as defined in (\ref{Orthonormality}) {\cite{Takagi:1986kn,Birrell:1982ix}}, yielding:
\begin{eqnarray}\label{BogoliubovTransformation}
\hat{c}^{>}_{k_{y},\Omega} & = & \int d^{2}k'~ \Bigg[\alpha^{+,>}_{\boldsymbol{k}',k_{y},\Omega}\hat{a}_{\boldsymbol{k}'} + \beta^{+,>}_{\boldsymbol{k}',k_{y},\Omega}\hat{b}^{\dagger}_{\boldsymbol{k}'}\Bigg], \nonumber \\
\hat{d}^{>\dagger}_{k_{y},\Omega} & = & \int d^{2}k'~ \Bigg[\beta^{-,>}_{\boldsymbol{k}',k_{y},\Omega}\hat{a}_{\boldsymbol{k}'} + \alpha^{-,>}_{\boldsymbol{k}',k_{y},\Omega}\hat{b}^{\dagger}_{\boldsymbol{k}'}\Bigg],
\end{eqnarray}
the Bogoliubov transformations that express the Rindler ladder operators for $x>0$ (denoted by superscript $>$) as a linear combination of the Minkowski ladder operators. Similar relations hold for $x<0$ region with the superscript $<$ at the appropriate places.
Following Takagi \cite{Takagi:1986kn}, the coefficients of this linear relationship $\alpha^{\pm}_{\boldsymbol{k},k'_{y},\Omega'}$ and $\beta^{\pm}_{\boldsymbol{k},k'_{y},\Omega'}$, known as Bogoliubov coefficients, are found to be:


\begin{eqnarray}\label{BogoliubovCoeffs}
\alpha^{+,>}_{\boldsymbol{k}',k_{y},\Omega} & = & \sqrt{n_{\text{F}}(-2\pi\Omega)}~\delta(k_{y}-k'_{y})~\mathcal{P}(\boldsymbol{k}',\Omega), \nonumber \\
\beta^{+,>}_{\boldsymbol{k}',k_{y},\Omega} & = & -i \sqrt{n_{\text{F}}(2\pi\Omega)}~\delta(k_{y}+k'_{y})~\mathcal{P}(\boldsymbol{k}',\Omega), \nonumber \\
\alpha^{-,>}_{\boldsymbol{k}',k_{y},\Omega} & = & \alpha^{+,<}_{\boldsymbol{k}',k_{y},\Omega} =  \big(\alpha^{+,>}_{\boldsymbol{k}',k_{y},\Omega}\big)^{*} = \big(\alpha^{-,<}_{\boldsymbol{k}',k_{y},\Omega}\big)^{*}, \nonumber \\
\beta^{-,>}_{\boldsymbol{k}',k_{y},\Omega} & = & \beta^{+,<}_{\boldsymbol{k}',k_{y},\Omega} =  \big(\beta^{+,>}_{\boldsymbol{k}',k_{y},\Omega}\big)^{*} = \big(\beta^{-,<}_{\boldsymbol{k}',k_{y},\Omega}\big)^{*},~~~~~~~
\end{eqnarray}
where the first Bogoliubov coefficient $\alpha^{+,>}$ for the right side of graphene is found by taking the inner product of the positive energy Rindler spinor for $x>0$ with the positive energy Minkowski modes, whereas the second coefficient $\beta^{+,>}$ is found using the negative energy Minkowski modes. Similarly, the other two coefficients $\alpha^{-,>}$ and $\beta^{-,>}$ can be found by using negative energy Rindler spinors. In the last two lines, we list how the rest of the coefficients are related to the first two via complex conjugation. These coefficients are written in terms of the Fermi-Dirac function $n_{\text{F}}(x)=(e^{x}+1)^{-1}$ and the projection operator:
\begin{eqnarray}\label{Projection Operator}
\mathcal{P}(\boldsymbol{k},\Omega) & = & \frac{1+i}{\sqrt{2}}\frac{1}{\sqrt{2\pi k}}\bigg(\frac{k+k_{x}}{k-k_{x}}\bigg)^{\frac{i\Omega}{2}} \nonumber \\ 
& \times & \bigg(\sqrt{\frac{k+k_{x}}{2k}}+i\sqrt{\frac{k-k_{x}}{2k}}\bigg).~~~~~~
\end{eqnarray}
The anticommutation relations for the Rindler operators $\hat{c}_{k_{y},\Omega}$ and $\hat{d}_{k_{y},\Omega}$, along with those of the Minkowski operators
$\hat{a}_{\boldsymbol{k}}$ and $\hat{b}_{\boldsymbol{k}}$ and the transformations
Eq.~(\ref{BogoliubovTransformation}) imply the following normalization condition for the Bogoliubov coefficients:
\begin{eqnarray}
&&\int d^{2}\tilde{k}~ \Big(\alpha^{\sigma,r}_{\tilde{\boldsymbol{k}},k_{y},\Omega}\alpha^{\sigma',r'*}_{\tilde{\boldsymbol{k}},k'_{y},\Omega'} + \beta^{\sigma,r}_{\tilde{\boldsymbol{k}},k_{y},\Omega}\beta^{\sigma',r'*}_{\tilde{\boldsymbol{k}},k'_{y},\Omega'}\Big) \nonumber \\
& = & \delta^{\sigma\sigma'}\delta^{rr'}\delta(k_{y}-k'_{y})\delta(\Omega-\Omega'),
\end{eqnarray}
where the superscript $\sigma=\pm$ labels the positive and negative energy solutions, and $r=>,<$ labels the right ($x>0$) or left ($x<0$) region of graphene. To find these Bogoliubov coefficients, we made use of the Fourier transform of the modified Bessel functions of the second kind {\cite{Jauregui:1991me,Gradshteyn}}:
\begin{eqnarray}\label{FourierBessel}
&& \int_{0}^{\infty}dx~K_{\nu}(ax)e^{ibx} \nonumber \\
& = & \frac{\pi}{4\sqrt{a^{2}+b^{2}}}\Bigg[\frac{(\sqrt{r^{2}+1}+r)^{\nu}+(\sqrt{r^{2}+1}-r)^{\nu}}{\cos(\pi\nu/2)} \nonumber \\
& + & i~\frac{(\sqrt{r^{2}+1}+r)^{\nu}-(\sqrt{r^{2}+1}-r)^{\nu}}{\sin(\pi\nu/2)}\bigg]~~~~~
\end{eqnarray}
where $r=b/a$. The conditions required for the validity of the sine transform are $\text{Re}~a>0$, $b>0$, $|\text{Re}~\nu|<2$ and $\nu\neq 0$. Whereas the conditions for the cosine transform are $\text{Re}~a>0$, $b>0$, $|\text{Re}~\nu|<1$. For our case, $a=k_{y}>0$ and $\nu=\frac{1}{2}\pm i\Omega$ satisfy the conditions. However, $b=k_{x}$ could be positive or negative. For $k_{x}>0$ case, the above Fourier transform can be used whereas for $k_{x}<0$, one needs to take the complex conjugate of the above transform. 

Note that the transformation in {(\ref{BogoliubovTransformation})} and the corresponding Bogoliubov coefficients in {(\ref{BogoliubovCoeffs})}, can be re-written 
in a much cleaner way {\cite{Takagi:1986kn}}:
\bse
\label{Bogoliubov Transformations Actual}
\begin{eqnarray}
\label{Bog1}
\hat{c}^{>}_{k_{y},\Omega} & = &
\sqrt{n_{\text{F}}(-2\pi\Omega)}\hat{A}_{k_{y},\Omega} -i
\sqrt{n_{\text{F}}(2\pi\Omega)}\hat{B}^{\dagger}_{-k_{y},\Omega},~~~~~ \\ \label{Bog2} \hat{d}^{>\dagger}_{k_{y},\Omega} & = &
i\sqrt{n_{\text{F}}(2\pi\Omega)}\hat{A}^{*}_{-k_{y},\Omega} +
\sqrt{n_{\text{F}}(-2\pi\Omega)}\hat{B}^{*\dagger}_{k_{y},\Omega}.~~~~~
\end{eqnarray}
\ese
where instead of using momentum integrations as in (\ref{BogoliubovTransformation}), the Rindler operators are expressed in terms of modified Minkowski $\hat{A}$ and $\hat{B}$, that are defined as a complex linear combination of the original Minkowski operators $\hat{a}$ and $\hat{b}$ as follows {\cite{Takagi:1986kn}}:
\begin{eqnarray}\label{Modified & Actual Minkwoski Operators}
\hat{A}_{k_{y},\Omega} & = & \int_{-\infty}^{\infty} dk_{x}~\mathcal{P}(\boldsymbol{k},\Omega) ~\hat{a}_{\boldsymbol{k}}, \nonumber \\
\hat{B}^{\dagger}_{k_{y},\Omega} & = & \int_{-\infty}^{\infty}dk_{x}~\mathcal{P}(\boldsymbol{k},\Omega)
~\hat{b}^{\dagger}_{\boldsymbol{k}},
\end{eqnarray}
that (like the operators $\{\hat{a},\hat{b}\}$)  also annihilate the Minkowski vacuum: 
\be
\label{Modified Minkwoski Annihilate Miknkoski vacuum}
\hat{A}_{k_{y},\Omega} |0_{\cal{M}}\rangle =   \hat{B}_{k_{y},\Omega} |0_{\cal{M}}\rangle=  0 ,
\ee
which follows from Eq.~(\ref{Minkowski Operators1}).   In addition, they satisfy the
anti-commutation relations:
\begin{eqnarray}
\big\{\hat{A}_{k_{y},\Omega},\hat{A}^{\dagger}_{k'_{y},\Omega'}\big\} 
& = & \big\{\hat{B}_{k_{y},\Omega},\hat{B}^{\dagger}_{k'_{y},\Omega'}\big\} \nonumber \\
& = & \delta(k_{y}-k'_{y})\delta(\Omega-\Omega').
\label{Anti-Commutation Modified Minkwoski}
\end{eqnarray}
%
As a result of these properties, the expectation value of modified operators in the Minkowski vacuum state $|0_{\cal{M}}\rangle$ become:
\begin{eqnarray}\label{VEV of Modified Minkowski}
\big\langle 0_{\cal{M}} \big| \hat{A}_{k_{y},\Omega}\hat{A}^{\dagger}_{k'_{y},\Omega'} \big| 0_{\cal{M}} \big\rangle  & = & \big\langle 0_{\cal{M}} \big| \hat{B}_{k_{y},\Omega}\hat{B}^{\dagger}_{k'_{y},\Omega'} \big| 0_{\cal{M}} \big\rangle \nonumber \\
& = & \delta(k_{y}-k'_{y})\delta(\Omega-\Omega'), \nonumber \\
\big\langle 0_{\cal{M}} \big| \hat{A}^{\dagger}_{k_{y},\Omega}\hat{A}_{k'_{y},\Omega'} \big| 0_{\cal{M}} \big\rangle  & = & \big\langle 0_{\cal{M}} \big| \hat{B}^{\dagger}_{k_{y},\Omega}\hat{B}_{k'_{y},\Omega'} \big| 0_{\cal{M}} \big\rangle = 0,~~~~~~~
\end{eqnarray}
where in order to derive the Dirac delta function in energies $\delta(\Omega-\Omega')$, in the above vacuum averages, the following identity was used {\cite{Takagi:1986kn}}:
\begin{eqnarray}\label{Integral-kx-Indentity}
\int_{-\infty}^{\infty} \frac{dk_{x}}{2\pi k} \bigg(\frac{k+k_{x}}{k-k_{x}}\bigg)^{i(\Omega-\Omega')/2} & = & \int_{-\infty}^{\infty} \frac{dy}{2\pi} e^{i(\Omega-\Omega')y} \nonumber \\
& = & \delta(\Omega-\Omega'),
\end{eqnarray}
where in the first equality we made the substitution $y=\frac{1}{2}\log\big(\frac{k+k_{x}}{k-k_{x}}\big)$.

The advantage of (\ref{Bogoliubov Transformations Actual}) emerges when we evaluate the expectation value of Rindler operators in the Minkowski vacuum, where we only need vacuum averages of modified Minkowski operators, 
simplifying our calculations. Interestingly,
when we compute expectation values of the Rindler operators with
respect to the Minkowski vacuum we find that such averages 
involve an emergent Fermi distribution:
\begin{eqnarray}\label{Vacuum Averages of Rindler Operators}
\langle0_{\cal{M}}|\hat{c}^{>\dagger}_{k_{y},\Omega}\hat{c}^{>}_{k'_{y},\Omega'}|0_{\cal{M}}\rangle & = & \langle0_{\cal{M}}|\hat{d}^{>\dagger}_{k_{y},\Omega	}
 \hat{d}^{>}_{k'_{y},\Omega'}|0_{\cal{M}}\rangle, \nonumber \\
& = & n_{\text{F}}(2\pi\Omega)\delta(k_{y}-k'_{y})\delta(\Omega-\Omega'),~~~~~~~
\end{eqnarray}
that arise solely due to strains in the material, rather than due to any real heat bath. This implies that although the occupancy of Rindler electrons and holes in the Rindler vacuum is zero, in the Minkowski vacuum state it is proportional to the Fermi function. Thus surprisingly, spontaneous particle creation here has a spectrum that turns out to be \emph{thermal} in nature. This is known as the Fulling-Davies-Unruh effect 
which, in the conventional setting, says that an accelerating observer views the Minkowski spacetime as a thermal bath of particles at the Unruh temperature $T_{\text{U}}=\frac{\hbar a}{2\pi k_{\text{B}} c}$. 
We note that this result is of course well known in the relativity literature~\cite{Takagi:1986kn,Crispino:2007eb}.
However, its derivation here is
included both for completeness of presentation and because our mechanism for the Unruh effect
(a sudden switch on of a spatially-inhomogeneous strain) is different than the standard case
(an accelerating observer).   While a coordinate transformation relates these two pictures, 
our presentation in this section makes it clear how the Unruh effect (characterized by the
strain-dependent Unruh temperature $T_{\text{U}}=\frac{\hbar\omega_{c}}{2\pi k_{\text{B}}}=\frac{\hbar v_{0}}{2\pi k_{\text{B}}\lambda}$) emerges in strained graphene, providing a prologue for computing specific observables (as 
we do in subsequent sections).

To see how this thermality arises in a concrete way, we re-write Eq.~(\ref{Bogoliubov Transformations Actual}) for electrons in the right side of graphene $(x>0)$ and holes on the left side $(x<0)$:
\bse
\label{Bogoliubov Transformations ><}
\begin{eqnarray}
\label{Bog >}
\hat{c}^{>}_{k_{y},\Omega} & = &
\sqrt{n_{\text{F}}(-2\pi\Omega)}\hat{A}_{k_{y},\Omega} -i
\sqrt{n_{\text{F}}(2\pi\Omega)}\hat{B}^{\dagger}_{-k_{y},\Omega},~~~~~~~ \\ \label{Bog <} \hat{d}^{<\dagger}_{-k_{y},\Omega} & = &
-i\sqrt{n_{\text{F}}(2\pi\Omega)}\hat{A}_{k_{y},\Omega} +
\sqrt{n_{\text{F}}(-2\pi\Omega)}\hat{B}^{\dagger}_{-k_{y},\Omega}.~~~~~~~
\end{eqnarray}
\ese
where we made use of the symmetry properties of Bogoliubov coefficients in (\ref{BogoliubovCoeffs}) and we chose to evaluate the hole operator for $x<0$ region and with inverted momentum $-k_{y}$ with respect to the electrons. These can be inverted to write the modified operators in terms of Rindler operators:
\bse
\label{Inverse Bogoliubov Transformations ><}
\begin{eqnarray}
\label{Inverse Bogoliubov A}
\hat{A}_{k_{y},\Omega} & = &
\sqrt{n_{\text{F}}(-2\pi\Omega)}\hat{c}^{>}_{k_{y},\Omega} + i
\sqrt{n_{\text{F}}(2\pi\Omega)}\hat{d}^{<\dagger}_{-k_{y},\Omega},~~~~~~~ \\ \label{Inverse Bogoliubov B} \hat{B}^{\dagger}_{-k_{y},\Omega} & = &
i\sqrt{n_{\text{F}}(2\pi\Omega)}\hat{c}^{>}_{k_{y},\Omega} +
\sqrt{n_{\text{F}}(-2\pi\Omega)}\hat{d}^{<\dagger}_{-k_{y},\Omega}.~~~~~~~ 
\end{eqnarray}
\ese
Equation (\ref{Vacuum Averages of Rindler Operators}) suggests that what we see as the vacuum of a flat graphene sheet, may appear as a state filled with Rindler strained particles. Thus we can express the Minkowski vacuum in terms of Rindler excited states in the following way~\cite{Alsing:2006cj,Leon:2009uod}: 
\bea
\label{Ansatz Minkwoski in terms of Rindler}
&&|0_{\cal{M}}\rangle = \prod_{k_{y},\Omega} |0_{k_y,\Omega}\rangle_{\cal{M}}
\\
&&
|0_{k_y,\Omega}\rangle_{\cal{M}}
= \sum_{m,n=0}^{1} A_{mn} |m^{>}_{k_{y},\Omega}\rangle_{{\cal{R}}} ~|n^{<}_{-k_{y},\Omega}\rangle_{{\cal{R}}},
\eea
which expresses the Minkowski vacuum state in terms of a Rindler state with $m$ electrons on the right and $n$ holes on the left side. Note that the sum has only two entries because of the Pauli principle for fermions which according to (\ref{Modified Minkwoski Annihilate Miknkoski vacuum}), means that the electron annihilation operator (also true for holes) acting on the state with no electrons as well as the corresponding electron creation operator acting on a state with one electron will yield zero, i.e. $\hat{c}|0_{\cal{R}}\rangle=\hat{c}^{\dagger}|1_{\text{R}}\rangle=0$. Dropping the quantum labels $k_{y}$ and $\Omega$, and the subscript $\cal{R}$, and applying the modified Minkowski electron annihilation operator $\hat{A}_{k_{y},\Omega}$ to the above Minkowski state in Eq.~(\ref{Ansatz Minkwoski in terms of Rindler}), we get (\cite{Alsing:2006cj,Leon:2009uod}):
\begin{eqnarray}
0  &=&  \hat{A}
|0_{k_y,\Omega}\rangle_{\cal{M}}
\\
&=& \big[n^{\frac{1}{2}}_{\text{F}}(2\pi\Omega)A_{11} +  in^{\frac{1}{2}}_{\text{F}}(-2\pi\Omega)A_{00}\big]|0^{>}\rangle |1^{<}\rangle \nonumber \\
& + & n^{\frac{1}{2}}_{\text{F}}(-2\pi\Omega)A_{10}|0^{>}\rangle |0^{<}\rangle +  in^{\frac{1}{2}}_{\text{F}}(2\pi\Omega)A_{10}\big]|1^{>}\rangle |1^{<}\rangle. \nonumber 
\end{eqnarray}
If the right hand side vanishes for arbitrary Rindler Fock states, then this yields the summation coefficients as $A_{10}=A_{01}=0$, and $A_{11}=-iA_{00}e^{-\pi\Omega}$. Also normalizing the ansatz in (\ref{Ansatz Minkwoski in terms of Rindler}) yields $|A_{00}|^{2}+|A_{11}|^{2}=1$. Combing these ideas, we get $A_{00}=n_{\text{F}}^{\frac{1}{2}}(2\pi\Omega)$ and $A_{11}=-in_{\text{F}}^{\frac{1}{2}}(-2\pi\Omega)$, and therefore the flat graphene vacuum state can be expressed as a two-mode squeezed state of Rindler-strained fermions:
\begin{eqnarray}
&& |0_{\cal{M}}\rangle = \prod_{k_{y},\Omega} n_{\text{F}}^{\frac{1}{2}}(-2\pi\Omega) \nonumber \\
& \hspace{-0.2cm} \times & \Big[|0^{>}_{k_{y},\Omega}\rangle_{\cal{R}}~ |0^{<}_{-k_{y},\Omega}\rangle_{\cal{R}} -ie^{-\pi\Omega}|1^{>}_{k_{y},\Omega}\rangle_{\cal{R}}~|1^{<}_{-k_{y},\Omega} \rangle_{\cal{R}}\Big],~~~~~~~
\end{eqnarray}
similar to the Bardeen-Cooper-Schrieffer (BCS) state \cite{BCS Short,BCS Long} for electrons that form a Cooper pair \cite{Cooper} inside a superconductor or superfluid. From this, a density matrix can be constructed $\hat{\rho}=|0_{\cal{M}}\rangle\langle0_{\cal{M}}|$ representing the pure state of the flat graphene sheet, and when traced over the left side ($x<0$) Rindler particle states, we get a reduced density matrix in terms of the Rindler Hamiltonian expressed in terms of modes pertaining to the right side only ($x>0$):
\begin{equation}\label{Density Matrix Reduced Gibbs Form}
\hat{\rho}^{>} = \frac{e^{-2\pi\hat{H}^{>}}}{\text{Tr}~e^{-2\pi\hat{H}^{>}}} 
\end{equation}
where the normal ordered Hamiltonian constrained to the right side should be understood in terms of a sum in modes $\hat{H}^{>}=\sum_{k_{y},\Omega}\Omega\{\hat{c}^{>\dagger}_{k_{y},\Omega}\hat{c}^{>}_{k_{y},\Omega} + \hat{d}^{>\dagger}_{k_{y},\Omega}\hat{d}^{>}_{k_{y},\Omega}\}$. This density matrix is clearly of the Gibbs' thermal ensemble form. 

In the case of the conventional Unruh effect with an accelerating observer, the Rindler horizon that bars any communication between the two wedges presents a natural trace of the density matrix. In the present setting
of a strained honeycomb lattice, the low-energy and long-wavelength
modes see the point $x=0$ as 
an analogue horizon and thus leakage of such modes between the two sides is either zero or minuscule (see Appendix \ref{SEC:Appendix Horizon Graphene} for more details).
Hence, even though the global state of the honeycomb system might be a pure state, when we make measurements on one side of the sheet, the degrees of freedom on the other side are not available to us and hence get \emph{naturally} traced out from the density matrix giving us a reduced mixed thermal state as in Eq.~{(\ref{Density Matrix Reduced Gibbs Form})} {\cite{Fabbri:2005mw,Takagi:1986kn,DeWitt:1979,Crispino:2007eb}}. This is known as the thermalization theorem which says that the presence of horizons in a spacetime is sufficient for thermality to emerge. It is intimately connected to the Kubo-Martin-Schwinger (KMS) condition {\cite{Kubo:1957,Martin & Schwinger:1959}} and the principle of detailed balance which we shall discuss in the next section. 
Thus any strain pattern that realizes an analogue spacetime with a natural horizon such as black holes, de-Sitter or Rindler, can lead to the appearance of such thermal effects.

So far we have discussed how a Rindler Hamiltonian
{(\ref{Rindler Hamiltonian})} forms from assuming a
linear-in-position Fermi velocity $v(x)=v_{0}\frac{|x|}{\lambda}$, how
this leads to the Bogoliubov transformations
{(\ref{Bogoliubov Transformations Actual})} between the
strained (Rindler) and flat (modified Minkowski) honeycomb operators,
giving rise to the vacuum averages in 
{(\ref{Vacuum Averages of Rindler Operators})} that behave as thermal averages over an
ensemble represented by the density matrix
{(\ref{Density Matrix Reduced Gibbs Form})}.
As emphasized by Rodr\'iguez-Laguna et al~\cite{Rodriguez-Laguna:2016kri}, an important aspect of the Unruh effect 
is the fact that the Minkowski vacuum is stationary with respect to the Rindler Hamiltonian. This is reflected 
in the form of the density matrix (\ref{Density Matrix Reduced Gibbs Form}) 
and in averages like Eq.~(\ref{Vacuum Averages of Rindler Operators}).  Thus,
while one might expect rapid post-quench dynamics to \lq\lq wash out\rq\rq\ the Unruh effect after a sudden 
switch-on of the strain, here (in the limit of an exact Rindler Hamiltonian after the quench), such dynamics are not
expected.
%

These
results, collectively  termed the Unruh effect,  emerge due
to the presence of a natural demarcation in the material. Before we
discuss the implications of this spontaneous electron-hole formation
on observables like the electronic conductivity and internal energy, in
the next section we will present the Green's functions pertaining to
the strained graphene system to discuss in what sense is the Unruh
effect a genuine thermal phenomena. We will also discuss how the
dimensionality of graphene leads to the violation of Huygens'
principle and the inversion of statistics which could possibly be seen
in photo-emission experiments.

\section{Green's Functions}\label{SEC:six}
In this section, we describe properties of single-particle Green's functions
that will help us explain how thermal behavior emerges, how the Huygens' principle is violated and how this leads to the phenomena of apparent statistics inversion in the excitation spectrum of fermions. Towards the end of this section, we discuss how these properties can be detected in experiments like photoemission spectroscopy (PES) and scanning tunneling microscopy (STM). 

Following Ooguri {\cite{Ooguri:1985nv}}, we introduce two fundamental single particle Green's functions defined with respect to the flat graphene vacuum state $|0_{\cal{M}}\rangle$:
\bse
\begin{eqnarray}\label{G+}
G_{+}(\boldsymbol{r},t;\boldsymbol{r}',t') & = & \langle 0_{\cal{M}}|\hat{\psi}_{\text{R}}(x,y,t)\hat{\psi}^{\dagger}_{\text{R}}(x',y',t')|0_{\cal{M}}\rangle,~~~~~~~ \\ \label{G-}
G_{-}(\boldsymbol{r},t;\boldsymbol{r}',t') & = & \langle 0_{\cal{M}}|\hat{\psi}^{\dagger}_{\text{R}}(x,y,t)\hat{\psi}_{\text{R}}(x',y',t')|0_{\cal{M}}\rangle.~~~~~~~
\end{eqnarray}
\ese
Here $G_{+}$ creates a particle at location $\boldsymbol{r}'=(x',y')$ and time $t'$, and then annihilates it at another location $\boldsymbol{r}=(x,y)$ and time $t$, whereas $G_{-}$ does the opposite. In the condensed-matter context, these are called the $>$ and $<$ Green's functions,
 respectively~\cite{Coleman} (up to factors of $i$), and their physical  interpretation will become clear when we discuss their Fourier transforms below. 

Interestingly, despite the intrinsically nonequilibrium nature of this setup, i.e., a sudden switch-on of the system strain that changes the system Hamiltonian from the Dirac to the Rindler Hamiltonian, these Green's functions have simple forms, at least in the local real-space limit.  To see this, we
set the positions equal i.e. $x'=x$ and $y'=y$. Making use of mode expansion {(\ref{ModeExpansionRightNode})} for the right node fields and the vacuum averages {(\ref{Vacuum Averages of Rindler Operators})} for Rindler ladder operators with respect to Minkowski vacuum, and taking the spinor trace, we find:
\begin{eqnarray}\label{G+ special}
&&\text{Tr}~G_{+}(x,y,\Delta t) = \frac{1}{2\pi x^{2}}\int_{0}^{\infty} d\Omega~\Omega\coth\pi\Omega \nonumber \\
& \times & \Big[e^{i\Omega\Delta t}n_{\text{F}}(2\pi\Omega) + e^{-i\Omega\Delta t}n_{\text{F}}(-2\pi\Omega)\Big],~~~~~~ \\ \label{G- special}
&&\text{Tr}~G_{-}(x,y,\Delta t) = \frac{1}{2\pi x^{2}}\int_{0}^{\infty} d\Omega \Omega\coth\pi\Omega \nonumber \\
& \times & \Big[e^{i\Omega\Delta t}n_{\text{F}}(-2\pi\Omega) + e^{-i\Omega\Delta t}n_{\text{F}}(2\pi\Omega)\Big],~~~~~~
\end{eqnarray}
where $\Delta t=(t-t')$. 
If we define a typical timescale associated with the Unruh temperature, $\hbar/(k_{\rm B} T_{\rm U})$ (equal to $2\pi$ in our
units) then it can be shown that the above Green's functions are periodic in imaginary shifts by this timescale:  
\begin{equation}\label{KMS Condition}
\text{Tr}~G_{+}(x,y,\Delta t-2\pi i) = \text{Tr}~G_{-}(x,y,\Delta t).
\end{equation}
This is known as the Kubo-Martin-Schwinger (KMS) condition {\cite{Kubo:1957,Martin & Schwinger:1959}} which in the conventional equilibrium case at 
temperature $T$, guarantees that the thermal average of any two operators $\hat{A}$ and $\hat{B}$ for a system kept in contact with a heat bath at temperature $\beta=(k_{\text{B}}T)^{-1}$ is also periodic in imaginary time, i.e. $\langle\hat{A}(t)\hat{B}(t')\rangle=\langle\hat{B}(t')\hat{A}(t+i\beta)\rangle$. For example, if we take the operators $\hat{A}$ and $\hat{B}$ as the graphene right-node field operators, then we get the following KMS condition for the Green's functions in (\ref{G+}) and (\ref{G-}):
\begin{equation}\label{KMS General}
G_{+}(\boldsymbol{r},\boldsymbol{r}',\Delta t-2\pi i) = G_{-}(\boldsymbol{r},\boldsymbol{r}',\Delta t).
\end{equation}
Note that here we have assumed that the Green's functions depend solely on the time difference $\Delta t$ because the system exhibits time translation invariance when it is in thermal equilibrium.
In an isolated strained graphene sheet, this condition implies that the vacuum (pure state) average of field operators behaves as a legitimate thermal (mixed state) average with respect to the reduced density operator (\ref{Density Matrix Reduced Gibbs Form}) (that can be thought of as an evolution operator \cite{Coleman,Matsubara:1955ws}), as if it is kept in contact with a real heat bath set at the Unruh temperature, i.e. $T=T_{\text{U}}$.

To further understand the meaning of the KMS condition, we take the Fourier transforms of the above Green's functions {(\ref{G+ special})} and {(\ref{G- special})} defined as 
\be
F_{\pm}(x,\omega)=\int_{-\infty}^{\infty}d(\Delta t)~e^{-i\omega\Delta t}\text{Tr}~ G_{\pm}(x,y,\Delta t),
\ee
from which we obtain:
\begin{eqnarray}\label{Power Spectrum}
F_{+}(x,\omega) & = & \frac{\omega}{x^{2}}n_{\text{B}}(2\pi\omega), \\
F_{-}(x,\omega) & = & -\frac{\omega}{x^{2}}n_{\text{B}}(-2\pi\omega).
\end{eqnarray}
As discussed by Coleman~\cite{Coleman}, $F_{+}(x,\omega)$ is the photo-emission spectra that gives the total excitation of electrons when graphene is illuminated by light. Similarly,  $F_-(x,\omega)$ measures the de-excitation spectra. The ratio of these two power spectra turns out to be:
\begin{eqnarray}
\label{eq:ratioPS}
\frac{F_{+}(x,\omega)}{F_{-}(x,\omega)} = e^{-2\pi\omega}
\end{eqnarray}
which says that the rate of excitation versus de-excitation is of the Boltzmann form with the strain frequency $1/2\pi$ playing the role of temperature. This is the principle of detailed balance which originates from Boltzmann's principle of microscopic reversibility {\cite{Boltzmann,Tolman}}, but was first applied to quantum systems by Einstein in {\cite{Einstein}} that predicted the phenomena of stimulated emission. He studied the set up where an atom with two energy levels $E_{1}<E_{2}$ is in thermal contact with a bath of photons such that when equilibrium sets in the ratio of number of particles in the excited state $|E_{2}\rangle$ versus $|E_{1}\rangle$ is $e^{-\beta(E_{2}-E_{1})}$. Then by demanding that the excitation probability should match de-excitation (spontaneous and stimulated) at equilibrium, the number distribution of photons will be given by a Planck distribution $\rho(\omega)=(\exp(\beta\omega)-1)^{-1}$, where $\omega=(E_{2}-E_{1})$ is the energy of the photon wave-packet that is absorbed by the two-level atoms. Such two-level systems are termed Unruh-DeWitt detectors in the relativistic context~\cite{Unruh:1976db,DeWitt:1979}. Thus the Fourier transform of the KMS condition, i.e., the principle of detailed balance, tells us that accelerated fermionic fields have a Fermi-Dirac spectrum and when they are in contact with a two-level or more atom or detector, then the latter comes into global thermal equilibrium with the field with the Unruh temperature defined everywhere on the real or analogue spacetime. 

The discussion above can be summarized by stating the thermalization theorem. For a comprehensive account of its various versions, see {\cite{Takagi:1986kn}}. It states that if the spacetime (or the analogue system) has a causal horizon (like the Minkowski spacetime in Rindler coordinates), then any quantum field on that spacetime will spontaneously emit particles in a thermal distribution characterized by a Bose or Fermi function which is captured by the reduced density matrix {(\ref{Density Matrix Reduced Gibbs Form})} in Gibbs' ensemble form. Once this density operator is achieved, then the KMS condition {(\ref{KMS Condition})}, or more generally Eq.~(\ref{KMS General}), guarantees that the field will also thermalize any other system (like an atom or a detector) in its contact, that has energy levels, thus establishing a global thermal equilibrium with temperature $T=1/2\pi$.

\begin{figure}[h!]
	\includegraphics[width=1.0\columnwidth]{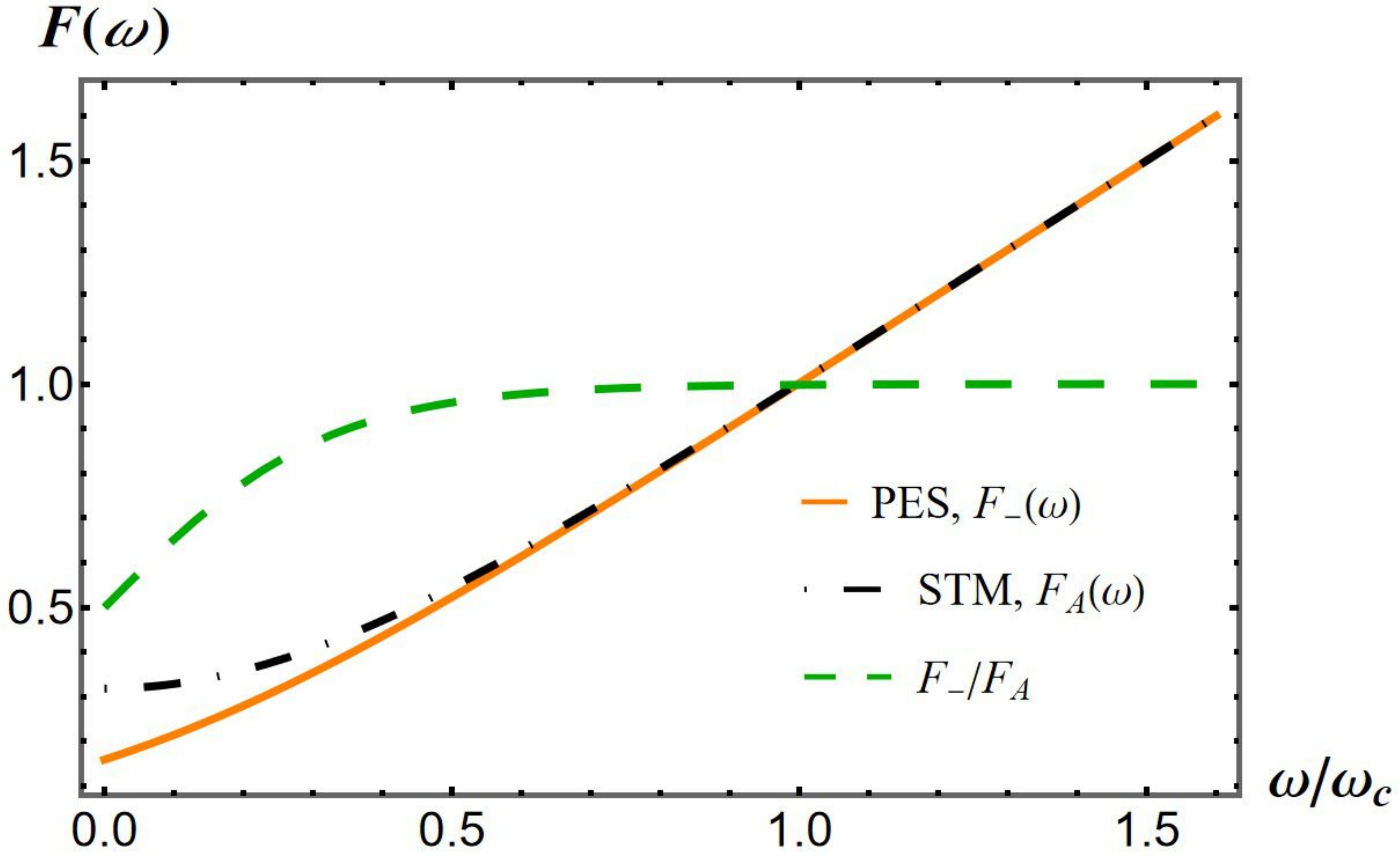}
	\caption{(Color Online) The orange solid curve 
	is a plot of the power spectrum 
	$F_{-}(\omega)$, that can be measured in 
	photo-emission spectroscopy (PES) experiments.
	The black dot-dashed curve is a plot of $F_{A}(\omega)$ that
	can be measured in scanning-tunneling microscopy (STM) experiments that measure the density of states.
	Their ratio (in green), yields the expected Fermi-Dirac spectrum in
accordance with the Unruh effect predictions. 	
	}
	\label{ARPES & STM}
\end{figure}



To discuss Huygens' principle and how its violation leads to statistics inversion, we now consider two other fundamental Green's functions pertaining to the commutator and the anti-commutator of fermionic fields, that are similar to the Green's functions defined in {(\ref{G+ special})} and {(\ref{G- special})}. The former is related to the Keldysh Green's function \cite{RammerSmith1986} and the 
latter is related to the retarded Green's function that takes causality into account:
\bse
\begin{eqnarray}\label{GC}
\hspace{-0.2cm} G_{C}(\boldsymbol{r},t;\boldsymbol{r}',t') & = & \langle 0_{\cal{M}}|\Big[\hat{\psi}_{\text{R}}(x,y,t),\hat{\psi}^{\dagger}_{\text{R}}(x',y',t')\Big]|0_{\cal{M}}\rangle,~~~~~~~ \\ \label{GA}
\hspace{-0.2cm} G_{A}(\boldsymbol{r},t;\boldsymbol{r}',t') & = & \langle 0_{\cal{M}}|\Big\{\hat{\psi}^{\dagger}_{\text{R}}(x,y,t),\hat{\psi}_{\text{R}}(x',y',t')\Big\}|0_{\cal{M}}\rangle.~~~~~~~
\end{eqnarray}
\ese
After setting $x= x'$ and $y=y'$, computing these Green's functions, and taking the trace, we get:
\begin{eqnarray}
\label{G-C}
\text{Tr}~G_{C}(x,y,\Delta t) & = & -\frac{i}{\pi x^{2}} \int_{0}^{\infty} d\Omega~\Omega\sin(\Omega\Delta t),
\\
\label{G-A}
\text{Tr}~G_{A}(x,y,\Delta t) & = & \frac{1}{\pi x^{2}} \int_{0}^{\infty} d\Omega~\Omega\coth(\pi\Omega)\cos(\Omega\Delta t).~~~~~~~ 
\end{eqnarray}
Conventionally, the Huygens' principle states that, if we have a source in even spacetime dimensions, then its wave-fronts can be constructed by drawing circles (appropriate to the dimensions) with the source at the center {\cite{Takagi:1986kn}. This means that the retarded Green's function that describes the propagation of waves to any point $(x,y,t)$ with the source at $(x',y',t')$ has support only on the light cone, i.e. it vanishes when $(x',y',t')$ and $(x,y,t)$ are either timelike or spacelike separated. This implies that the retarded Green's function in even spacetime dimensions is proportional to a Dirac delta function and its derivatives. 
However, strained graphene mimics an odd-dimensional spacetime where we find that the anticommutator in Eq.~(\ref{G-A}) is not a Dirac delta function. This is the manifestation of the well-known violation of Huygens' principle {\cite{Takagi:1986kn},\cite{Courant-Hilbert}}. It says that in odd spacetime dimensions, 
our intuition for wave propagation breaks down, i.e. 
a sharp source of wave does not lead to a single spherical wavefront, and instead the observer notices a continuously decreasing tail.

Curiously, although the anticommutator Green's function violates Huygens' principle, from Eq.~(\ref{G-C}) we see that
the commutator Green's function $G_C$ amounts  
 to $\frac{2i}{x^{2}}\delta'(\Delta t)$, i.e. it has support on the light cone. As a result, it is expected that the Fourier transform of the $G_{C}$ will be a polynomial in $\omega$, whereas for $G_{A}$ it will lead to the following: 
\begin{eqnarray}
\label{Commutator Polynomial}
F_{C}(x,\omega) & = & -\frac{\omega}{x^{2}},
\\
\label{Density of States}
F_{A}(x,\omega) & = & \frac{\omega}{x^{2}}\coth\pi\omega.
\end{eqnarray}
To see the connection of this violation of Huygens' principle with statistics inversion, we need the fluctuation-dissipation theorem. They can be derived in general by writing {(\ref{G-A})} and {(\ref{G-C})} in terms of {(\ref{G+})} and {(\ref{G-})}, i.e. $G_{A}=G_{+}+G_{-}$ and $G_{C}=G_{+}-G_{-}$, Fourier transforming them, and finally applying the KMS condition or the principle of detailed balance i.e. $F_{-}=e^{2\pi\omega}F_{+}$, yields two different but equivalent versions of the theorem:
\begin{eqnarray}\label{Fluctuation-Dissipation Version 1}
F_{+}(x,\omega) & = & n_{\text{F}}(2\pi\omega) \times F_{\text{A}}(x,\omega), \\ \label{Fluctuation-Dissipation Version 2}
& = & n_{\text{B}}(2\pi\omega) \times -F_{\text{C}}(x,\omega).
\end{eqnarray}
The excitation or power spectrum $F_{+}(x,\omega)$ is related to the rate at which an accelerated detector senses Rindler particles, and shows inversion of statistics depending on the dimension of the spacetime~\cite{Takagi:1986kn,Unruh:1986tc,Ooguri:1985nv,Terashima:1999xp,Sriramkumar:2002dn,Sriramkumar:2002nt,Pascazio_Huygens,Arrechea:2021szl}.
 Following Ooguri {\cite{Ooguri:1985nv}}, there are two interpretations for this. The first makes use of {(\ref{Fluctuation-Dissipation Version 1})}, which says that the excitation spectrum is basically the Fermi-Dirac function coming from the real statistics of the fermions, multiplied with the spectral density of states coming from the Fourier transform of anti-commutator which we know violates Huygens' principle and thus is not simply a polynomial in $\omega$. This, coupled with the particular form of the mode functions in {(\ref{ModeExpansionRightNode})} gives us a hyperbolic cotangent which coincidentally inverts the Fermi to a Bose function. The other interpretation comes from {(\ref{Fluctuation-Dissipation Version 2})}, where one can argue that since we are in odd spacetime dimensions in graphene, therefore we would expect the Fourier transform of the commutator to be polynomial in 
$\omega$ (see {(\ref{Commutator Polynomial})}). Thus the excitation spectrum should be expected to be a Bose-Einstein distribution multiplied by a factor which is polynomial in $\omega$, thereby removing the need to invoke any inversion. \\

To see how these power spectra could manifest themselves in experiments, we focus at the first version (\ref{Fluctuation-Dissipation Version 1}) of the Fluctuation-Dissipation theorem, but instead for $F_{-}$, i.e. $F_{-}(\omega)=n_{\text{F}}(-2\pi\omega)F_{\text{A}}(\omega)$. To do this, the experimenter will first obtain the photo-emission data from the Photo Emission Spectroscopy or the PES experiment {\cite{Rodriguez-Laguna:2016kri}}. For low-energies and long wavelengths, this will give us a plot of fermion occupancy in graphene's lowest energy band which in this limit, should mimic the Planck distribution $F_{-}=-\frac{\omega}{x^{2}}n_{\text{B}}(-2\pi\omega)$. As can be seen from Fig.~\ref{ARPES & STM}, $F_{-}(\omega)$ increases with energy, which is due to the fact that the PES-experiment measures the occupancy of valence band electrons by extracting them by shining light. If the intensity of light is increased, then more electrons residing in the lower valence energy levels will be detected. The experimenter can then obtain the data regarding the local density of states by performing the Scanning Tunneling Microscopy or the STM experiment {\cite{Iorio:2011yz,Iorio:2013ifa}} which, in the low-energy
and long-wavelength limit (where our calculations are valid), will be given by the statistics inversion factor $F_{A}=\frac{\omega}{x^{2}}\coth\pi\omega$, which implies that Huygens' principle is being violated in strained graphene. Now if we take the ratio of the PES and STM data, we will find:
\begin{equation}
\frac{\text{PES data}}{\text{STM data}} = \frac{F_{-}(x,\omega)}{F_{A}(x,\omega)} = n_{\text{F}}(-2\pi\omega),
\end{equation}
we will obtain the Fermi-Dirac distribution as expected from the Unruh effect of fermions, as can be seen in Fig.~\ref{ARPES & STM}. For the case of cold atom honeycomb setups, radio-frequency spectroscopy could be performed where photons transfer atoms from an occupied to an unoccupied auxiliary band \cite{Dao:2007,Stewart:2008}. After this, the optical laser trap would be turned off and a time of flight absorption imaging would be performed which will yield the energy and momentum resolved photoemission data. When this is summed over all possible momenta, it will give us the density of states of the occupied band \cite{Stewart:2008}.
%
%

Note that the experiments for photo-emission and density of states use interactions between photons and electrons, and thus could affect the state of the system $|0_{\mathcal{M}}\rangle$ by evolving it into an interacting vacuum state Ref.\cite{Rodriguez-Laguna:2016kri}. However, since these interactions are perturbative in nature, therefore to first order, all observables, i.e., vacuum expectation values such as the Green’s functions and conductivity will be the same as the results presented in this paper. Also note that, the timescale of these interactions should be less than or equal to timescale of Unruh effect, which for strains of $\lambda=1$mm size results in $1$ns. This way, the PES photons will feebly interact with the emergent electrons and excite them out of the material.

Equipped with the Bogoliubov transformations {(\ref{Bogoliubov Transformations Actual})} between the strained (Rindler) and flat (modified Minkowski) honeycomb operators, that lead to the vacuum averages in {(\ref{Vacuum Averages of Rindler Operators})} and the statistics inversion  in Eqs.~{(\ref{Fluctuation-Dissipation Version 1})}-{(\ref{Fluctuation-Dissipation Version 2})}, we are now ready to discuss in the next two sections, the implications of this spontaneous electron-hole formation on observables like the electronic conductivity and total internal energy.

\section{Electronic Conductivity}
\label{SEC:seven}
In this section, we consider another observable that is sensitive to the Unruh effect in
strained graphene, the frequency-dependent conductivity.  For this calculation, we shall require the 
Bogoliubov transformations {(\ref{Bogoliubov Transformations Actual})}, derived
in Sec.~\ref{SEC:five}, that establish the relationship between the Rindler operators $\{\hat{c},\hat{d}\}$ in a strained honeycomb system with the modified Minkowski operators $\{\hat{A},\hat{B}\}$ in a flat (unstrained) honeycomb system. This led us to the expectation value {(\ref{Vacuum Averages of Rindler Operators})} of the Rindler operators with respect to the Minkowski vacuum.
To use these results, we will need the Kubo formula that 
 relates the frequency-dependent conductivity to an associated current-current correlation function.  %
 For generality, we'll briefly recall the Kubo formula derivation for both the setups
 considered here, i.e., the case of
 electronic graphene  (in which the fermions are charged electrons) and the case of neutral cold atoms in an optical lattice. 
 
 For the electronic graphene case, we can start with the Rindler Hamiltonian {(\ref{Rindler Hamiltonian})} minimally coupled to an electromagnetic vector potential $\boldsymbol{A}(\boldsymbol{r},t)$, i.e. we can make the replacement $-i\hbar\boldsymbol{\nabla}\rightarrow-i\hbar\boldsymbol{\nabla}-e\boldsymbol{A}$ in the derivative operators giving us the following new Hamiltonian {\cite{Mahan}}:
\begin{eqnarray}\label{Electromagnetic Coupling}
\hat{H}(t) & = & \hat{H}_{\text{R}} + \hat{H}^{1}(t), \nonumber \\
\hat{H}^{1}(t) & = & - \int d^{2}r~ \boldsymbol{\hat{j}}(\boldsymbol{r},t)\cdot \boldsymbol{A}(\boldsymbol{r},t),
\end{eqnarray}
where $\hat{H}_{\text{R}}$ is the Rindler Hamiltonian {(\ref{Rindler Hamiltonian})}.  Here, the conserved current operator in the strained (Rindler) system is: 
\be
\label{eq:rindlercurrent}
\boldsymbol{\hat{j}}(\boldsymbol{r},t) \equiv ev_{0}\frac{|x|}{\lambda} \hat{\psi}^{\dagger}_{\text{R}}(\boldsymbol{r},t)\boldsymbol{\sigma}\hat{\psi}_{\text{R}}(\boldsymbol{r},t).
\ee
Within linear response theory, we can treat the vector potential term as a perturbation, and to linear order  the response of the average current is given by:
\begin{eqnarray}\label{Kubo Formula}
&&\langle\hat{j}_{\mu}(\boldsymbol{r},t)\rangle = -\frac{i}{\hbar}\int_{-\infty}^{t}dt'~\big\langle\big[\hat{j}_{\mu}(\boldsymbol{r},t),\hat{H}^{1}(t')\big]\big\rangle,  \\ 
& = & \frac{i}{\hbar}\int_{-\infty}^{t}dt'\int d^{2}r'~\big\langle\big[\hat{j}_{\mu}(\boldsymbol{r},t),\hat{j}_{\nu}(\boldsymbol{r}',t')\big]\big\rangle A_{\nu}(\boldsymbol{r}',t').\nonumber 
\end{eqnarray}
The time-dependent vector potential can be written as 
$A_{\nu}(\boldsymbol{r}',t')=\frac{1}{i\omega^{+}}E_{\nu}(\boldsymbol{k},\omega)e^{-i(\boldsymbol{k}\cdot\boldsymbol{r}'+\omega^{+}t')}$, where $\omega^{+}=\omega+i\delta$ , with $\delta=0^{+}$.  Here, $E_{\nu}(\boldsymbol{k},\omega)$ is the electric
field at wavevector $\boldsymbol{k}$ and frequency $\omega$.  Upon plugging this into Eq.~(\ref{Kubo Formula}), multiplying
both sides by 
 $e^{-i\boldsymbol{q}\cdot\boldsymbol{r}}$ and integrating over $\br$ in the limit of   $\boldsymbol{q}\rightarrow0$ (corresponding to spatial averaging), we obtain:
\begin{eqnarray}\label{Current q0}
\langle\hat{j}_{\mu}(\boldsymbol{q}\rightarrow0,t)\rangle & = & \frac{1}{\hbar\omega^{+}}\int_{-\infty}^{\infty}dt'\Theta(t-t')e^{-i\omega^{+}t'} \nonumber \\
& \times & \big\langle\big[\hat{j}_{\mu}(0,t),\hat{j}_{\nu}(0,t')\big]\big\rangle E_{\nu}(0,\omega).~~~
\end{eqnarray}
Noting that the time-dependent electric field $E_{\nu}(0,t) = E_{\nu}(0,\omega) e^{-i\omega^{+}t}$, 
redefining the variable of integration to $T=(t-t')$ and taking the ratio of current and electric field, we find the average conductivity tensor $\sigma_{\mu\nu}=\frac{\langle\hat{j}_{\mu}(0,t)\rangle}{E_{\nu}(0,t)}$:
\begin{equation}\label{ConductivityTensor}
\sigma_{\mu\nu} = \frac{1}{\hbar\omega^{+}}\int_{-\infty}^{\infty}dT~\Theta(T)e^{i\omega^{+}T}\big\langle\big[\hat{j}_{\mu}(0,T),\hat{j}_{\nu}(0,0)\big]\big\rangle.
\end{equation}
The preceding derivation depends on the use of the vector potential as an external stimulus.   But, for a system that is not made of charged particles such as neutral ultracold atomic gases, we must use a different approach.  In this case, 
a change in the local chemical potential creates a pressure difference and hence affects the density of fermions. Instead of Eq.~(\ref{Electromagnetic Coupling}),
the perturbing Hamiltonian involves a coupling of the atom density $\hat{n}(\boldsymbol{r},t)=\hat{\psi}^{\dagger}(\boldsymbol{r},t)\hat{\psi}(\boldsymbol{r},t)$ to a spatially and temporally varying chemical potential: 
\be
\hat{H}^{1}(t) = - \int d^{2}r~ \mu(\boldsymbol{r},t)\hat{n}(\boldsymbol{r},t).
\ee
%
%
Plugging this into Eq.~(\ref{Kubo Formula}) with $\mu(\boldsymbol{r},t)=\mu(\boldsymbol{r})e^{-i\omega t}$,
integrating by parts in the $t'$ integral and also 
integrating by parts in space using the equation of continuity $0 = \frac{\partial}{\partial t}\hat{n}(\boldsymbol{r},t)+\boldsymbol{\nabla}
\cdot\hat{\boldsymbol{j}}(\boldsymbol{r},t)$ ,
we finally arrive at the Kubo formula
for neutral atoms, with the average atom current 
related to the chemical potential gradient as
 $\boldsymbol{j}=-\sigma\boldsymbol{\nabla}\mu$ where $\sigma$ is given by ({\ref{ConductivityTensor}}). 

Thus, in either case we require the current-current correlation function, with the averages being
performed with respect to the Minkowski vacuum.   
We start with the computation of $\sigma_{xx}$.
Instead of directly using Eq.~(\ref{ConductivityTensor}) that involves the spatially Fourier-transformed current correlator, we 
start with the real-space current-current correlation function,
%
perform spatial averages (on $\boldsymbol{r}$ and $\boldsymbol{r}'$), and finally Fourier transform to frequency space. The average current correlation function at the right node has the following form :
\begin{eqnarray}\label{Current-Current xx-Correlation}
\hspace{-0.2cm} \bar{C}^{xx}(t-t') & = & \int d^{2}r\int d^{2}r' \langle 0_{\cal{M}}| \hat{j}^{x}(\boldsymbol{r},t) \hat{j}^{x}(\boldsymbol{r}',t') |0_{\cal{M}} \rangle,~~~~~~~
\end{eqnarray}
where we are evaluating the correlations only between fields on the right node. In what follows, we will set $e\rightarrow1$, $\hbar\rightarrow1$ and $\omega_{c}\rightarrow1$.
%


\begin{figure}[h]
	\centering
	\includegraphics[width=1.0\columnwidth]{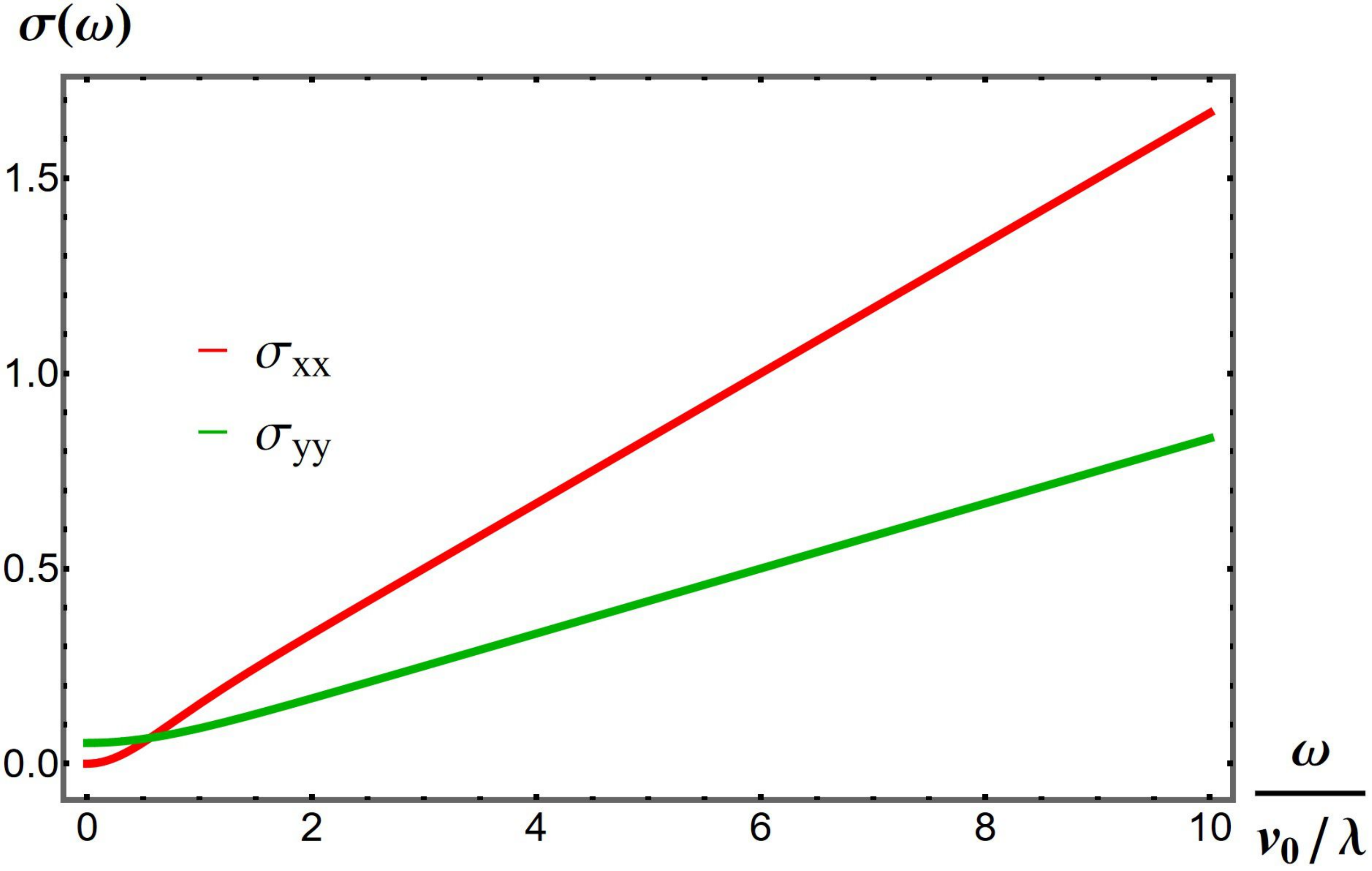}
	\caption{(Color Online) A plot showing how the average dissipative conductivity (in units of $e^{2}/\hbar$) grows approximately linearly as a function of AC-frequency (in units of strain frequency $\omega_{c}=v_{0}/\lambda$). The longitudinal component $\sigma'_{xx}(\omega)$ (in {red}) vanishes in the DC-limit $\omega\rightarrow0$, whereas $\sigma'_{yy}(\omega)$ (in green) reaches a value of $1/(6\pi)$.
	}
	\label{Average Conductivity}
\end{figure}

We performed spatial integrals on (\ref{Current-Current xx-Correlation}) along the coordinates $\boldsymbol{r}$ and $\boldsymbol{r}'$ because the conductivity Eq.~(\ref{ConductivityTensor}) requires the current-current correlation in the reciprocal space in the limit $\boldsymbol{q}\rightarrow0$. 
Integration along $y$ and $y'$ will yield Dirac delta functions in wavevectors $\delta(k_{y}-k'_{y})$,
after which integration of spinor products is performed over $x$ and $x'$ directions using the following identity:
\begin{eqnarray}
&&\int_{0}^{\infty}dx~x\Big[K_{\frac{1}{2}+i\Omega}(x)K_{\frac{1}{2}-i\Omega'}(x) - K_{\frac{1}{2}-i\Omega}(x)K_{\frac{1}{2}+i\Omega'}(x)\Big] \nonumber \\
& = & \frac{i\pi^{2}(\Omega^{2}-\Omega'^{2})}{2[\sinh(\pi\Omega)+\sinh(\pi\Omega')]}.
\end{eqnarray}
Thus the average current-current correlator as a function of time for the right handed fermions looks as follows:
\begin{eqnarray}\label{Correlator t-t'}
&&\bar{C}^{xx}(\Delta t) = \frac{1}{2}\int_{-\infty}^{\infty}d\Omega\int_{-\infty}^{\infty}d\Omega'~
\cosh\pi\Omega\cosh\pi\Omega' \nonumber \\ &&n_{\text{F}}(2\pi\Omega)n_{\text{F}}(-2\pi\Omega')e^{i(\Omega-\Omega')\Delta t}~ \frac{(\Omega^{2}-\Omega'^{2})^{2}}{[\sinh(\pi\Omega)+\sinh(\pi\Omega')]^{2}}, \nonumber \\
\end{eqnarray}
%
where $\Delta t=(t-t')$. 
We now subtract from this the current correlator with time coordinates interchanged, $t\leftrightarrow t'$ i.e. $\bar{C}^{xx}(t'-t)=\bar{C}^{xx}(-\Delta t)$, to obtain the vacuum average of the commutator of current-current correlation. Plugging this 
into the expression for conductivity tensor {(\ref{ConductivityTensor})}, where we perform the Fourier transform of a retarded function in time using the Plemelj formula $\lim\limits_{\delta\rightarrow0^{+}}\frac{1}{x+i\delta}=\mathcal{P}\frac{1}{x}-i\pi\delta(x)$, and extracting the imaginary part, we finally obtain the $xx$-component of the dissipative average conductivity as follows:
%
%
\begin{eqnarray}\label{Conductivity-XX}
&&\bar{\sigma}'_{xx}(\omega)  =  \frac{\pi\omega}{2}\int_{-\infty}^{\infty} d\Omega~ \cosh\pi\Omega\cosh\pi(\Omega+\omega)  \\
&& \hspace{-0.5cm}\times  \frac{(2\Omega+\omega)^{2}}{(\sinh\pi\Omega+\sinh\pi(\Omega+\omega))^{2}} \big[n_{\text{F}}(2\pi\Omega)-n_{\text{F}}(2\pi(\Omega+\omega))\big],
\nonumber 
\end{eqnarray}
where the prime $'$ denotes the real
part of conductivity that leads to dissipation of electronic current. 
%
In this formula, 
we have dropped dimensionful prefactors (such as $e^2/\hbar$,
the typical units of conductivity), and we have dropped an extensive factor
\be
\mathcal{A}=\int_{0}^{\infty}\frac{dk_y}{2\pi}\frac{1}{k_y^{2}}
\int_{-\infty}^{\infty} dy = L_y \int_{0}^{\infty}\frac{dk_y}{2\pi}\frac{1}{k^{2}_y},
\ee
with $L_y$ the size of the system in the $y$ direction.  Properly handling
the remaining integral would require analyzing our problem in a finite
system along $x$, a task we leave for future work. 

We have plotted Eq.~(\ref{Conductivity-XX}) in Fig.~\ref{Average Conductivity} which shows that the conductivity grows approximately linearly with the probing frequency and vanishes in the DC-limit ($\omega\rightarrow0$). As we discussed in Sec.~\ref{SEC:two}, the Rindler Hamiltonian with Fermi velocity $v(x)\simeq v_{0}|x|/\lambda$, can be achieved for modes with low energies and long wavelengths. 
%
%
Hence, the results for conductivity (and for internal energy) for strained honeycomb lattices are valid if we choose to probe long-wavelength modes $k_{y}\lambda\ll1$. This is valid because in order to evaluate these observables, spatial averages need to be performed equivalent to setting $\boldsymbol{q}\rightarrow0$ in $\sigma(\omega,\boldsymbol{q}\rightarrow0)$ as we discussed in Eq.~(\ref{Current q0}). 


To understand the result in Eq.~(\ref{Conductivity-XX}), we revisit the electronic conductivity of flat graphene (per node and per spin) in the collisionless limit and at a finite environment temperature $\beta=(k_{\text{B}}T)^{-1}$ \cite{Stauber:2008}:
\begin{eqnarray}\label{Flat Graphene Conductivity}
\bar{\sigma}'_{xx}(\omega) & = & \frac{1}{16}\big[n_{\text{F}}\Big(-\frac{\beta\omega}{2}\Big)-n_{\text{F}}\Big(\frac{\beta\omega}{2}\Big)\big] \\ \label{Flat Graphene Conductivity Alt}
& = & \frac{1}{16}\big[1-2n_{\text{F}}\Big(\frac{\beta\omega}{2}\Big)\big],
\end{eqnarray}
where the left hand side is measured in units of $e^{2}/\hbar$. The right hand side vanishes in the DC-limit $\omega\rightarrow0$. This happens because in this limit, only the energy levels close to the Dirac point participate in electronic transitions due to switching on the vector potential in (\ref{Electromagnetic Coupling}). However, here the electron occupancy in conduction band, given by $n_{\text{F}}(\beta\omega)\sim0.5$, is equal to the electron occupancy in the valence band, given by $n_{\text{F}}(-\beta\omega)\sim0.5$. Thus the rate of excitation and de-excitation are equal and hence the electrons near the Dirac point (DC-limit) do not participate in conductivity. On the other hand, as the probing frequency is increased, the electron occupancy in the valence band starts exceeding the conduction band, thus giving us a net rate of excitation of electrons that give non-zero conductivity. In the opposite limit of $\omega\gg(\hbar\beta)^{-1}$, the high energy modes are unaffected by the thermal scale and hence, the electron occupancy here is approximately unity, i.e. the de-excitation is minuscule and conductivity reaches it maximum value of $e^{2}/16\hbar$. The density of states in a two-dimensional material such as graphene, is expected to be linear in energy. However this gets cancelled out due to the $1/\omega$ in the expression for conductivity (\ref{ConductivityTensor}), and therefore only the Fermi functions are needed to physically understand the behavior of conductivity.

Since the strained graphene system is effectively at an Unruh temperature $T_U$, by analogy with the preceding argument 
we might also expect to find that $\sigma(\omega) \to 0$ for $\omega \to 0$, as we indeed find in Fig.~\ref{Average Conductivity}. 
To derive the approximate linear behavior, we use the fact that the factor multiplying the Fermi functions in square brackets in
Eq.~(\ref{Conductivity-XX}) is sharply peaked at $\Omega=-\omega/2$.  Then, we are allowed to make this replacement in
the square brackets, yielding $\big[n_{\text{F}}(-\pi\omega)-n_{\text{F}}(\pi\omega)\big]$ that can be pulled outside the integral.
Upon evaluating the remaining $\Omega$ integration over the peak region finally gives
\be
\bar{\sigma}'_{xx}(\omega) \simeq \frac{\sqrt{3}}{2\pi^{3/2}} \omega \tanh \frac{\pi \omega}{2} ,
\ee
which agrees with our numerical result in Fig.~\ref{Average Conductivity}.

We can also interpret these results by focusing on the second version similar to (\ref{Flat Graphene Conductivity Alt}) and noting that the conductivity is reduced due to the presence of emergent Fermi functions. This happens due to stimulated particle reduction \cite{Parker:1966,Parker:1971pt,Hu:1986jd,Hu:1986jj,Kandrup:1988sg}. The process of straining the honeycomb lattice leads to creation of fermions in the conduction band with a Fermi distribution $n_{\text{F}}(2\pi\Omega)$ characterized by Unruh temperature (here $1/2\pi$), yielding a thermally excited state. To study the linear electronic response of this system, a vector potential stimulus is provided because of which more electrons from the valence band jump to the conduction band. Pauli's exclusion principle does not allow the strained electrons to co-exist with the electronically excited ones, hence leading to an overall reduction in the response. Since particle creation is maximum at zero energy where the two bands meet (as the Unruh-Fermi function is maximum at low energies), it is easiest for strains to create electrons at this zero-energy level, and hence the stimulated reduction is maximum for zero probing frequency i.e. the DC-limit $\omega\rightarrow0$. In contrast higher energies overpower the strains making the Fermi functions small, and hence maximum conductivity is achieved.

Next we turn to the conductivity $\sigma_{yy}$ for directions perpendicular to
the strain fields, which, following the same procedure, leads to the similar result:
\begin{eqnarray}
\label{Conductivity-YY}
&&\bar{\sigma}'_{yy}(\omega) = \frac{\pi \omega}{2}\int_{-\infty}^{\infty} d\Omega~ \cosh\pi\Omega\cosh\pi(\Omega+\omega) 
\\
&& \hspace{-0.5cm}\times   \frac{(2\Omega+\omega)^{2}}{(\sinh\pi\Omega-\sinh\pi(\Omega+\omega))^{2}} \big[ n_{\text{F}}(2\pi\Omega)-n_{\text{F}}(2\pi(\Omega+\omega))\big],
\nonumber 
\end{eqnarray}
the only difference being a minus sign in the denominator of one factor in the
integrand. In this case the factor multiplying the Fermi functions in square brackets is not a narrow peak at $-\omega/2$; nonetheless
the qualitative behavior is similar as seen in Fig.~\ref{Average Conductivity} which shows that just like the $xx$-component, the $yy$-component of conductivity also grows approximately linearly with the probing frequency. Two key differences are that $\bar{\sigma}'_{yy}(\omega)$ is smaller in magnitude, and does not vanish in the DC-limit ($\omega\rightarrow0$) approaching a value of $1/(6\pi)$.
The reason is that in $\hat{x}$-direction the atoms have been forced to come closer to each other using strains of type ${(\ref{RindlerStrainPattern})}$ thereby increasing the Fermi velocity, and thus hopping becomes easier. Whereas in the $\hat{y}$-direction, the strains do not depend on coordinate $y$, and thus the atoms are further apart in this direction as compared to $\hat{x}$, hence the hopping and the conductivity here are lower. 

The results for $\bar{\sigma}'_{xx}$ in Eq.~(\ref{Conductivity-XX}) and for $\bar{\sigma}'_{yy}$ in Eq.~(\ref{Conductivity-YY}) imply that (when the dimensionful Unruh temperature $T_{\text{U}}$ is reintroduced),
$\bar{\sigma}'(\omega)$ is suppressed at low frequencies with increasing $T_{\text{U}}$.  Since the conductivity
is expected to obey the f-sum rule \cite{Coleman,Gusynin:2007} 
that conserves the integral of $\bar{\sigma}'(\omega)$ over $\omega$, we expect
this suppression to be accompanied by an additional $\delta(\omega)$ peak (as occurs in flat graphene~\cite{Fritz:2008,Stauber:2008} at nonzero real temperature $T$).  In the presence of disorder or interactions,
we furthermore expect this peak to be broadened into a Drude peak~\cite{Fritz:2008}, but with a weight controlled by $T_{\text{U}}$.


The transverse or off-diagonal components of conductivity tensor are anti-symmetric i.e. $\sigma_{xy}(\omega)=-\sigma_{yx}(\omega)$, which can be readily inferred from the commutator in Eq.~{(\ref{ConductivityTensor})}. This means that knowledge of one, determines the other. 
Performing similar calculations as the longitudinal case, yields a vanishing transverse conductivity:
\begin{equation}\label{Conductivity-XY}
\sigma'_{xy}(\omega)=-\sigma'_{yx}(\omega)=0.
\end{equation}
This can be expected because if $\sigma^{xy}\neq0$, then an electric field in the $x$-direction $E_{x}$ would be able to create a current in the $y$-direction. However due to translation symmetry, there is no reason why $+\hat{y}$ would be favored over $-\hat{y}$, and thus the current is expected to be zero by symmetry. This symmetry gets broken when there is a real magnetic field in the system. 


In this section we showed how the Rindler Hamiltonian {(\ref{Rindler Hamiltonian})} leads to a linear in probing frequency behavior of longitudinal components of the electronic conductivity {(\ref{Conductivity-XX})}, {(\ref{Conductivity-YY})}, and that the transverse {(\ref{Conductivity-XY})} components simply vanish. These results for average dissipative conductivity are summarized in Fig.~\ref{Average Conductivity}, where both the longitudinal components scale linearly for frequencies.
In the next section, we will take a look at the consequence of Rindler Hamiltonian on the internal energy of such honeycomb systems.

\section{Internal Energy}\label{SEC:eight}
As we saw in the previous section, that spontaneous particle creation due to us assuming a linear-in-position Fermi velocity had a profound effect on the behavior of electronic conductivity which scaled linearly in the probing frequency, as opposed to the flat honeycomb case where it has a constant value for all frequencies. In this section, we will be looking at how this Rindler-Unruh particle creation affects the response of honeycomb systems when brought in contact with a thermal heat bath, i.e. we will find the total electronic energy in the system $U$, which can be calculated using the expectation value of the Rindler Hamiltonian {(\ref{Rindler Hamiltonian})} with respect to a Minkowski thermal density matrix labeled by the temperature parameter $\beta=(k_{\text{B}}T)^{-1}$ as a subscript:
\begin{eqnarray}
U_{\text{M}} & = & \langle\hat{H}_{\text{R}}\rangle_{\beta}, \nonumber \\
& = & i\hbar~\bigg\langle \int d^{2}x~\hat{\psi}_{\text{R}}^{\dagger}(x)\cdot \partial_{t}\hat{\psi}_{\text{R}}(x) \bigg\rangle_{\beta,\cal{M}},
\end{eqnarray}
where 
to get the second line we made use of the Dirac equation {(\ref{Dirac Equation Rindler Graphene})} to simplify further calculations. Equivalently, this can also be calculated using the energy-momentum tensor operator as discussed in Ref.~\cite{Takagi:1986kn}. However, the above Minkowski thermal average is divergent and thus requires normal ordering. This involves subtracting off the Rindler thermal average (i.e. the limit of zero strains $\lambda\rightarrow\infty$) of the Rindler Hamiltonian from the Minkowski average as follows:
\begin{equation}\label{Renormalized Internal Energy}
U = \langle\hat{H}_{\text{R}}\rangle_{\beta,\cal{M}} - \langle\hat{H}_{\text{R}}\rangle_{\beta,\cal{R}}.
\end{equation}
This renormalization is needed because the Hamiltonian is quadratic in the fields $\hat{\psi}^{2}(x)$ \cite{Birrell:1982ix,Fulling:1989nb,Wald:1995yp,Fabbri:2005mw,Mukhanov:2007zz,Parker and Toms}, and thus the expectation value has a genuine divergence since even smearing the field operators will not cure this divergence, unlike the case of two-point functions which are bi-distributions and their divergences at short distances can be cured by smearing.

To evaluate these expectation values, the physical picture that we will be needing is that the honeycomb lattice is initially in a thermal state (due to contact with a heat bath or the surroundings), and then strains are put on it. As a result, the initial state of the flat honeycomb lattice is described by the eigenstates of the standard Dirac Hamiltonian {(\ref{Dirac Hamiltonian})}, whose excitations are described by Minkowski operators $\{\hat{a}_{\boldsymbol{k}},\hat{b}_{\boldsymbol{k}}\}$ in Eq.~{(\ref{MinkowskiModeExpansionStandardRight1})} labeled by momentum vector $\boldsymbol{k}$. Since this system is also kept in contact with a heat bath at temperature $\beta=(k_{\text{B}}T)^{-1}$, the thermal averages of Minkowski operators will be given by the Fermi distributions:
\begin{equation}\label{Thermal Averages Minkwoski}
\langle\hat{a}_{\boldsymbol{k}}^{\dagger}\hat{a}_{\boldsymbol{k}}\rangle_{\beta,\cal{M}} = \langle\hat{b}_{\boldsymbol{k}}^{\dagger}\hat{b}_{\boldsymbol{k}}\rangle_{\beta,\cal{M}} = n_{\text{F}}(\beta\epsilon_{k}) \equiv \frac{1}{e^{\beta\epsilon_{k}}+1},
\end{equation}
as a function of the Minkowski energy dispersion relation $\epsilon_{k}=\hbar\omega_{k}=\hbar v_{0}|\boldsymbol{k}|$. When the strains are turned on, then the system is described the Rindler Hamiltonian {(\ref{Rindler Hamiltonian})}, whose excitations are governed by the Rindler operators $\{\hat{c}_{k_{y},\Omega},\hat{d}_{k_{y},\Omega}\}$, labeled by the independent pair of momenta $\hbar k_{y}$ and energy $\hbar \Omega$. We have seen in Sec.~\ref{SEC:four}, that the Bogoliubov transformations {(\ref{Bogoliubov Transformations Actual})} help express these Rindler operators in terms of modified Minkowski operators $\{\hat{A}^{\pm}_{k_{y},\Omega},\hat{B}^{\pm}_{k_{y},\Omega}\}$, which are in turn complex linear combinations of the standard ones $\{\hat{a}_{\boldsymbol{k}},\hat{b}_{\boldsymbol{k}}\}$ as given in Eq.~{(\ref{Modified & Actual Minkwoski Operators})}. Thus making use of this transformation between operators, and the thermal averages in Eq.~{(\ref{Thermal Averages Minkwoski})}, we obtain the thermal averages of the Rindler operators in the Minkowski vacuum as follows:
\begin{eqnarray}
&&\big\langle \hat{c}^{\dagger}_{k_{y},\Omega}\hat{c}_{k'_{y},\Omega'} \big\rangle_{\beta,\cal{M}}  = \big\langle \hat{d}^{\dagger}_{k_{y},\Omega}\hat{d}_{k'_{y},\Omega'} \big\rangle_{\beta,\cal{M}} \nonumber \\ 
& = & \delta(k_{y}-k'_{y})\Bigg[\delta(\Omega-\Omega')\sqrt{n_{\text{F}}(2\pi\Omega)}\sqrt{n_{\text{F}}(2\pi\Omega')} \nonumber \\
& + & \bigg\{\sqrt{n_{\text{F}}(-2\pi\Omega)}\sqrt{n_{\text{F}}(-2\pi\Omega')} \nonumber \\
& - & \sqrt{n_{\text{F}}(2\pi\Omega)}\sqrt{n_{\text{F}}(2\pi\Omega')}\bigg\} Z_{k_{y}}(\Delta)\Bigg],
\label{Thermal Average Rindler Operators}
\end{eqnarray}
where $\Delta\equiv\Omega-\Omega'$ and we define the function
$Z_{k_{y}}(\Delta)$:
\begin{equation}
Z_{k_{y}}(\Delta) = \int_{-\infty}^{\infty}\frac{dk_{x}}{2\pi k} \bigg(\frac{k+k_{x}}{k-k_{x}}\bigg)^{-i\Delta/2}n_{\text{F}}(\beta\epsilon_{k}),
\end{equation}
which we emphasize is real (i.e., $Z^{*}_{k_{y}}(\Delta)=Z_{k_{y}}(\Delta)$).
Note the difference between the two types of Fermi distributions being used here. The first $n_{\text{F}}(\beta\epsilon_{k})$, is due to a heat bath labeled by the environment temperature parameter $\beta$ and is a function of the Minkowski energy $\epsilon_{k}$. The second one $n_{\text{F}}(2\pi\Omega)$ is an emergent thermal distribution governed by the strain frequency $\omega_{c}=v_{0}/\lambda$.

The thermal averages in {(\ref{Thermal Average Rindler Operators})} have a temperature-independent part proportional to a delta function in energy $\delta(\Omega-\Omega')$ and a temperature-dependent part having the function $Z_{k_{y}}(\Omega-\Omega')$. 
 To get an intuition for this formula, we discretize the wavevector and frequency delta functions to Kronecker delta functions,
  effectively smearing the Rindler operators~\cite{Takagi:1986kn}. Then taking $k'_{y}=k_{y}$ and $\Omega'=\Omega$, the electron (or hole) thermal averages take the following form:
\begin{eqnarray}\label{Thermal Average Rindler Operators Smeared}
&& N_{k_{y},\Omega} \equiv \big\langle \hat{c}^{\dagger}_{k_{y},\Omega}\hat{c}_{k_{y},\Omega} \big\rangle_{\beta,\cal{M}} - Z_{k_{y}}(0) \nonumber \\
& = & n_{\text{F}}(2\pi\Omega)\Bigg[1-\frac{1}{\pi}\int_{-\infty}^{\infty}\frac{d\hat{k}_{x}}{\sqrt{\hat{k}^{2}_{x}+\hat{k}^{2}_{y}}}	n_{\text{F}}\Big(\sqrt{\hat{k}^{2}_{x}+\hat{k}^{2}_{y}}\Big)\Bigg],~~~~~~~
\end{eqnarray}
where $\Omega$ is the dimensionless frequency used elsewhere (in which the Unruh temperature is $1/(2\pi)$) and 
the wavevectors $\hat{\boldsymbol{k}}=\frac{\hbar v_{0}\boldsymbol{k}}{k_{B}T}$ are normalized using the real system temperature $T$. We have also renormalized the number average by subtracting off the Rindler vacuum contribution which can be found by setting $T_{\text{U}}=0$ ($\lambda\rightarrow\infty$) in (\ref{Thermal Average Rindler Operators}), or equivalently subtracting off $Z_{k_{y}}(0)$ from the expectation value in the first line. This is the same procedure as was discussed in (\ref{Renormalized Internal Energy}) without which the integrals inside the expectation values diverge.

In Fig.~\ref{Thermal Averages}, we plot the renormalized occupancy as a function of frequency for various values of 
the normalized wavevector $\hat{k}_{y}$.  This figure shows that nonzero environment temperature leads to a stimulated 
reduction of fermions~\cite{Parker:1966,Parker:1971pt,Hu:1986jd,Hu:1986jj,Kandrup:1988sg}, i.e., a smaller Unruh effect.  However, this reduction is dependent on the momentum $k_y$, with the 
$\hat{k}_{y}\rightarrow\infty$ curve (dashed line) identical to the zero-temperature Unruh effect, and an increasing
stimulated reduction with decreasing $\hat{k}_y$.  
This happens because we start with an initial thermal state of fermions and Pauli's exclusion principle does not allow new fermions to co-exist with them that are spontaneously created via strains, hence leading to reduction. The higher the initial temperature, the lower the value of $\hat{k}_{y}$ and therefore the further away the spectrum is from the Fermi-Dirac. In other words, if we keep the environment temperature fixed, then in the limit of small wavelength we recover the Unruh effect and for larger wavelengths the average fermion number strays away from the perfect Fermi-Dirac distribution. 

\begin{figure}[h]
	\centering
	\includegraphics[width=1.0\columnwidth]{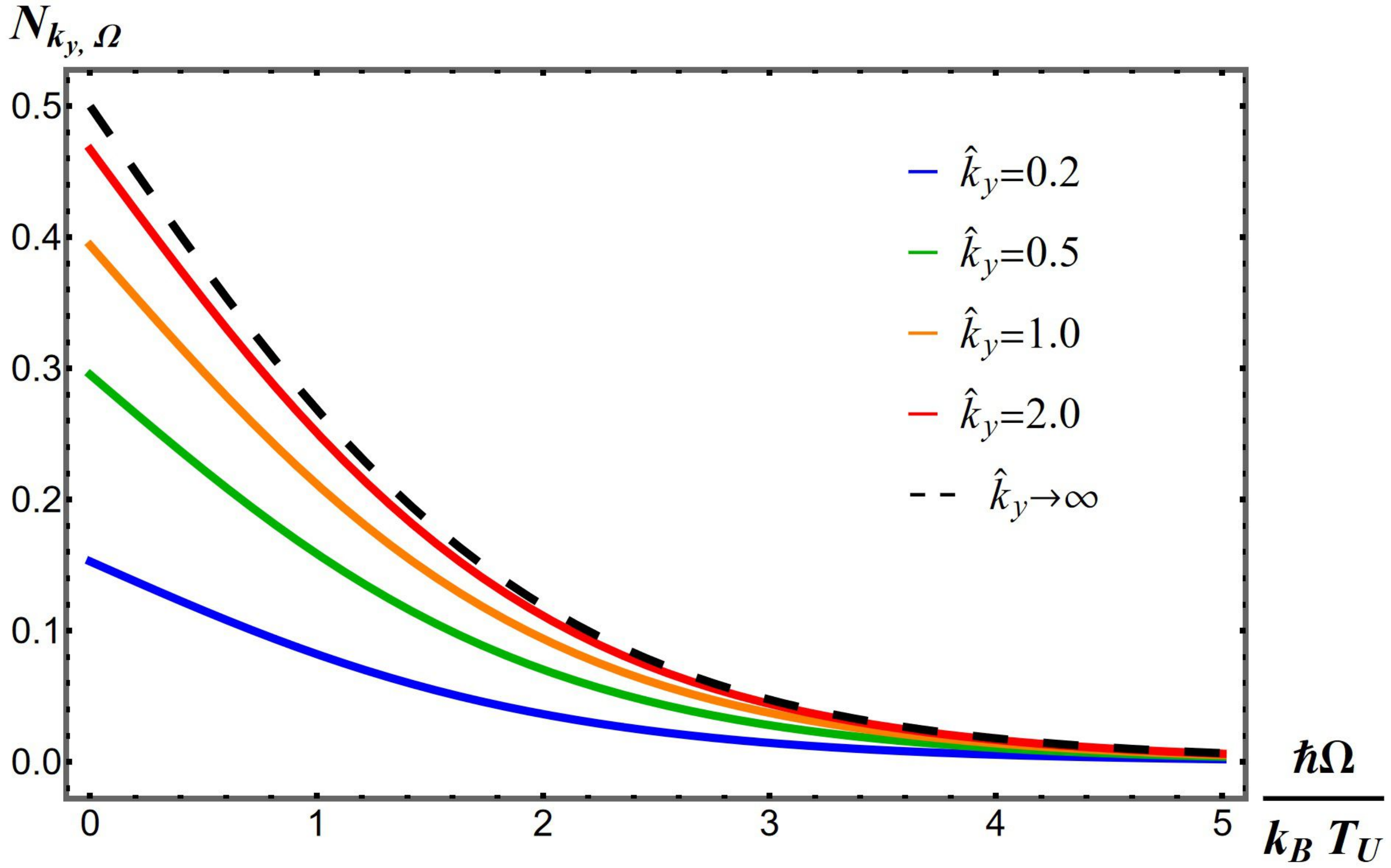}
	\caption{(Color Online) A figure showing the average number of fermions (plotted with respect to mode energy normalized with Unruh temperature, see Eq.(\ref{Thermal Average Rindler Operators Smeared})) in a graphene sheet which is initially in a thermal state and is then strained leading to stimulated particle reduction, for various values of momenta $\hat{k}_{y}$ normalized with real temperature. The dashed black curve represents the Unruh effect with a perfect Fermi-Dirac distribution which could be achieved in the limit of $\hat{k}_{y}\rightarrow\infty$, i.e. large $k_{y}$ or zero environment temperature. As the temperature rises, the Fermi-Dirac distribution gets reduced due to Pauli's principle.}
\label{Thermal Averages}
\end{figure}

Next, we turn to the direct calculation of the internal energy $U$ using Eq.~(\ref{Renormalized Internal Energy}).
For this task we shall use  Eq.~(\ref{Thermal Average Rindler Operators}) without making the abovementioned discretization that 
was used for Fig.~\ref{Thermal Averages}.
We find that the internal energy has two contributions $U=U^{0}+U^{\beta}$. For the zero temperature part $U^{0}$, the energy and momentum integrals inside the thermal averages can be simplified by using the Dirac delta functions $\delta(\Omega-\Omega')$ and $\delta(k_{y}-k_{y}')$ that pin $\Omega'=\Omega$ and $k_{y}=k_{y}'$. Then integration can be performed over momenta $k_{y}$ using the following identity:
\begin{eqnarray}
\hspace{-0.5cm}	\int_{0}^{\infty}dk_{y}~k_{y}~  K_{\frac{1}{2}+i\Omega}(k_{y}x)K_{\frac{1}{2}-i\Omega}(k_{y}x) & = & \frac{\pi^{2}}{4x^{2}}\frac{\Omega}{\sinh\pi\Omega},~~~~~
\end{eqnarray}
which leads to: 
\begin{eqnarray}\label{Temperature Independent Part}
U^{0} & = & \frac{L_{y}}{\pi}\int \frac{dx}{x^{2}} \int_{0}^{\infty}d\Omega~ \hbar\Omega~ \Omega\coth\pi\Omega~ n_{\text{F}}(2\pi\Omega), \nonumber \\
& = & \frac{L_{y}}{\pi}\int_a^\infty \frac{dx}{x^{2}} \int_{0}^{\infty}d\Omega~ \hbar\Omega~ \Omega~ n_{\text{B}}(2\pi\Omega),
\end{eqnarray}
where in going from first to second line we used the identity $\coth x\cdot n_{\text{F}}(2x)=n_{\text{B}}(2x)$, and we cutoff the $x$ spatial integral
at the lattice scale $a$.  
We also note that the energy labels $\Omega$ that are not associated with an $\hbar$, need to be understood as being normalized with $\omega_{c}$. This result is for the right side of the honeycomb lattice per node and per spin state. This temperature independent contribution $U^{0}$ is made up of three elements: the mode energy $\hbar\Omega$, the density of states $\Omega\coth\pi\Omega$ and the occupancy of energy levels given by a Fermi-Dirac distribution $n_{\text{F}}(2\pi\Omega)$. 
In the last line, however, we see that the product of the last two factors in the first line effectively 
yields a linear-in-energy density of states multiplied by the Bose-Einstein distribution.
 This is Takagi's apparent \emph{statistics inversion} {\cite{Takagi:1986kn}} that we discussed in equations {(\ref{Fluctuation-Dissipation Version 1})} and {(\ref{Fluctuation-Dissipation Version 2})}. 
 Thus, although Eq.~{(\ref{Temperature Independent Part})} pertains to fermions, the final result looks like 
Planck's black body result for photons.


The temperature-dependent part of the total internal energy, $U^{\beta}$, depends on the  temperature-dependent terms of the thermal averages in Eq.~(\ref{Thermal Average Rindler Operators}). For this contribution, 
the momentum integrals inside the thermal averages can be simplified by using the Dirac delta function $\delta(k_{y}-k_{y}')$, pinning $k'_{y}=k_{y}$. Then,  integrating over $x$  using Eq.~(\ref{Normalization Bessel}) gives us a Dirac delta function  $\delta(\Omega-\Omega')$. 
%
%
This along with the finite temperature renormalization discussed in Eq.~(\ref{Renormalized Internal Energy}) gives the temperature dependent part of internal energy:
\begin{equation}
U^{\beta} = -\frac{L_{y}}{\pi}\int_{0}^{\infty}d\Omega~ \hbar\Omega~ n_{\text{F}}(2\pi\Omega) \int_{-\infty}^{\infty}dk_{y}~Z_{k_{y}}(0),
\end{equation}
The momentum integral can be simplified by switching to polar coordinates, i.e. $(k_{x},k_{y})\rightarrow(k,\theta)$, and using the identity $\int_{0}^{\infty}dx~ n_{\text{F}}(x)=\log2$, thus yielding:
\begin{equation}
\int_{-\infty}^{\infty}dk_{y}~Z_{k_{y}}(0) = \frac{k_{\text{B}}T\log2}{\hbar v_{0}}.
\end{equation}
%
%
Compiling the results for the temperature independent and dependent cases we find that the
renormalized total internal energy $U=U^{0} + U^{\beta}$ for a strained graphene sheet kept in an environment with finite temperature is:
\begin{eqnarray}
U & = & \frac{L_{y}}{\pi a} \int_{0}^{\infty}d\Omega~ \Omega\coth\pi\Omega ~\hbar\Omega~ n_{\text{F}}(2\pi\Omega) \nonumber \\
& - & \frac{L_{y}}{\pi \lambda} \frac{\log2}{\beta\epsilon_{c}} \int_{0}^{\infty}d\Omega~ \hbar\Omega~ n_{\text{F}}(2\pi\Omega),
\end{eqnarray}
%
%
where $\epsilon_{c}=\hbar\omega_{c}$. This result depends linearly on temperature and manifestly shows that because we started with an initial thermal state of fermions in flat graphene, then the process of straining leads to stimulated particle reduction due to Pauli's exclusion principle {\cite{Parker:1966,Parker:1971pt,Hu:1986jd,Hu:1986jj,Kandrup:1988sg}}.

\section{Concluding remarks}
\label{SEC:nine}
In this paper, we have discussed how a honeycomb lattice that is strained inhomogenously can act as an arena where analogue Rindler physics associated with accelerating observers can be realized. We broke this problem into two stages. The first stage is that of an unstrained flat graphene sheet that possesses
(discrete) translation symmetry, and leads to an emergent Dirac equation for low energy modes. This mimics the evolution of fermions in flat Minkowski spacetime. We then solved the evolution equation to obtain the mode expansion in terms of plane waves in space and time. This choice helps us define a structure of ladder operators $\hat{a}_{\boldsymbol{k}}$ and $\hat{b}_{\boldsymbol{k}}$ which, when acting on the Minkowski or flat graphene vacuum state $|0_{\cal{M}}\rangle$, lead to excitation of electrons and holes that obey a linear in energy-momentum dispersion relation.

The second stage starts when we suddenly switch on strains to create a Rindler Hamiltonian with a spatially varying Fermi velocity $v(x)\simeq v_{0}\frac{|x|}{\lambda}$ where the origin $x=0$ acts as an analogue of Rindler horizon separating the $x<0$ and $x>0$ regions and forbidding low-energy and long-wavelength electrons to tunnel through. Thus the two disconnected sides of 
strained graphene mimic the causally disconnected left and right Rindler wedges. Then we solved the Dirac equation for right handed Weyl fermions and obtained the solutions in terms of Bessel functions that blow up at the horizon and asymptotically vanish at large $x$. Here the plane wave basis in Rindler time helps us choose the structure of Rindler creation and annihilation operators $\hat{c}_{k_{y},\Omega}$ and $\hat{d}_{k_{y},\Omega}$ for electrons and holes with respect to the Rindler vacuum state $|0_{\cal{R}}\rangle$. However, unlike the Minkowski case, here due to broken translation symmetry there is no band dispersion and the energy and momentum are decoupled.

Since the same quantum field operator has two different representations in the flat and strained regimes, by projecting one onto the other we find that
the Minkowski vacuum
$|0_{\cal{M}}\rangle$ appears to operators of the strained system as
if it is at finite temperature, swarming with Rindler particles. 
This can be understood in terms of the Heisenberg picture where the state of the system remains the same, whereas the operators evolve, and thus in the sudden approximation the original state is viewed as a linear combination of the eigenstates of the new Hamiltonian. In fact, the Minkowski vacuum state corresponding to the flat system can be expressed as a two-mode squeezed state with respect to the Rindler vacuum, since one side of the lattice is unavailable to the modes residing on the opposite side.
Thus, expectation values on the right side effectively involve a trace over the left side, amounting to a mixed thermal density operator for the right side. This is similar to what happens in Rindler spacetime because when an observer picks a certain acceleration say $a>0$, then they are naturally causally disconnected from the observers accelerating opposite to them. As a result of this, the Minkowski vacuum averages of Rindler ladder operators pertaining to one side appear as thermal averages, which is known as the Fulling-Davies-Unruh effect. 

After discussing this thermal-like creation of particles, we looked into the properties of the strained Green's functions which satisfy the KMS condition that ensures that if the analogue spacetime has a horizon in it, then the spectrum of particles it creates is bound to be thermal in nature. Another feature of these Green's functions was that the Huygens' principle gets violated due to graphene being a two-dimensional material and thus leads to a Bose-Einstein spectrum for electron-hole pairs created by strains, 
a manifestation of Takagi's statistics inversion.

We then discussed how the Unruh thermality (for low-energy and long wavelength modes) could be measured in photo-emission spectroscopy (PES) experiments  and the inversion factor could be seen in scanning tunneling microscopy experiments that measure the density of states. In PES, shining photons on graphene would excite fermions to higher states according to a Bose distribution and therefore, in this sense, these experiments are related to the Unruh-DeWitt detectors that also get excited with a Bose-Einstein response when interacting with acceleration radiation. We also found that a similar thermal like behavior could be seen in measurements of the spatially averaged electronic conductivity of an isolated strained honeycomb lattice, which at low energies, exhibits a frequency dependence
that is similar to that found in the case of a flat graphene sheet kept at finite environment temperature, hence signaling emergence of Unruh-like thermality. Finally, we ended our discussion with a calculation of the total system energy due to strains at finite environment temperature and found that it has a zero temperature portion which resembles the black body spectrum of photons thus signaling statistics inversion, and a finite temperature part whose contribution is negative. This is due to the fact that if we start with an initially excited (thermal) state in flat graphene, then strains lead to stimulated particle reduction due to the Pauli principle not allowing newly created fermions to occupy the energy levels already occupied by thermal fermions.

A possible future direction of research is to consider a more realistic temporal profile for the strains $\lambda(t)$. Tuning the speed of this parameter can give rise to three possible regimes. First is the rapid quenching regime that gives rise to turbulence, which when coupled with dissipation could lead to exponential growth and decay of perturbations. For this, a Lindbladian analysis \cite{Chen:2021xhd} could be performed over the thermal density matrix in Eq.~(\ref{Density Matrix Reduced Gibbs Form}). Second is the slow or adiabatic phase where the system will end up in $|0_{\mathcal{R}}\rangle$, with the Hamiltonian being given by (\ref{Rindler Hamiltonian}). Since the system Hamiltonian has changed, this could have potentially interesting effects on observables like conductivity. Thirdly, tuning between these two limits could provide information about the appropriate time-window in which the sudden approximation is valid
and how observables related to the Unruh effect are modified away from the sudden regime.  Another possible direction is to look at the effects of electron-electron interactions on the Unruh thermality. It is well-known that the KMS condition is valid even for interacting field theories \cite{Unruh:1983ac}. This could be especially interesting for conductivity at $\omega=0$ where the expected Dirac delta peak broadens due to interactions and leads to hydrodynamic behavior of graphene quasiparticles.

\section{Acknowledgments}
The authors are grateful to Jorma Louko, Boris Narozhny, J\"org
Schmalian, Mette Gaarde, Álvaro Jiménez Galán, Lun Yue, and Dana Browne for useful comments and discussions.  AB and DES acknowledge
support from the National Science Foundation under Grant PHY-2208036.
AB acknowledges
financial support from the Department of Physics and Astronomy at
LSU.
DES acknowledges funding from the European Union's Horizon 2020
research and innovation programme under the Marie Sklodowska-Curie
grant agreement No 873028.

\bibliography{Bibliography.bib}

\begin{thebibliography}{99}
\bibitem{Birrell:1982ix} 
N.~D.~Birrell and P.~C.~W.~Davies,
\emph{Quantum Fields in Curved Space}, Cambridge University Press (1984). \\

\bibitem{Fulling:1989nb} 
S.~A.~Fulling,
\emph{Aspects of Quantum Field Theory in Curved Space-time},
London Math.\ Soc.\ Student Texts {\bf 17}, 1 (1989). \\

\bibitem{Wald:1995yp}
R.~M.~Wald,
\emph{Quantum Field Theory in Curved Space-Time and Black Hole Thermodynamics}, University of Chicago Press (1994). \\

\bibitem{Fabbri:2005mw}
A.~Fabbri and J.~Navarro-Salas,
\emph{Modeling black hole evaporation}, Imperial College Press (2005). \\

\bibitem{Mukhanov:2007zz} 
V.~Mukhanov and S.~Winitzki,
\emph{Introduction to quantum effects in gravity},
Cambridge Univ. Pr. (2007). \\

\bibitem{Parker and Toms} 
L.~Parker and D.~Toms,
\emph{Quantum Field Theory in Curved Spacetime: Quantized Fields and Gravity}, Cambridge University Press (2009). \\	

\bibitem{Schrodinger:1939}
E.~Schrodinger,
\emph{The proper vibrations of the expanding universe},
Physica \textbf{6}, 899 (1939). \\

\bibitem{Parker:1966}
L.~Parker,
\emph{The Creation of Particles in an Expanding Universe}, Ph.D. thesis, Harvard University (1966). \\

\bibitem{Parker:1968mv}
L.~Parker,
\emph{Particle creation in expanding universes},
\href{https://journals.aps.org/prl/abstract/10.1103/PhysRevLett.21.562}{{Phys. Rev. Lett. \textbf{21}, 562 (1968)}}. \\
	
\bibitem{Rindler:1966zz}
W.~Rindler,
\emph{Kruskal Space and the Uniformly Accelerated Frame},
\href{https://aapt.scitation.org/doi/10.1119/1.1972547}{{Am. J. Phys. \textbf{34}, 1174 (1966).}} \\
	
\bibitem{Fulling:1972md}
S.~A.~Fulling,
\emph{Nonuniqueness of canonical field quantization in Riemannian space-time},
 \href{https://journals.aps.org/prd/abstract/10.1103/PhysRevD.7.2850}{{Phys. Rev. D \textbf{7}, 2850 (1973).}} \\
	
\bibitem{Davies:1974th}
P.~C.~W.~Davies,
\emph{Scalar particle production in Schwarzschild and Rindler metrics},
\href{https://iopscience.iop.org/article/10.1088/0305-4470/8/4/022}{{J. Phys. A \textbf{8}, 609 (1975).}} \\
	
\bibitem{Unruh:1976db}
W.~G.~Unruh,
\emph{Notes on black hole evaporation},
\href{https://journals.aps.org/prd/abstract/10.1103/PhysRevD.14.870}{{Phys. Rev. D \textbf{14}, 870 (1976).}} \\

\bibitem{Kubo:1957}
R.~Kubo,
\emph{Statistical-mechanics theory of irreversible processes. I. General theory and simple application to magnetic and conduction problems},
\href{https://journals.jps.jp/doi/10.1143/JPSJ.12.570}{{J. Phys. Soc. Jpn. \textbf{12}, 570, (1957).}} \\

\bibitem{Martin & Schwinger:1959}
P.~C.~Martin, and J.~Schwinger,
\emph{Theory of many-particle systems.~I},
\href{https://journals.aps.org/pr/abstract/10.1103/PhysRev.115.1342}{{Phys. Rev. \textbf{115}, 1342 (1959).}} \\

\bibitem{Hawking:1974rv} 
S.~W.~Hawking,
\emph{Black hole explosions},
\href{https://www.nature.com/articles/248030a0}{{Nature {\bf 248}, 30 (1974).}} \\

\bibitem{Hawking:1974sw} 
S.~W.~Hawking,
\emph{Particle Creation by Black Holes},
\href{https://link.springer.com/article/10.1007/BF02345020}{{Commun.\ Math.\ Phys.\  {\bf 43}, 199 (1975)}}, 
Erratum: \href{https://link.springer.com/article/10.1007/BF01608497}{{ Commun.\ Math.\ Phys.\  {\bf 46}, 206 (1976).}} \\

\bibitem{Gibbons:1977mu}
G.~W.~Gibbons and S.~W.~Hawking,
\emph{Cosmological Event Horizons, Thermodynamics, and Particle Creation},
\href{https://journals.aps.org/prd/abstract/10.1103/PhysRevD.15.2738}{{Phys. Rev. D \textbf{15}, 2738 (1977).}} \\

\bibitem{Takagi:1986kn}
S.~Takagi,
\emph{Vacuum Noise and Stress Induced by Uniform Acceleration: Hawking-Unruh Effect in Rindler Manifold of Arbitrary Dimension},
\href{https://academic.oup.com/ptps/article/doi/10.1143/PTP.88.1/1938595?login=true}{{Prog. Theor. Phys. Suppl. \textbf{88}, 1 (1986).}} \\

\bibitem{Ooguri:1985nv}
H.~Ooguri,
\emph{Spectrum of Hawking Radiation and Huygens' Principle},
\href{https://journals.aps.org/prd/abstract/10.1103/PhysRevD.33.3573}{{Phys. Rev. D \textbf{33}, 3573 (1986).}} \\

\bibitem{Unruh:1986tc}
W.~G.~Unruh,
\emph{Accelerated Monopole Detector in Odd Space-time Dimensions},
\href{https://journals.aps.org/prd/abstract/10.1103/PhysRevD.34.1222}{{Phys. Rev. D \textbf{34}, 1222 (1986).}} \\

\bibitem{Terashima:1999xp}
H.~Terashima,
\emph{Fluctuation dissipation theorem and the Unruh effect of scalar and Dirac fields},
\href{https://journals.aps.org/prd/abstract/10.1103/PhysRevD.60.084001}{{Phys. Rev. D \textbf{60}, 084001 (1999).}} \\

\bibitem{Sriramkumar:2002nt}
L.~Sriramkumar,
\emph{Odd statistics in odd dimensions for odd couplings},
\href{https://www.worldscientific.com/doi/abs/10.1142/S0217732302007545}{{Mod. Phys. Lett. A \textbf{17}, 1059 (2002).}} \\

\bibitem{Sriramkumar:2002dn}
L.~Sriramkumar,
\emph{Interpolating between the Bose-Einstein and the Fermi-Dirac distributions in odd dimensions},
\href{https://link.springer.com/article/10.1023/A:1025791420706}{{Gen. Rel. Grav. \textbf{35}, 1699 (2003).}} \\


\bibitem{Pascazio_Huygens}
  S. Pascazio, F. V. Pepe, and J. M. P\'erez-Pardo,
  \lq\lq Huygens' principle and Dirac-Weyl equation'',
%
    \href{https://link.springer.com/article/10.1140/epjp/i2017-11593-6}{{Eur. Phys. J. Plus {\bf 132}, 287 (2017).}}\\

\bibitem{Arrechea:2021szl}
J.~Arrechea, C.~Barcel\'o, L.~J.~Garay and G.~Garc\'\i{}a-Moreno,
\emph{Inversion of statistics and thermalization in the Unruh effect},
\href{https://journals.aps.org/prd/abstract/10.1103/PhysRevD.104.065004}{{Phys. Rev. D \textbf{104}, 065004 (2021).}} \\

\bibitem{Kalinski:2005}
M.~Kalinski, 
\emph{Hawking radiation from Trojan states in muonic Hydrogen in strong laser field},
Laser Physics, \textbf{15}, No 10 (2005), \href{https://arxiv.org/abs/quant-ph/0501172}{arXiv:quant-ph/0501172.} \\

\bibitem{Crispino:2007eb}
L.~C.~B.~Crispino, A.~Higuchi and G.~E.~A.~Matsas,
\emph{The Unruh effect and its applications},
\href{https://journals.aps.org/rmp/abstract/10.1103/RevModPhys.80.787}{{Rev. Mod. Phys. \textbf{80}, 787 (2008).}} \\

\bibitem{Martin-Martinez:2010gnz}
E.~Martin-Martinez, I.~Fuentes and R.~B.~Mann,
\emph{Using Berry's phase to detect the Unruh effect at lower accelerations},
\href{https://journals.aps.org/prl/abstract/10.1103/PhysRevLett.107.131301}{{Phys. Rev. Lett. \textbf{107}, 131301 (2011).}} \\

\bibitem{Nation:2011dka}
P.~D.~Nation, J.~R.~Johansson, M.~P.~Blencowe and F.~Nori,
\emph{Stimulating Uncertainty: Amplifying the Quantum Vacuum with Superconducting Circuits},
\href{https://journals.aps.org/rmp/abstract/10.1103/RevModPhys.84.1}{{Rev. Mod. Phys. \textbf{84}, 1 (2012).}} \\

\bibitem{Gooding:2020scc}
C.~Gooding, S.~Biermann, S.~Erne, J.~Louko, W.~G.~Unruh, J.~Schmiedmayer and S.~Weinfurtner,
\emph{Interferometric Unruh detectors for Bose-Einstein condensates},
\href{https://journals.aps.org/prl/abstract/10.1103/PhysRevLett.125.213603}{{Phys. Rev. Lett. \textbf{125}, 213603 (2020).}} \\

\bibitem{Retzker:2008}
A.~Retzker, J.~I.~Cirac, M.~B.~Plenio, and B.~Reznik,
\emph{Methods for Detecting Acceleration Radiation in a Bose-Einstein Condensate},
\href{https://journals.aps.org/prl/abstract/10.1103/PhysRevLett.101.110402}{{Phys. Rev. Lett. \textbf{101}, 110402 (2008).}} \\

\bibitem{Barcelo:2005fc} 
C.~Barcelo, S.~Liberati and M.~Visser,
\emph{Analogue gravity},
\href{https://link.springer.com/article/10.12942/lrr-2011-3}{{Living Rev.\ Rel.\  {\bf 14}, 3 (2011).}} \\

\bibitem{Unruh:1980cg} 
W.~G.~Unruh,
\emph{Experimental black hole evaporation},
\href{https://journals.aps.org/prl/abstract/10.1103/PhysRevLett.46.1351}{{Phys.\ Rev.\ Lett.\  {\bf 46}, 1351 (1981).}} \\

\bibitem{Philbin:2007ji} 
T.~G.~Philbin, C.~Kuklewicz, S.~Robertson, S.~Hill, F.~Konig and U.~Leonhardt,
\emph{Fiber-optical analogue of the event horizon},
\href{https://science.sciencemag.org/content/319/5868/1367}{{Science {\bf 319}, 1367 (2008).}} \\

\bibitem{Belgiorno:2010wn} 
F.~Belgiorno {\it et al.},
\emph{Hawking radiation from ultrashort laser pulse filaments},
\href{https://journals.aps.org/prl/abstract/10.1103/PhysRevLett.105.203901}{{Phys.\ Rev.\ Lett.\  {\bf 105}, 203901 (2010).}} \\

\bibitem{Weinfurtner:2010nu} 
S.~Weinfurtner, E.~W.~Tedford, M.~C.~J.~Penrice, W.~G.~Unruh and G.~A.~Lawrence,
\emph{Measurement of stimulated Hawking emission in an analogue system},
\href{https://journals.aps.org/prl/abstract/10.1103/PhysRevLett.106.021302}{{Phys.\ Rev.\ Lett.\  {\bf 106}, 021302 (2011).}} \\

\bibitem{Steinhauer:2015saa} 
J.~Steinhauer,
\emph{Observation of quantum Hawking radiation and its entanglement in an analogue black hole},
\href{https://www.nature.com/articles/nphys3863}{{Nature Phys.\  {\bf 12}, 959 (2016).}} \\

\bibitem{Eckel:2017uqx}
S.~Eckel, A.~Kumar, T.~Jacobson, I.~B.~Spielman and G.~K.~Campbell,
\emph{A rapidly expanding Bose-Einstein condensate: an expanding universe in the lab},
\href{https://journals.aps.org/prx/abstract/10.1103/PhysRevX.8.021021}{{Phys. Rev. X \textbf{8}, 021021 (2018).}} \\

\bibitem{Banik:2021xjn}
S.~Banik, M.~Gutierrez Galan, H.~Sosa-Martinez, M.~Anderson, S.~Eckel, I.~B.~Spielman and G.~K.~Campbell,
\emph{Accurate Determination of Hubble Attenuation and Amplification in Expanding and Contracting Cold-Atom Universes},
\href{https://journals.aps.org/prl/abstract/10.1103/PhysRevLett.128.090401}{{Phys. Rev. Lett. \textbf{128}, 090401 (2022).}} \\

\bibitem{Llorente:2019rbs} 
J.~M.~Gomez Llorente and J.~Plata,
\emph{Expanding ring-shaped Bose-Einstein condensates as analogs of cosmological models: Analytical characterization of the inflationary dynamics},
\href{https://journals.aps.org/pra/abstract/10.1103/PhysRevA.100.043613}{{Phys.\ Rev.\ A {\bf 100}, 043613 (2019).}} \\

\bibitem{Bhardwaj:2020ndh}
A.~Bhardwaj, D.~Vaido and D.~E.~Sheehy,
\emph{Inflationary Dynamics and Particle Production in a Toroidal Bose-Einstein Condensate},
\href{https://journals.aps.org/pra/abstract/10.1103/PhysRevA.103.023322}{{Phys. Rev. A \textbf{103}, 023322 (2021).}} \\

\bibitem{Eckel:2020qee}
S.~Eckel and T.~Jacobson,
\emph{Phonon redshift and Hubble friction in an expanding BEC},
\href{https://scipost.org/10.21468/SciPostPhys.10.3.064}{{SciPost Phys. \textbf{10}, 064 (2021).}} \\

\bibitem{Ghorashi:2020}
S.~A.~A.~Ghorashi, J.~F.~Karcher, S.~M.~Davis, and M. S. Foster,
\emph{Criticality across the energy spectrum from random artificial gravitational lensing in two-dimensional Dirac superconductors},
\href{https://journals.aps.org/prb/abstract/10.1103/PhysRevB.101.214521}{Phys. Rev. B \textbf{101}, 214521 (2020).} \\

\bibitem{Davis:2022}
S.~M.~Davis and M.~S.~Foster,
\emph{Geodesic geometry of 2+1-D Dirac materials subject to artificial, quenched gravitational singularities},
\href{https://scipost.org/SciPostPhys.12.6.204}{SciPost Phys. 12, 204 (2022).} \\

\bibitem{Hu:2018psq}
J.~Hu, L.~Feng, Z.~Zhang and C.~Chin,
\emph{Quantum simulation of Unruh radiation},
\href{https://www.nature.com/articles/s41567-019-0537-1}{{Nature Phys. \textbf{15}, 785 (2019).}} \\

\bibitem{Rodriguez-Laguna:2016kri} 
J.~Rodr\'iguez-Laguna, L.~Tarruell, M.~Lewenstein and A.~Celi,
\emph{Synthetic Unruh effect in cold atoms},
\href{https://journals.aps.org/pra/abstract/10.1103/PhysRevA.95.013627}{{Phys.\ Rev.\ A {\bf 95}, 013627 (2017).}} \\

\bibitem{Kosior:2018vgx}
A.~Kosior, M.~Lewenstein and A.~Celi,
\emph{Unruh effect for interacting particles with ultracold atoms},
\href{https://scipost.org/10.21468/SciPostPhys.5.6.061}{{SciPost Phys. \textbf{5}, 061 (2018).}} \\
	
\bibitem{Boada:2010sh}
O.~Boada, A.~Celi, J.~I.~Latorre and M.~Lewenstein,
\emph{Dirac Equation For Cold Atoms In Artificial Curved Spacetimes},
\href{https://iopscience.iop.org/article/10.1088/1367-2630/13/3/035002}{{New J. Phys. \textbf{13}, 035002 (2011).}} \\

\bibitem{Iorio:2011yz}
A.~Iorio and G.~Lambiase,
\emph{The Hawking-Unruh phenomenon on graphene},
\href{https://www.sciencedirect.com/science/article/pii/S037026931200860X?via%3Dihub}{{Phys. Lett. B \textbf{716}, 334 (2012).}} \\

\bibitem{Cvetic:2012vg}
M.~Cvetic and G.~W.~Gibbons,
\emph{Graphene and the Zermelo Optical Metric of the BTZ Black Hole},
\href{https://www.sciencedirect.com/science/article/pii/S0003491612000814?via%3Dihub}{{Annals of Phys. \textbf{327}, 2617 (2012).}} \\

\bibitem{Iorio:2013ifa}
A.~Iorio and G.~Lambiase,
\emph{Quantum field theory in curved graphene spacetimes, Lobachevsky geometry, Weyl symmetry, Hawking effect, and all that},
\href{https://journals.aps.org/prd/abstract/10.1103/PhysRevD.90.025006}{{Phys. Rev. D \textbf{90}, 025006 (2014).}} \\

\bibitem{Hegde:2018xub}
S.~S.~Hegde, V.~Subramanyan, B.~Bradlyn and S.~Vishveshwara,
\emph{Quasinormal Modes and the Hawking-Unruh Effect in Quantum Hall Systems: Lessons from Black Hole Phenomena},
\href{https://journals.aps.org/prl/abstract/10.1103/PhysRevLett.123.156802}{{Phys. Rev. Lett. \textbf{123}, 156802 (2019).}} \\

\bibitem{Subramanyan:2020fmx}
V.~Subramanyan, S.~S.~Hegde, S.~Vishveshwara and B.~Bradlyn,
\emph{Physics of the Inverted Harmonic Oscillator: From the lowest Landau level to event horizons},
\href{https://www.sciencedirect.com/science/article/abs/pii/S0003491621000762?via%3Dihub}{{Annals Phys. \textbf{435}, 168470 (2021).}} \\

\bibitem{Volovik:2016kid}
G.~E.~Volovik,
\emph{Black hole and Hawking radiation by type-II Weyl fermions},
\href{https://link.springer.com/article/10.1134/S0021364016210050}{{JETP Lett. \textbf{104}, (2016)}} \\

\bibitem{Wallace:1947} 
P.~R.~Wallace,
\emph{The Band Theory of Graphite},
\href{https://journals.aps.org/pr/abstract/10.1103/PhysRev.71.622}{{Phys. Rev. \textbf{71}, 622 (1947).}} \\

\bibitem{Semenoff} 
G.~W.~Semenoff,
\emph{Condensed-Matter Simulation of a Three-Dimensional Anomaly},
\href{https://journals.aps.org/prl/abstract/10.1103/PhysRevLett.53.2449}{{Phys. Rev. Lett. {\bf 53}, 2449 (1984).}} \\

\bibitem{Novoselov 2004} 
K.~S.~Novoselov, A.~K.~Geim, S.~V.~Morozov, D.~Jiang, Y.~Zhang, S.~V.~Dubonos, I.~V.~Grigorieva, and A.~A.~Firsov ,
\emph{Electric Field Effect in Atomically Thin Carbon Films},
\href{https://www.science.org/doi/10.1126/science.1102896}{{Science {\bf 306}, 666, (2004).}} \\

\bibitem{Novoselov 2005} 
K.~S.~Novoselov, A.~K.~Geim, S.~V.Morozov, D.~Jiang, M.~I.~Katsnelson, I.~V.~Grigorieva, S.~V.~Dubonos \& A.~A.~Firsov,
\emph{Two-dimensional gas of massless Dirac fermions in graphene},
\href{https://www.nature.com/articles/nature04233}{{Nature {\bf 438}, 197 (2005)}} \\

\bibitem{Lee:2009}
K.~L.~Lee, B.~Gremaud, R.~Han, B.~Englert and C.~Miniatura,
\emph{Ultracold fermions in a graphene-type optical lattice},
\href{https://journals.aps.org/pra/abstract/10.1103/PhysRevA.80.043411}{{Phys. Rev. A \textbf{80}, 043411 (2009).}} \\

\bibitem{Soltan-Panahi 2011}
P.~Soltan-Panahi, J.~Struck, P.~Hauke, A.~Bick, W.~Plenkers, G.~Meineke, C.~Becker, P.~Windpassinger, M.~Lewenstein \& K.~Sengstock,
\emph{Multi-component quantum gases in spin-dependent hexagonal lattices},
\href{https://www.nature.com/articles/nphys1916#citeas}{{Nature Phys. {\bf 7}, 434 (2011).}} \\

\bibitem{Soltan-Panahi 2012}
P.~Soltan-Panahi, D.-S.~L\"{u}hmann, J.~Struck, P.~Windpassinger, \& K.~Sengstock,
\emph{Quantum phase transition to unconventional multi-orbital superfluidity in optical lattices},
\href{https://www.nature.com/articles/nphys2128}{{Nature Phys. {\bf 8}, 71 (2012).}} \\

\bibitem{Tarruell:2012zz}
L.~Tarruell, D.~Greif, T.~Uehlinger, G.~Jotzu and T.~Esslinger,
\emph{Creating, moving and merging Dirac points with a Fermi gas in a tunable honeycomb lattice},
\href{https://www.nature.com/articles/nature10871}{{Nature {\bf 483}, 302 (2012).}} \\

\bibitem{Li:2016}
J.~Li, W.~Huang, B.~Shteynas, S.~Burchesky, F.~Çağrı Top, E.~Su, J.~Lee, A.~Jamison, and W.~Ketterle,
\emph{Spin-Orbit Coupling and Spin Textures in Optical Superlattices}, 
\href{https://journals.aps.org/prl/abstract/10.1103/PhysRevLett.117.185301}{{Phys. Rev. Lett. \textbf{117}, 185301 (2016).}} \\

\bibitem{Nair2008}
R.~R.~Nair, P.~Blake, A.~N.~Grigorenko, K.~S.~Novoselov, T.~J.~Booth, T.~Stauber, N.~M.~R.~Peres, A.~K.~Geim, 
\emph{Fine Structure Constant Defines Visual Transparency of Graphene},
\href{https://www.science.org/doi/10.1126/science.1156965}{{Science \textbf{320}, 1308 (2008).}} \\

\bibitem{Mishchenko2008}
E.~G.~Mishchenko,
\emph{Minimal conductivity in graphene: Interaction corrections and ultraviolet anomaly},
\href{https://iopscience.iop.org/article/10.1209/0295-5075/83/17005}{{Europhys. Lett. \textbf{83}, 17005 (2008).}} \\

\bibitem{Sheehy2009}
D.~E.~Sheehy, J.~Schmalian,
\emph{Optical transparency of graphene as determined by the fine-structure constant},
\href{https://journals.aps.org/prb/abstract/10.1103/PhysRevB.80.193411}{{Phys. Rev. B \textbf{80}, 193411 (2009).}} \\

\bibitem{Link2016}
J.~M.~Link, P.~P.~Orth, D.~E.~Sheehy, and J.~Schmalian,
\emph{Universal collisionless transport of graphene},
\href{https://journals.aps.org/prb/abstract/10.1103/PhysRevB.93.235447}{{Phys. Rev. B \textbf{93}, 235447 (2016).}} \\

\bibitem{deJuan:2012hxm}
F.~de Juan, M.~Sturla and M.~A.~H.~Vozmediano,
\emph{Space dependent Fermi velocity in strained graphene},
\href{https://journals.aps.org/prl/abstract/10.1103/PhysRevLett.108.227205}{{Phys. Rev. Lett. \textbf{108}, 227205 (2012).}} \\

\bibitem{Jimenez-Galan:2020}
Á.~Jiménez-Galán, R.~E.~F.~Silva, O.~Smirnova and M.~Ivanov,
\emph{Lightwave control of topological properties in 2D materials for sub-cycle and non-resonant valley manipulation},
\href{https://www.nature.com/articles/s41566-020-00717-3}{Nat. Photonics 14, 728–732 (2020).} \\

\bibitem{Castro Neto:2009}
A.~H.~Castro Neto, F.~Guinea, N.~M.~R.~Peres, K.~S.~Novoselov, A.K.~Geim,
\emph{The electronic properties of graphene}, 
\href{https://journals.aps.org/rmp/abstract/10.1103/RevModPhys.81.109}{{Rev. Mod. Phys. \textbf{81}, 109 (2009).}} \\

\bibitem{Das:2008zze}
A.~Das,
\emph{Lectures on quantum field theory}, World Scientific Publishing Company (2008). \\

\bibitem{Misner:1973prb}
C.~W.~Misner, K.~S.~Thorne and J.~A.~Wheeler,
\emph{Gravitation}, Princeton University Press (1972). \\

\bibitem{Wald:1984rg}
R.~M.~Wald,
\emph{General Relativity}, University of Chicago Press (1984). \\

\bibitem{Rindler:2006km}
W.~Rindler,
\emph{Relativity: Special, general, and cosmological}, Oxford University Press (2006) \\

\bibitem{Unruh:1974}
W.~G.~Unruh, 
\emph{Alternative Fock Quantization of Neutrinos in Flat Space-Time},
\href{https://royalsocietypublishing.org/doi/10.1098/rspa.1974.0100}{{Proc. Roy. Soc. Lond. Series A, Mathematical and Physical Sciences \textbf{338}, (1974): 517. }} \\

\bibitem{Candelas:1978gg}
P.~Candelas and D.~Deutsch,
\emph{Fermion Fields in Accelerated States},
\href{https://royalsocietypublishing.org/doi/10.1098/rspa.1978.0132}{{Proc. Roy. Soc. Lond. A \textbf{362}, 251 (1978).}} \\

\bibitem{Soffel:1980kx}
M.~Soffel, B.~Muller and W.~Greiner,
\emph{Dirac Particles in Rindler Space},
\href{https://journals.aps.org/prd/abstract/10.1103/PhysRevD.22.1935}{{Phys. Rev. D \textbf{22}, 1935 (1980).}} \\

\bibitem{Hughes:1983ch}
R.~J.~Hughes,
\emph{Uniform Acceleration and the Quantum Field Theory Vacuum. 1},
\href{https://www.sciencedirect.com/science/article/abs/pii/0003491685902246?via%3Dihub}{{Annals of Phys. \textbf{162}, 1 (1985).}} \\

\bibitem{Iyer:1985ufr}
B.~R.~Iyer,
\emph{Dirac equation in Kasner spacetime with local rotational symmetry},
\href{https://www.sciencedirect.com/science/article/abs/pii/0375960185903482?via%3Dihub}{{Phys. Lett. A \textbf{112}, 313 (1985).}} \\

\bibitem{Jauregui:1991me}
R.~J\'auregui, M.~Torres and S.~Hacyan,
\emph{Dirac vacuum: Acceleration and external field effects},
\href{https://journals.aps.org/prd/abstract/10.1103/PhysRevD.43.3979}{{Phys. Rev. D \textbf{43}, 3979 (1991).}} \\

\bibitem{Gradshteyn} 
I.~S.~Gradshteyn, I.~M.~Ryzhik, Alan Jeffrey, and Daniel Zwillinger,
\emph{Table of Integrals, Series, and Products}, Academic Press; 6th edition (2000). \\

\bibitem{Hung:2012nc}
C.~L.~Hung, V.~Gurarie and C.~Chin,
\emph{From Cosmology to Cold Atoms: Observation of Sakharov Oscillations in Quenched Atomic Superfluids},
\href{https://www.science.org/doi/10.1126/science.1237557}{Science \textbf{341}, 1213-1215 (2013).} \\

\bibitem{Chen:2021xhd}
C.~A.~Chen, S.~Khlebnikov and C.~L.~Hung,
\emph{Observation of Quasiparticle Pair Production and Quantum Entanglement in Atomic Quantum Gases Quenched to an Attractive Interaction},
\href{https://journals.aps.org/prl/abstract/10.1103/PhysRevLett.127.060404}{Phys. Rev. Lett. \textbf{127}, no.6, 060404 (2021).} \\

\bibitem{Alsing:2006cj}
P.~M.~Alsing, I.~Fuentes-Schuller, R.~B.~Mann and T.~E.~Tessier,
\emph{Entanglement of Dirac fields in non-inertial frames},
\href{https://journals.aps.org/pra/abstract/10.1103/PhysRevA.74.032326}{{Phys. Rev. A \textbf{74}, 032326 (2006).}} \\

\bibitem{Leon:2009uod}
J.~Leon and E.~Martin-Martinez,
\emph{Spin and occupation number entanglement of Dirac fields for noninertial observers},
\href{https://journals.aps.org/pra/abstract/10.1103/PhysRevA.80.012314}{{Phys. Rev. A \textbf{80}, 012314 (2009).}} \\

\bibitem{BCS Short}
J.~Bardeen, L.~N.~Cooper, and J.~R.~Schrieffer,
\emph{Microscopic Theory of Superconductivity},
\href{https://journals.aps.org/pr/abstract/10.1103/PhysRev.106.162}{{Phys. Rev. \textbf{106}, 162 (1957).}} \\

\bibitem{BCS Long}
J.~Bardeen, L.~N.~Cooper, and J.~R.~Schrieffer,
\emph{Theory of Superconductivity},
\href{https://journals.aps.org/pr/abstract/10.1103/PhysRev.108.1175}{{Phys. Rev. \textbf{108}, 1175 (1957).}} \\

\bibitem{Cooper}
Leon N.~Cooper,
\emph{Bound Electron Pairs in a Degenerate Fermi Gas},
\href{https://journals.aps.org/pr/abstract/10.1103/PhysRev.104.1189}{{Phys. Rev. \textbf{104}, 1189 (1956).}} \\

\bibitem{DeWitt:1979} 
B.~S.~DeWitt, \emph{Quantum gravity: the new synthesis},
in General relativity: an Einstein centenary survey, S.W. Hawking and W. Israel eds., Cambridge University Press, Cambridge U.K. (1979), pg. 680. \\

\bibitem{Coleman} 
Piers Coleman, 
\emph{Introduction to Many-Body Physics}, Cambridge University Press; 1st edition (February 1, 2016) . \\

\bibitem{Matsubara:1955ws}
T.~Matsubara,
\emph{A New approach to quantum statistical mechanics},
\href{https://academic.oup.com/ptp/article/14/4/351/1869035?login=true}{{Prog. Theor. Phys. \textbf{14}, 351 (1955)}}. \\

\bibitem{Boltzmann}
L.~Boltzmann,
\emph{Lectures on gas theory},
Berkeley, CA, USA: U. of California Press, (1964). \\

\bibitem{Tolman} 
R.~C.~Tolman, 
\emph{The Principles of Statistical Mechanics}, Oxford University Press, London, UK, (1938). \\

\bibitem{Einstein}
A.~Einstein,
Zur Quantentheorie der Strahlung [\emph{On the quantum theory of radiation}], Physikalische Zeitschrift 18 (1917), 121-128. English translation: D. ter Haar (1967): The Old Quantum Theory. Pergamon Press, pp. 167. \\

\bibitem{RammerSmith1986}
J.~Rammer and H.~Smith,
\emph{Quantum field-theoretical methods in transport theory of metals},
\href{https://journals.aps.org/rmp/abstract/10.1103/RevModPhys.58.323}{{Rev. Mod. Phys. \textbf{58}, 323 (1986).}} \\

\bibitem{Courant-Hilbert} 
R.~Courant and D.~Hilbert (1962), 
\emph{Methods of Mathematical Physics}, Interscience Publishers, vol. II . \\

\bibitem{Mahan} 
G.~D.~Mahan,
\emph{Many Particle Physics, Plenum}, New York, (1990). \\

\bibitem{Gusynin:2007}
V.~P.~Gusynin, S.~G.~Sharapov, and J.~P.~Carbotte,
\emph{Sum rules for the optical and Hall conductivity in graphene}, 
\href{https://journals.aps.org/prb/abstract/10.1103/PhysRevB.75.165407}{Phys. Rev. B \textbf{75}, 165407 (2007).} \\

\bibitem{Fritz:2008}
L.~Fritz, J.~Schmalian, M.~Müller, and S.~Sachdev,
\emph{Quantum critical transport in clean graphene}, 
\href{https://journals.aps.org/prb/abstract/10.1103/PhysRevB.78.085416}{Phys. Rev. B \textbf{78}, 085416 (2008).} \\

\bibitem{Stauber:2008}
T.~Stauber, N.~M.~R.~Peres, and A.~K.~Geim,
\emph{Optical conductivity of graphene in the visible region of the spectrum}, 
\href{https://journals.aps.org/prb/abstract/10.1103/PhysRevB.78.085432}{Phys. Rev. B \textbf{78}, 085432 (2008).} \\

\bibitem{Dao:2007}
T.~Dao, A.~Georges, J.~Dalibard, C.~Salomon, and I.~Carusotto,
\emph{Measuring the One-Particle Excitations of Ultracold Fermionic Atoms by Stimulated Raman Spectroscopy}, 
\href{https://journals.aps.org/prl/abstract/10.1103/PhysRevLett.98.240402}{Phys. Rev. Lett. \textbf{98}, 240402 (2007).} \\

\bibitem{Stewart:2008}
J.~T.~Stewart, J.~P.~Gaebler, and D.~S.~Jin,
\emph{Using photoemission spectroscopy to probe a strongly interacting Fermi gas}, 
\href{https://www.nature.com/articles/nature07172}{Nature \textbf{454}, 744–747 (2008).} \\

\bibitem{Parker:1971pt}
L.~Parker,
\emph{Quantized fields and particle creation in expanding universes. 2},
\href{https://journals.aps.org/prd/abstract/10.1103/PhysRevD.3.346}{{Phys. Rev. D \textbf{3}, 346 (1971)}}. \\

\bibitem{Hu:1986jd}
B.~L.~Hu and D.~Pavon,
\emph{Intrinsic Measures of Field Entropy in Cosmological Particle Creation},
\href{https://www.sciencedirect.com/science/article/abs/pii/0370269386911974?via%3Dihub}{{Phys. Lett. B \textbf{180}, 329 (1986)}}. \\

\bibitem{Hu:1986jj}
B.~L.~Hu and H.~E.~Kandrup,
\emph{Entropy Generation in Cosmological Particle Creation and Interactions: A Statistical Subdynamics Analysis},
\href{https://journals.aps.org/prd/abstract/10.1103/PhysRevD.35.1776}{{Phys. Rev. D \textbf{35}, 1776 (1987)}.} \\

\bibitem{Kandrup:1988sg}
H.~E.~Kandrup,
\emph{Entropy Generation, Particle Creation, and Quantum Field Theory in a Cosmological Space-time: When Do Number and Entropy Increase?},
\href{https://journals.aps.org/prd/abstract/10.1103/PhysRevD.37.3505}{{Phys. Rev. D \textbf{37}, 3505 (1988).}} \\

\bibitem{Unruh:1983ac}
W.~G.~Unruh and N.~Weiss,
\emph{Acceleration Radiation in Interacting Field Theories},
\href{https://journals.aps.org/prd/abstract/10.1103/PhysRevD.29.1656}{Phys. Rev. D \textbf{29}, 1656 (1984).}







	
\end{thebibliography}

\appendix

\section{The Dirac Equation}

\label{SEC:Appendix Dirac Eqn}

In this section, we will investigate how fermionic quantum fields evolve in a $(2+1)$-dimensional spacetime equipped with the following line element:
\begin{eqnarray}\label{General Metric}
	ds^{2} = -\bigg(1+\frac{|x|}{\lambda}\bigg)^{2}c^{2}dt^{2} + dx^{2} + dy^{2},
\end{eqnarray}
which is written in some coordinates $(t,x,y)$ whose interpretation depends on the choice of parameter $\lambda$. The limit in which it diverges, i.e. $\lambda\rightarrow\infty$, we recover the flat Minkowski metric expressed in inertial coordinates $(t,x,y)$, which we could also re-label with $(T,X,Y)$ as was done in {(\ref{Minkowski Metric})}:
\begin{eqnarray}\label{Inertial Coordinates}
\lim\limits_{\lambda\rightarrow\infty}	ds^{2} = -c^{2}dt^{2} + dx^{2} + dy^{2},
\end{eqnarray}
whereas in the opposite limit where this parameter is small, i.e. $\lambda\rightarrow0$, we recover the flat Minkowski metric written in terms of the Rindler coordinates $(t,x,y)$:
\begin{eqnarray}\label{Rindler Coordinates}
\lim\limits_{\lambda\rightarrow0}	ds^{2} = -\frac{x^{2}}{\lambda^{2}}c^{2}dt^{2} + dx^{2} + dy^{2},
\end{eqnarray}
and thus here $\lambda$ plays the role of $x_{\text{min}}=\frac{c^{2}}{a}$ which is the closest distance of approach from the origin at $x=0$, of a Rindler observer accelerating with $a$. To derive the Dirac equation in these two limits, we will write it using the most general metric {(\ref{General Metric})}. The Dirac equation describing the evolution for massless or Weyl fermions in arbitrary spacetime is as follows:
\begin{equation}\label{Dirac Equation Curved Spacetime}
	i\gamma^{a}e^{\mu}_{a}\nabla_{\mu}\hat{\psi}(x) = 0
\end{equation}
where $\hat{\psi}(x)$ is the massless Dirac spinor (or Weyl spinor) and can be written as a two component spinor, which due to zero rest mass are decoupled from each other $\hat{\psi}^{\text{T}}(x)=\begin{bmatrix}
	\hat{\psi}_{\text{R}}(x),~\hat{\psi}_{\text{L}}(x)
\end{bmatrix}$. Also, the covariant derivative is defined as  $\nabla_{\mu}=\partial_{\mu}-\frac{i}{4}\omega^{ab}_{\mu}\sigma_{ab}$, where $\sigma_{ab}=\frac{i}{2}\big[\gamma_{a},\gamma_{b}\big]$, where the Dirac matrices satisfy the Clifford algebra $\big\{\gamma^{a},\gamma^{b}\big\}=2\eta^{ab}$. From the line element in {(\ref{General Metric})}, we can write down the metric components as follows:
\begin{eqnarray}\label{Metric Components}
g_{\mu\nu}=\text{diag}\bigg[-c^{2}\bigg(1+\frac{|x|}{\lambda}\bigg)^{2},1,1\bigg],
\end{eqnarray}
which is diagonal and hence simplifies our derivation. Tetrads are objects that take us from an arbitrary metric to the local flat metric of the tangent space at a point. They are as defined as:
\begin{equation}\label{Tetrads&Minkowski}
	g_{\mu\nu}=e^{a}_{\mu}e^{b}_{\nu}\eta_{ab},~~~\eta_{ab} = \begin{bmatrix}
		-1 & 0 & 0 & 0 \\
		0 & 1 & 0 & 0 \\
		0 & 0 & 1 & 0 \\
		0 & 0 & 0 & 1 \\
	\end{bmatrix}.
\end{equation}
where $\eta_{ab}$ is the Minkowski tensor. In our notation we make use of Greek $(\mu, \nu,..)$ indices to denote curved spacetime labels such as $(t,x,y)$, and Roman $(a, b,..)$ indices signify that we are in the tangent space at a particular point in spacetime and therefore can take values $(0,1,2)$. By comparing {(\ref{Metric Components})} and {(\ref{Tetrads&Minkowski})} we get the following tetrads:
\begin{eqnarray}\label{RindlerTetrads}
	e^{a}_{\mu}=\text{diag}\bigg[c\bigg(1+\frac{|x|}{\lambda}\bigg),1,1\bigg],
\end{eqnarray}
These tetrads can now be used to derive the spin connections $\omega_{\mu}^{ab}$ which take into account the spin-precession of fermions due to the curvature of spacetime. They are defined as follows:
\begin{eqnarray}
\omega_{\mu}^{ab} & = & \frac{1}{2}e^{\nu a}\big(\partial_{\mu}e^{b}_{\nu}-\partial_{\nu}e^{b}_{\mu}\big) - \frac{1}{2}e^{\nu b}\big(\partial_{\mu}e^{a}_{\nu}-\partial_{\nu}e^{a}_{\mu}\big) \nonumber \\
& - & \frac{1}{2}e^{\rho a}e^{\sigma b}\big(\partial_{\rho}e_{\sigma c}-\partial_{\sigma}e_{\rho c}\big)e^{c}_{\mu}.
\end{eqnarray}
which is manifestly anti-symmetric $\omega_{\mu}^{ab}=-\omega_{\mu}^{ba}$. The only surviving components of the spin connection in the metric of {(\ref{General Metric})} are as follows:
\begin{eqnarray}\label{Spin Connection}
	\omega_{t}^{01} = -\omega_{t}^{10} & = & \frac{c}{\lambda}\text{sgn}(x),
\end{eqnarray}
where $\text{sgn}(x)$ is the signum function, i.e. it returns $+1$ for positive values and $-1$ for negative entries. Finally, we will be needing the Dirac matrices in the Weyl or Chiral representation:
\begin{equation}\label{GammaMatrices}
	\gamma^{0} = \begin{bmatrix}
		0 & -I_{2} \\
		-I_{2} & 0 \\
	\end{bmatrix},~~~
	\gamma^{i} = \begin{bmatrix}
		0 & \sigma^{i} \\
		-\sigma^{i} & 0 \\
	\end{bmatrix},~~~
	\gamma^{5} = \begin{bmatrix}
		I_{2} & 0 \\
		0 & I_{2} \\
	\end{bmatrix},
\end{equation}
where $\sigma^{i}$ are the Pauli matrices:
\begin{equation}\label{PauliMatrices}
	\sigma^{1} = \begin{bmatrix}
		0 & 1 \\
		1 & 0 \\
	\end{bmatrix},~~~
	\gamma^{2} = \begin{bmatrix}
		0 & -i \\
		i & 0 \\
	\end{bmatrix},~~~
	\sigma^{3} = \begin{bmatrix}
		1 & 0 \\
		0 & -1 \\
	\end{bmatrix}.
\end{equation}
Plugging in the tetrads {(\ref{RindlerTetrads})} and the spin-connection {(\ref{Spin Connection})} pertaining to the metric {(\ref{General Metric})} into the massless Dirac equation {(\ref{Dirac Equation Curved Spacetime})}, we get two decoupled Weyl equations for the left and right handed fermions:
\begin{eqnarray}\label{WeylEquations}
\partial_{t}\hat{\psi}_{\text{L}} & = & c\bigg(1+\frac{|x|}{\lambda}\bigg)\boldsymbol{\sigma}\cdot\boldsymbol{\nabla}\hat{\psi}_{\text{L}} + \frac{c~\text{sgn}(x)}{2\lambda}\sigma^{x}\hat{\psi}_{\text{L}}, \nonumber \\
\partial_{t}\hat{\psi}_{\text{R}} & = &  -c\bigg(1+\frac{|x|}{\lambda}\bigg)\boldsymbol{\sigma}\cdot\boldsymbol{\nabla}\hat{\psi}_{\text{R}} -\frac{c~\text{sgn}(x)}{2\lambda}\sigma^{x}\hat{\psi}_{\text{R}}.~~~~~~~~~~
\end{eqnarray}
In the limit of $\lambda\rightarrow\infty$, the above set reduces to the Weyl equations for fermions in inertial frames {(\ref{Inertial Coordinates})}:
\begin{eqnarray}\label{Weyl Equations Inertial Coordinates}
\partial_{t}\hat{\psi}_{\text{L}} & = & c\boldsymbol{\sigma}\cdot\boldsymbol{\nabla}\hat{\psi}_{\text{L}}, \nonumber \\
\partial_{t}\hat{\psi}_{\text{R}} & = &  -c\boldsymbol{\sigma}\cdot\boldsymbol{\nabla}\hat{\psi}_{\text{R}},
\end{eqnarray}
which is the same as {(\ref{Dirac Equation Flat Graphene})} describing massless fermions in a flat graphene sheet. In the opposite limit $\lambda\rightarrow0$, we recover the Weyl equations for massless fermions in uniformly accelerating frames:
\begin{eqnarray}\label{Weyl Equations Rindler Coordinates}
\partial_{t}\hat{\psi}_{\text{L}} & = & \frac{c|x|}{\lambda}\boldsymbol{\sigma}\cdot\boldsymbol{\nabla}\hat{\psi}_{\text{L}} + \frac{c~\text{sgn}(x)}{2\lambda}\sigma^{x}\hat{\psi}_{\text{L}}, \nonumber \\
\partial_{t}\hat{\psi}_{\text{R}} & = &  -\frac{c|x|}{\lambda}\boldsymbol{\sigma}\cdot\boldsymbol{\nabla}\hat{\psi}_{\text{R}} -\frac{c~\text{sgn}(x)}{2\lambda}\sigma^{x}\hat{\psi}_{\text{R}},
\end{eqnarray}
which is the same as Eq.~{(\ref{Dirac Equation Rindler Graphene})} that describes how electrons and holes evolve in a Rindler strained graphene sheet. \\

\section{Rindler Horizon in Strained Graphene}

\label{SEC:Appendix Horizon Graphene}

In this section, we will discuss how an effective Rindler horizon
forms at
$x=0$ in a graphene sheet that is strained according to
Eq.~(\ref{RindlerStrainPattern}). Our aim is to show that, for
long-wavelength excitations ($k_y\lambda \ll 1$, with $\lambda$ the
parameter characterizing the strain), excitations in graphene
obey an effective Schr\"odinger equation with a potential that diverges
for $x\to 0$.

To illustrate this, we start with the
Hamiltonian in Eq.~(\ref{Eq:fullHAM}) that results in the following
Dirac equation for right-handed massless Dirac fermions:
\begin{equation}\label{Dirac Equation Full}
	i\hbar\partial_{t}\hat{\psi}_{\text{R}}(\boldsymbol{r}) =
	\sqrt{v(x)} \boldsymbol{\sigma}\cdot\hat{\boldsymbol{p}} \sqrt{v(x)}
	~\hat{\psi}_{\text{R}}(\boldsymbol{r}),
\end{equation}
where $v(x)=v_{0}\big(1+\frac{|x|}{\lambda}\big)$ is the Fermi velocity.
Next, we assume the following ansatz:
\begin{equation}
	\hat{\psi}_{\text{R}} = \begin{pmatrix}
		f(x) \\
		g(x)
	\end{pmatrix}e^{i(k_{y}y-\Omega t)}, 
\end{equation}
which is a product of a spinor with component functions $f(x)$ and
$g(x)$, and a plane wave in $y$ and $t$. Substituting this ansatz into
Eq.~(\ref{Dirac Equation Full}), we get a system of coupled
differential equations:
\bse
\label{ODE for f & g}
\bea
\label{ODE for f}
\bigg[\big(|x|+\lambda\big)\frac{d}{dx}+k_{y}\big(|x|+\lambda\big)+\frac{\text{sgn}(x)}{2}\bigg]g & = & i\bar{\Omega} f,~~~~~ \\ \label{ODE for g}
\bigg[\big(|x|+\lambda\big)\frac{d}{dx}-k_{y}\big(|x|+\lambda\big)+\frac{\text{sgn}(x)}{2}\bigg]f & = & i\bar{\Omega} g,~~~~~
\eea
\ese
where we have defined $\bar{\Omega}=\Omega/\omega_{c}$.
%
%
Assuming $x>0$, we eliminate
$g(x)$ in favor of $f(x)$.  Defining 
a new function $F(x)=f(x)/(x+\lambda)$, we arrive at
the following second order
differential equation:
\begin{equation}\label{Schrodinger Like Eqn}
	\frac{d^{2}F(x)}{dx^{2}} - \big[k_{y}^{2} - V(x)\big]F(x) = 0,
\end{equation}
%
%
taking the form of an effective Schr\"odinger equation with
the single-particle potential: 
%
%
\begin{equation}
	V(x) = -\frac{k_{y}}{(x+\lambda)} + \frac{1/4+\bar{\Omega}^{2}}{(x+\lambda)^{2}}.
\end{equation}
We now re-scale the coordinates $x\to x/k_{y}$ and take the
long-wavelength limit, i.e., $k_{y}\lambda\ll1$, which makes
Eq.~(\ref{Schrodinger Like Eqn}) take the following form:
\begin{equation}\label{Rescaled Schrodinger}
	\frac{d^{2}F(x)}{dx^{2}} - \bigg[1 + \frac{1}{x} -
	\frac{1/4+\bar{\Omega}^{2}}{x^{2}}\bigg]F(x) = 0,
\end{equation}
exhibiting a divergence at $x\to 0$.  This shows that in the presence
of an appropriate strain field, electronic quasiparticles in graphene
experience an effective infinite potential barrier reflecting the
Rindler horizon.


\end{document}